\documentclass[numberedappendix]{emulateapj}

\usepackage{xspace}
\usepackage{color}
\usepackage{amsmath}
\usepackage{rotating}
\usepackage{subfigure}
\usepackage{afterpage}
\usepackage{hyperref}
\usepackage[ampersand]{easylist}

\PassOptionsToPackage{hyphens}{url}\usepackage{hyperref}
\newcommand{\occulted}[0]{\ensuremath{\mathcal{P}}\xspace}
\newcommand{\occultor}[0]{\ensuremath{\mathcal{O}}\xspace}
\newcommand{\batman}[0]{\texttt{batman}\xspace}

\newcommand{\nudge}[0]{\vspace*{0.1in}}
\newcommand{\github}{\href{https://github.com/rodluger/planetplanet}{\ttfamily\fontseries{b}\selectfont github}\xspace}
\newcommand{\planetplanet}{\href{https://github.com/rodluger/planetplanet}{\ttfamily\fontseries{b}\selectfont planetplanet}\xspace}

\newcommand{\edited}[1]{{{#1}}}
\newcommand{\bugfix}[1]{{{#1}}}

\slugcomment{Draft version \today}

\shorttitle{Planet-Planet Occultations}
\shortauthors{Luger, Lustig-Yaeger and Agol (2017)}

\begin{document}

\title{Planet-Planet Occultations in TRAPPIST-1 and Other Exoplanet Systems}

\author{Rodrigo Luger\altaffilmark{1,2,3,4}}
\author{Jacob Lustig-Yaeger\altaffilmark{1,2,3}}
\author{Eric Agol\altaffilmark{1,3\edited{,5}}}

\altaffiltext{1}{Astronomy Department, University of Washington, Box 951580, Seattle, WA 98195, USA}
\altaffiltext{2}{Astrobiology Program, University of Washington, 3910 15th Ave. NE, Box 351580, Seattle, WA 98195, USA}
\altaffiltext{3}{NASA Astrobiology Institute -- Virtual Planetary Laboratory Lead Team, USA}
\altaffiltext{4}{\email{rodluger@uw.edu}}
\edited{\altaffiltext{5}{Guggenheim Fellow}}


\begin{abstract}
We explore the occurrence and detectability of planet-planet occultations (PPOs) in exoplanet systems. These are events during which a planet occults the disk of another planet in the same system, imparting a small photometric signal as its thermal or reflected light is blocked. We focus on the planets in TRAPPIST-1, whose orbital planes we show are aligned to $<0.3^\circ$ at 90\% confidence. We present a photodynamical model for predicting and computing PPOs in TRAPPIST-1 and other systems for various assumptions of the planets' atmospheric states. When marginalizing over the uncertainties on all orbital parameters, \edited{we find that the rate of PPOs in TRAPPIST-1 is about 1.4 per day}. We investigate the prospects for detection of these events with the James Webb Space Telescope, \bugfix{finding that ${\sim}10-20$ occultations per year of b and c should be above the noise level at $12-15\ \mu\mathrm{m}$. Joint modeling of several of these PPOs could lead to a robust detection}. Alternatively, observations with the proposed Origins Space Telescope should be able to detect individual PPOs at high signal-to-noise. We show how PPOs can be used to break transit timing variation degeneracies, imposing strong constraints on the eccentricities and masses of the planets, as well as to constrain the longitudes of nodes and thus the complete three-dimensional structure of the system. We further show how modeling of these events can be used to reveal a planet's day/night temperature contrast and construct crude surface maps. We make our photodynamical code available on \github.
\end{abstract}


\section{Introduction}
\label{sec:intro}
Among the most remarkable accomplishments of the \emph{Kepler} telescope was the discovery of a large population of tightly-packed, close-in, and highly coplanar multi-planet exoplanet systems. These include Kepler-32 \citep{Swift2013}, Kepler-444 \citep{Campante2015}, and Kepler-80 \citep{MacDonald2016}, all of which host five transiting planets, although dozens of other such systems with three or more planets are known \citep{Borucki2011,Lissauer2011}.

These Kepler systems were bested by the discovery of the TRAPPIST-1 system \citep{Gillon2016,Gillon2017}, which hosts {\it seven} transiting terrestrial-size planets within less than 0.1 AU of their star, an ultracool dwarf only 12 pc away. The system is so packed that a transit occurs, on average, about 6\% of the time, placing the TRAPPIST-1 light curves among the most information-rich transiting exoplanet datasets. Transit timing variation (TTV) analyses of these light curves have so far placed strong constraints on the masses and eccentricities of these planets \citep{Gillon2017} and dynamical studies of resonances in the system led to the precise prediction of the period of TRAPPIST-1h, the farthest-out planet \citep{Luger2017}. A spectroscopic analysis of the transit light curves of TRAPPIST-1b and c with the Hubble Space Telescope (HST) has further ruled out cloud-free hydrogen-dominated atmospheres for those planets \citep{deWit2016}. In the near future, the James Webb Space Telescope (JWST) is expected to detect secondary eclipses of the shortest-period TRAPPIST-1 planets and to reveal more detailed information about their atmospheres via transit transmission and secondary eclipse spectroscopy (\citealp{Barstow2016}; \edited{\citealp{Morley2017}}; Lustig-Yaeger et al., in preparation).

As new generations of telescopes continue to enable the detection of smaller signals, it is worthwhile to explore new methods to study exoplanets. In this work, we consider the detectability of planet-planet occultations \citep[PPOs;][]{Ragozzine2010} in the TRAPPIST-1 system and in other systems of multiple, close-in transiting planets. PPOs occur when one planet transits the disk of another planet in the same system as seen by a distant observer, producing a dip in the light curve due to the interception of light emitted from or reflected off of the occulted planet. While PPOs are in this sense analogous to transits, their signals are orders of magnitude weaker, given the large star/planet surface brightness ratio. PPOs are also in general quite rare, given the small planet sizes (relative to the star) and the low probability of the overlap of their disks on the sky plane. For this reason they are also intermittent and typically short-lived.

\begin{figure*}[!ht]
\centering
\includegraphics[width=\textwidth]{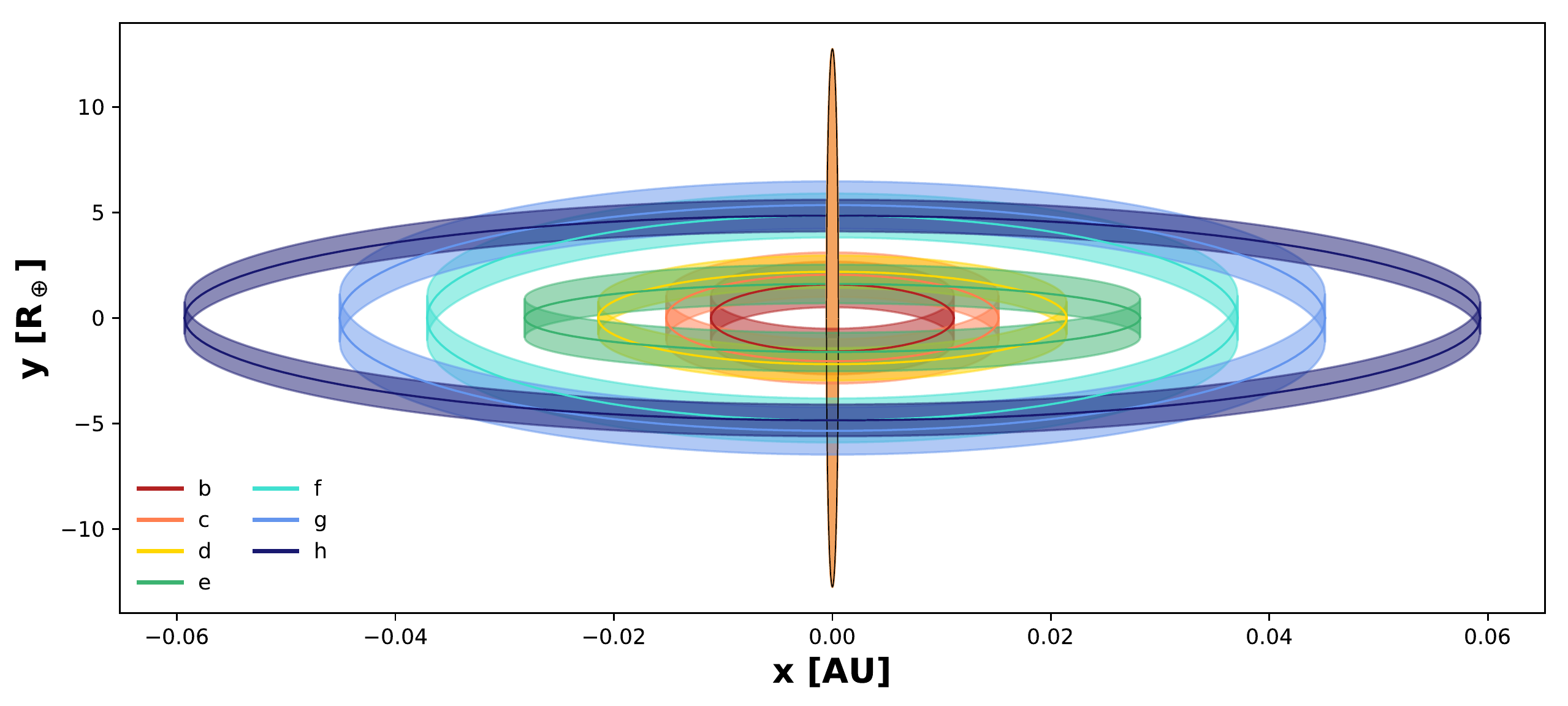}
\caption{The orbits of the seven planets in the TRAPPIST-1 system as seen from Earth, assuming the mean orbital parameters from \citet{Gillon2017} and \citet{Luger2017}. The thickness of each orbital track is the planet diameter. The aspect ratio of the plot is 100:1, but all horizontal and vertical distances are to scale. The star is shown in orange for reference. Because of the compactness of the system, its near edge-on orientation, and the dynamical coldness of the disk, the orbital paths of all planets overlap with those of their neighbors over a significant fraction of their orbits. For this particular configuration, planet-planet occultations occur between the set of planets $\{b, c, d, e\}$ and the set $\{f, g, h\}$.}
\label{fig:orbits}
\end{figure*}

However, the geometry of the orbits in TRAPPIST-1, and in several of the known multi-planet systems, is particularly favorable for planet-planet occultations. Because of the extreme coplanarity of the system, the orbital tracks of the seven TRAPPIST-1 planets overlap over considerable portions of their orbits as seen from Earth. In Figure~\ref{fig:orbits} we plot these tracks for the mean orbital parameters reported in \citet{Gillon2017} and \citet{Luger2017}. To make the orbital paths apparent, this figure is plotted with an aspect ratio of 100:1, but horizontal and vertical distances are each to scale. The star is shown in orange at the center and appears as a thin ellipse due to the vertical stretch. The widths of the orbital tracks are the planet diameters.

It is evident from the figure that the orbital paths of all seven TRAPPIST-1 planets overlap with those of their neighbors over large fractions of their orbits. Even when the observational uncertainties in the planet inclinations, radii, semi-major axes, and eccentricities are accounted for, the orbital tracks of all planets and those of their neighbors overlap for virtually any configuration of the system allowed by the orbital constraints. The figure assumes the planets have the same longitude of ascending node $\Omega$, which for an edge-on system is the polar angle of the orbit in the $x$-$y$ plane. While large scatter in the values of $\Omega$ for the TRAPPIST-1 planets could disrupt the alignment of their orbits, we show that such configurations can be ruled out at high confidence with minimal assumptions (see \S\ref{sec:results:dynamics:coplanarity}).

The crossing orbital tracks suggest PPOs may be common among the TRAPPIST-1 planets and among planets in other compact, coplanar multi-planet systems. In particular, orbital crossings on the same side of the star can lead to long-lived PPOs, given the low sky-projected relative velocity of prograde, neighboring planets---in many cases, PPOs among these systems last significantly longer than transit or secondary eclipse. Furthermore, because the signal of a PPO event is the missing flux from a planet as it is occulted, complete planet-planet occultations can be as deep as secondary eclipses. For the TRAPPIST-1 system, PPOs are most detectable in emission at long wavelengths, where the planet/star contrast is most favorable\edited{, approaching} the linear relationship between flux and temperature in the Rayleigh-Jeans limit.


The detection of one or more PPOs in a multi-planet system pins down the relative orbital positions of pairs of planets to extremely high precision, placing strong constraints on the mutual inclinations of the planets as well as their eccentricities and the relative orientation of their orbits on the sky. Moreover, because PPOs occur in general off the face of the star, the timing of a PPO event is affected by the planets' orbital eccentricities (\S \ref{sec:dynamics:eccentricities}), but also probes a planet's TTV curve at an orbital phase inaccessible to either transit or secondary eclipse, potentially breaking mass and/or eccentricity degeneracies inherent to traditional TTV measurements \citep{Lithwick2012,Deck2015}\edited{, and thus better constraining the masses and densities of the planets}. Similarly to secondary eclipses, PPOs can provide constraints on the albedo and/or temperature of an occulted planet, with the added benefit that they allow one to sample both the day side \emph{and} night side of the occulted planet when the occultation occurs far from the disk of the star. Finally, at high time resolution, the shape of a PPO light curve can constrain the two-dimensional surface reflectance or emission map, potentially at a range of wavelengths. Because PPOs are aperiodic, different occultations of a given planet will occur at different phases and with different impact parameters, in principle allowing one to construct crude multi-wavelength surface maps of the entire planet surface.

In this work we assess the frequency and dynamical properties of PPOs in the TRAPPIST-1 system and their detectability with current and future instruments, including the James Webb Space Telescope (JWST) and the Origins Space Telescope (OST). We develop a framework to predict and model PPOs with the new, open-source, photodynamical model \planetplanet. In \S\ref{sec:literature} we review other works that have previously considered PPOs or similar events, and in \S\ref{sec:methods} we describe our methodology for predicting, modeling, and extracting orbital information and crude surface maps from PPOs. We present our results for TRAPPIST-1 in \S\ref{sec:results} and discuss our findings and their applicability to other systems in \S\ref{sec:discussion}.

\section{Planet-planet occultations in the literature}
\label{sec:literature}

Planet-planet occultations have previously been considered in other works as a method to detect or characterize planets. \citet{Ragozzine2010} first introduced the term, citing previous work by \citet{Cabrera2007}, \citet{Sato2009}, and \citet{Sato2010} on the detectability of mutual transits of binary planets and planets with large moons. \citet{Ragozzine2010} extended the idea to occultations among planets on independent astrocentric orbits. The authors pointed out that these so-called ``planet-planet occultations'' could place strong constraints on the three-dimensional architecture of exoplanet systems and even permit surface mapping with future telescopes, citing JWST as a potential facility to detect PPOs in emitted light. More recently, \cite{Brakensiek2016} developed code to calculate geometric probabilities of transits in exoplanet systems, with applications to PPOs and mutual transits. \citet{Veras2017} studied eclipses in exoplanet systems, with applications to TRAPPIST-1 and other compact systems. The coplanarity of TRAPPIST-1 enhances the likelihood that planets in that system will eclipse one another when seen from an observer situated on the surface of one of the planets. In principle, the shadow cast on the eclipsed planet could be detected from Earth, but the signal would be orders of magnitude weaker than a PPO observed in thermal light \citep{Ragozzine2010}.

Other relevant studies include those of \citet{Kipping2011} and \citet{Pal2012}, who developed algorithms to model light curves of planet--moon and mutual planet--planet transits, respectively. The latter case, in which two planets occult each other as they transit their host star, has been studied by several other authors \citep[e.g.,][]{Hirano2012, Masuda2013, Masuda2014}. \citet{Hirano2012} presented the first claimed detection of such an event, seen as a brightening in the light curve of Kepler-89 during a simultaneous transit of Kepler-89b and Kepler-89d. During a mutual transit the planet-planet occultation occurs on the face of the star, resulting in a signal that is typically orders of magnitude stronger than that of a typical PPO; these events are therefore detectable in white light with photometers like \emph{Kepler}.

Closer to home, occultations have been studied, predicted, and observed among solar system bodies. The first ever recorded occultation of two planets in the solar system was that of Jupiter by Mars in the year 1170, which was observed by the monk Gervase of Canterbury and by Chinese astronomers \citep{Stubbs1879,Hilton1988}. \citet{Albers1979} computed the ephemerides of past and future planet-planet occultations in the solar system, reporting two in the 19th century and five in the 21st century (but none in the 20th century!), all of which involve one of Mercury and Venus and one of the superior planets. Occultations among moons in the solar system have also been studied. Recently, \cite{deKleer2017} used the Large Binocular Telescope (LBT) to observe an occultation of Io by Europa, deriving high resolution maps of the volcano Loki Patera from interferometric imaging of the moon's thermal emission during the occultation.
\textbf{}

However, despite the considerable number of studies on planet-planet occultations, a detailed assessment of the detectability of PPOs in exoplanet systems and a framework to predict, model, and extract information from these events is still absent from the literature. In the next sections we discuss our approach to modeling and detecting PPOs in TRAPPIST-1 and in other exoplanet systems.


\section{Methods}
\label{sec:methods}

In order to model planet-planet occultation light curves, we developed the open source\footnote{\url{https://github.com/rodluger/planetplanet}} software package \planetplanet, a photodynamical code written in \texttt{C} and wrapped in \texttt{Python}. Given initial orbital, stellar, and planetary parameters, \planetplanet integrates the system forward in time using an $N$-body code and computes light curves for all planet-planet occultations, as well as all transits, secondary eclipses, and planet phase curves. \edited{In \S\ref{sec:dynamics} we discuss the dynamical theory of PPOs and its implementation in \planetplanet, in \S\ref{sec:photo} we describe our approach to the photometric modeling of occultations, and in \S\ref{sec:detect} we discuss the basics of our detectability calculations.}

\subsection{PPO Dynamics}
\label{sec:dynamics}

Unlike transits or secondary eclipses, planet-planet occultations are episodic. Unless the ratio of the orbital periods of two planets is \emph{exactly} the ratio of two integers, PPOs among the two planets will occur at different phases, with different impact parameters, and with varying durations. While all adjacent pairs of planets in TRAPPIST-1 are close to two-body resonances, departures from exact commensurability cause PPOs in this system to be aperiodic. Nonetheless, precise knowledge of the orbital parameters of pairs of planets can allow one to deterministically predict the times of future PPOs. By the same token, if the orbital parameters are not well known, the detection of PPOs can be used to impose strict constraints on their values. Below we describe techniques to constrain the eccentricity vectors and other three-dimensional orbital information via the detection of PPOs. In general, this is best done with an $N$-body code (\S\ref{sec:dynamics:nbody}), but it is instructive to first consider a few analytic methods.

\subsubsection{Constraining the eccentricities}
\label{sec:dynamics:eccentricities}
In the limit of two edge-on planets on plane-parallel orbits, and weak planet-planet perturbations, the timing of each PPO may be used to constrain a combination of the free eccentricity vectors of both planets involved. For a pair of planets, labelled $i,j$, with ephemerides $(t_{0,i},P_i)$ and $(t_{0,j},P_j)$ with $t_{0,i}$ the time of transit and $P_i$ the orbital period, sorted by $P_i < P_j$, we can compute the longitudes at which the times of planet-planet occultations would occur if they were on {\it circular}, coplanar, and edge-on orbits:
\begin{eqnarray}\label{ppo_condition}
\alpha \cos{\left(\lambda_{k,i}\right)} &=& \cos{\left(\lambda_{k,j}\right)}\\
\lambda_{k,i} &=& \frac{2\pi(t_{k}^0-t_{0,i})}{P_i}-\frac{\pi}{2},
\end{eqnarray}
where $\alpha = (P_i/P_j)^{2/3}$ and $t_{k}^0$ is the midpoint of the $k$th PPO between planets $i$ and $j$, assuming zero eccentricity, and $\lambda_{k,j}$ is the mean longitude at time $t_k^0$ for planet $j$, measured from the sky plane.  This is a transcendental equation that may be solved numerically for the roots $t_{k}^0$.  The superscript zero indicates that this assumes zero eccentricity and edge-on orbits for each planet.  These times {\it only} depend upon the ephemerides, which are measured precisely from the transit times (although they may have small uncertainties associated with each planet's TTVs).  Note that we have also neglected the masses of the planets in this equation.

\begin{figure}[!ht]
\centering
\includegraphics[width=0.47\textwidth]{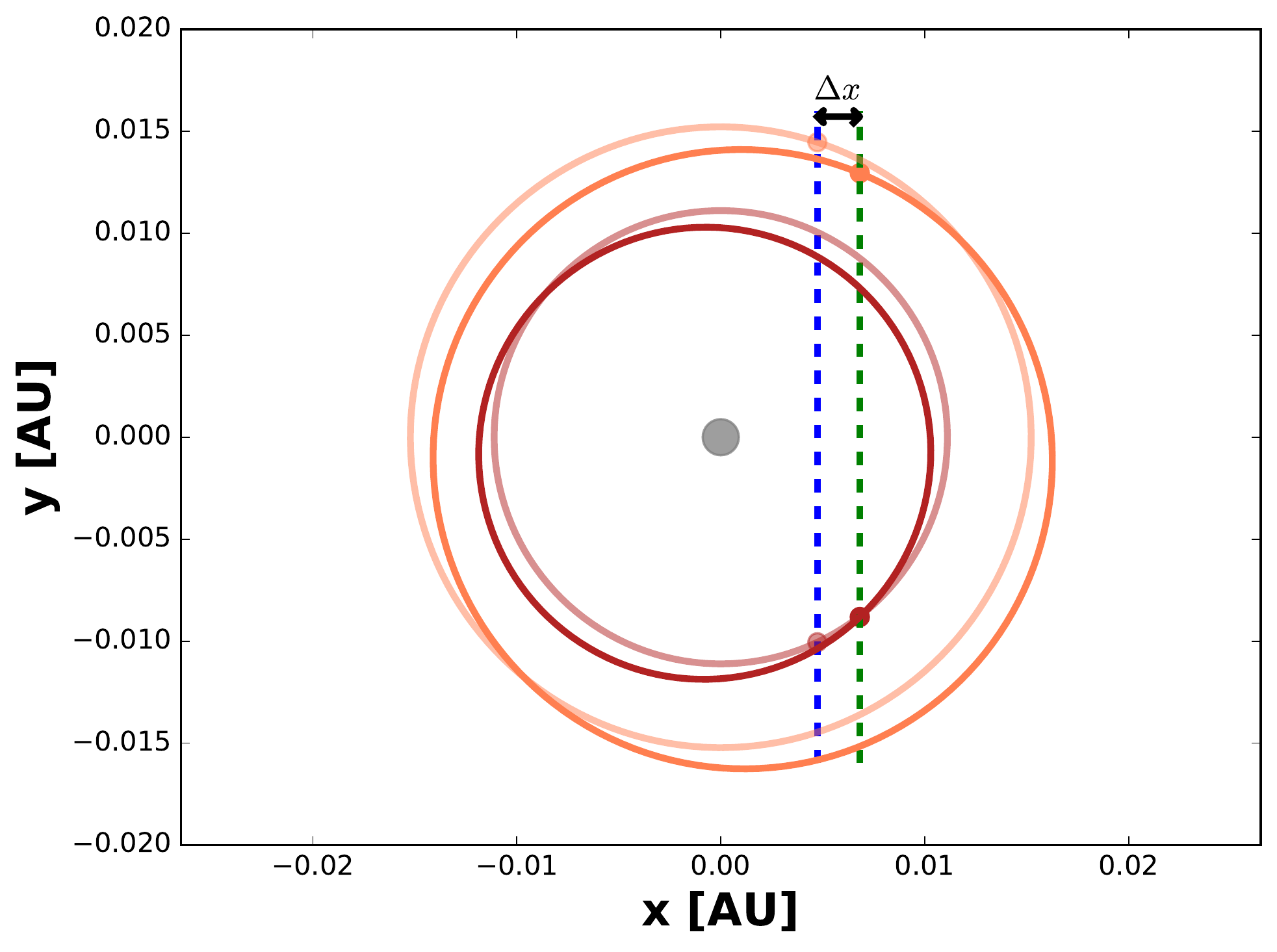}
\caption{Pole-on diagram of coplanar orbits for planets b and c.  The light colored circles (red and orange) are the orbits assuming zero eccentricity, while the darker ellipses are for $e_b=e_c=0.1$.  The position of the planets at the PPO in the circular case (blue dashed line) is offset from the eccentric case (green dashed line) by $\Delta x$. The observer is located at $(0,-\infty)$.}
\label{fig:timing_offset}
\end{figure}

We can compare this to the actual time of the $k$th PPO, $t_{k}$, for the more general case of non-zero eccentricity, albeit still ignoring inclination and TTVs. The difference between the eccentric and circular PPO times, $\delta t_{k} = t_{k}-t_{k}^0$,  may be expanded to first order in eccentricity;  assuming the eccentricities are small, as they are for packed planetary systems due to tidal damping and stability considerations, truncating second order eccentricity terms is well justified.  Figure \ref{fig:timing_offset} illustrates a method to calculate the timing offset analytically as a function of the eccentricity vectors of the planets.  Using the epicyclic approximation, we can compute the offset of the position of the planet relative to the guiding center, as well as the offset due to the fact that the transit time is shifted by the epicyclic motion of the planet.  The offset between the $x$-position of the planets in the eccentric and circular case, $\Delta x$, divided by the relative $x$-velocity, $\Delta v$, gives the timing offset, $\delta t_{k}=\Delta x /\Delta v$.

To first order in eccentricity, we find that
\begin{align}
\label{eqn:delta_t}
\delta t_{k} = \tau_{k} \left({\mathbf v}_j\cdot{\mathbf e}_j-\alpha{\mathbf v}_i\cdot{\mathbf e}_i\right),
\end{align}
where
\begin{eqnarray}
\tau_{k} &=&  \frac{P_j}{4\pi}\left(\sin{\lambda_j}-\alpha^{-1/2}\sin{\lambda_i}\right)^{-1},\\
{\mathbf v}_i &=& \left[\cos{(2\lambda_i)}-4\sin{\lambda_i}-3,\sin{(2\lambda_i)}\right],\\
{\mathbf e}_i &=& \left[e_i \cos{\omega_i},e_i\sin{\omega_i}\right].
\end{eqnarray}
This equation breaks down when $\sin\lambda_j \approx \alpha^{-1/2} \sin{\lambda_i}$ as the denominator approaches zero;  in this limit the sky-projected acceleration can be included to obtain an accurate expression for  $\delta t_{k}$, but the solution for the eccentricity vector in terms of the time offset is no longer linear.   The mean longitude, $\lambda_j$, assumes zero eccentricity, and is zero when the planet crosses the sky plane in the direction away from the observer.  Hence, transits occur when $\lambda_j \approx -\pi/2$ and secondary eclipses when $\lambda_j \approx \pi/2$. The cosine component of these vectors can be very similar in behavior due to the relation that causes PPOs to occur, Equation~(\ref{ppo_condition});  however, these components can be constrained by the timing of secondary eclipses, which should be as detectable as (or more than) the PPO events. Note that since the eccentric orbits have a different breadth than the circular orbits, there may be occasional events that occur in the eccentric case that do not occur in the circular case (or vice versa);  these will place a strict limit on the eccentricity vectors, but we ignore these for this analysis.

Equation~(\ref{eqn:delta_t}) depends on the four components of the eccentricity vectors of both planets, and so at least four planet-planet occultations are needed to obtain a unique solution for the eccentricities of the planets.  Given a set of measured PPO times, $\delta t_{k,obs}$, with uncertainties $\sigma_k$, then the eccentricities may be fit for directly by linear regression. Let ${\mathbf x} = \left[{\mathbf e}_i,{\mathbf e}_j\right]$ be the set of four eccentricity-vector parameters for the two planets.  Letting $\delta t_{k} = {\mathbf y}_k\cdot{\mathbf x}$, where ${\mathbf y}_k = \tau_{k}\left[-\alpha{\mathbf v}_i,{\mathbf v}_j\right]$, we may write our goodness of fit metric as
\begin{equation}
\chi^2 = \sum_{k=1}^K \frac{\left(\delta t_{k,obs}-{\mathbf y}_k \cdot {\mathbf x}\right)^2}{\sigma_k^2}
\end{equation}
for the observation of $K$ planet-planet events between planets $i$ and $j$.
Since this equation is quadratic in ${\mathbf x}$, $\chi^2$ has a unique minimum given by setting $d\chi^2/d{\mathbf x}=0$.
Taking the derivative with respect to $x_m$ for $m = $ 1 to 4, and setting these to zero, we get four equations for the four unknowns:
\begin{equation}
{\bf M} {\bf x} = {\bf b},
\end{equation}
where
\begin{align}
M_{mn} = \sum_{k=1}^K \frac{y_{kn} y_{km}}{\sigma_k^2}
\end{align}
and
\begin{align}
b_m = \sum_{k=1}^K \frac{y_{km} \delta t_{k,obs}}{\sigma_k^2}.
\end{align}
This may be inverted to solve for
\begin{align}
\hat{\mathbf x} = {\mathbf M}^{-1}{\mathbf b}, \label{eqn:eccentricity_solution}
\end{align}
where ${\mathbf \Sigma} = {\mathbf M}^{-1}$ is the covariance matrix with $\Sigma_{mm}$ the uncertainty on each eccentricity vector component.

\begin{figure}[!ht]
\centering
\includegraphics[width=0.47\textwidth]{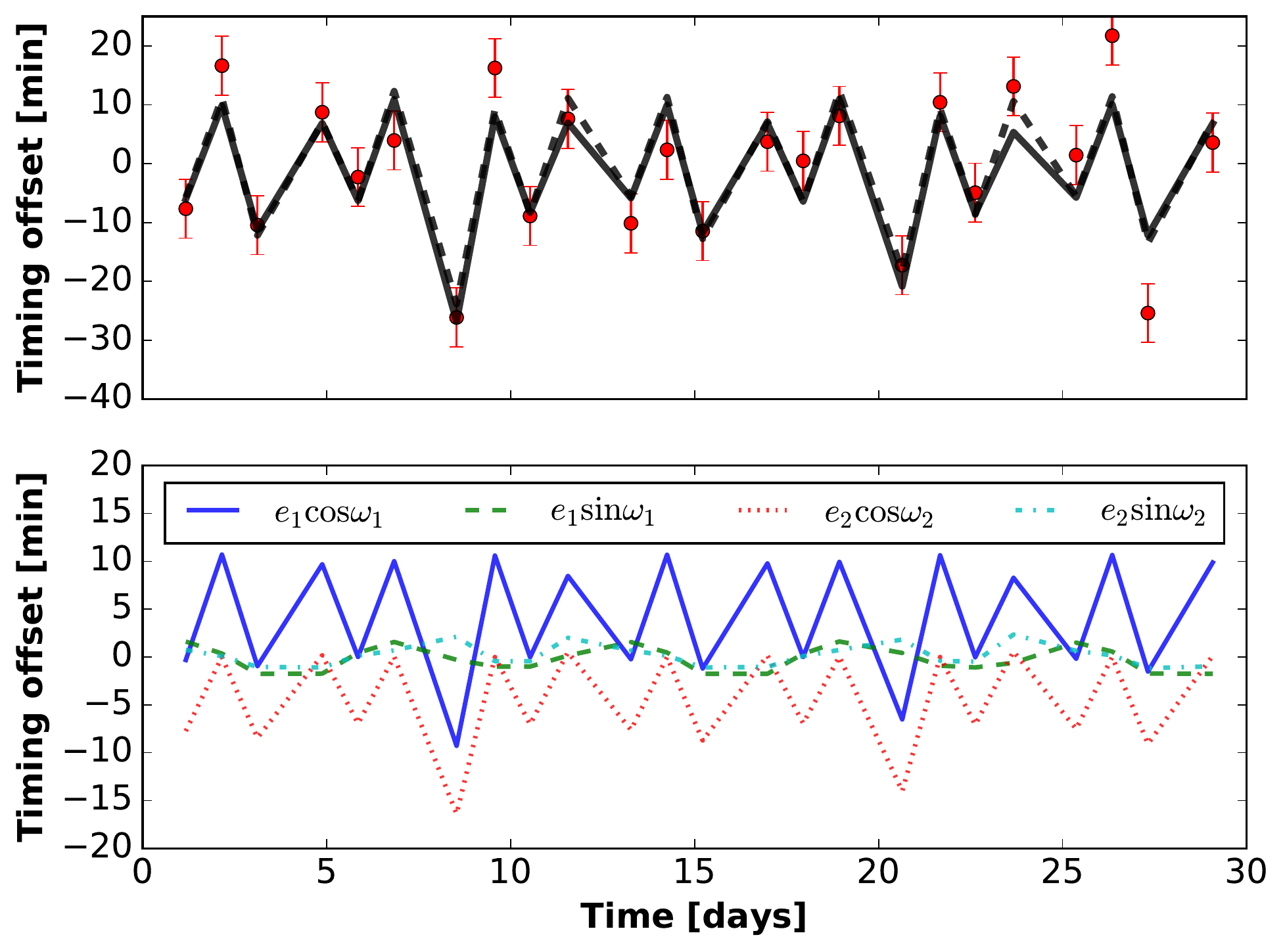}
\caption{(Top) Simulated timing offsets for TRAPPIST-1b,c with an arbitrary eccentricity (red dots) and 5-minute Gaussian noise; the \edited{solid} curve shows the noise-free computation.  The optimized linear fit captures the timing offsets well (dashed \edited{line}), and recovers the correct eccentricity vectors (Table~\ref{tab:eccentricity}).  (Bottom) The amplitude due to each eccentricity vector component is shown;  each of these has a different functional form, which is what allows the eccentricity vectors to be recovered.}
\label{fig:fit_timing_offset}
\end{figure}

An example of this technique is shown in Figure \ref{fig:fit_timing_offset}.  We simulated the timing offsets versus the time of planet-planet occultation events between TRAPPIST-1b and c for 30 days assuming edge-on Keplerian orbits, with 5-minute Gaussian random timing noise added to each PPO time. We then used these timing offsets to recover the eccentricity vectors of the planets with Equation~(\ref{eqn:eccentricity_solution}). \edited{Note that the $e_1\cos{\omega_1}$ and $e_2\cos{\omega_2}$ components have very similar shapes, but different constant offsets.  For this particular example, the mean timing offsets of these components helps to disentangle the degeneracy that can occur due to the similar shape, but for other choices of parameters there may be stronger degeneracy between these components.} The input and recovered eccentricity vectors are listed in Table~\ref{tab:eccentricity}; our solutions agree within the uncertainties. The uncertainties on the eccentricity vectors will depend upon the period ratios of the planets, the timing precision achieved, and the sampling of the times;  we provide code for carrying out these simulations to assist with experimental design in the \github repository.

Note that in this simulation the timing offsets are larger than the amplitude of TTVs for TRAPPIST-1b and c, which are of order 1 minute, so that TTVs can be approximately neglected for these planets. However, if the eccentricities were much smaller for these planets, as they may likely be due to tidal damping \citep{Luger2017}, then TTVs must be accounted for, which can be accomplished with a photodynamical code (see \S\ref{sec:dynamics:ttvs}). 

\begin{deluxetable}{lll}[ht!]
\tablewidth{0pt}
\tablecaption{Simulated recovery of the eccentricities of TRAPPIST-1b and c\label{tab:eccentricity}}
\tablehead{\colhead{} & \colhead{Input} & \colhead{Recovered}} 
\startdata
$e_b\cos\omega_b$ & $-0.0128$ & $-0.0102\pm0.0021$\\
$e_b\sin\omega_b$ & $-0.0225$ & $-0.0289\pm0.0142$\\
$e_c\cos\omega_c$ & $\phantom{-}0.0067$ & $\phantom{-}0.0078\pm 0.0012$\\
$e_c\sin\omega_c$ & $\phantom{-}0.0053$ & $\phantom{-}0.0107\pm0.0064$
\enddata
\end{deluxetable}

Thanks to the linear dependence, the eccentricity uncertainties are proportional to the timing uncertainties.  The timing uncertainties may be affected by the radiative intensity asymmetry of the planets, if accounted for inaccurately in the photometric analysis. As shown below, an airless body can mimic an occultation timing offset of several minutes;  this gives an upper limit on the expected timing error since the recirculation can be modeled, even if only approximately.  The impact of intensity asymmetry on timing uncertainty will require further simulations to quantify more precisely. \edited{Furthermore, additional constraints upon the eccentricity may be derived from the shape and depth of these events. We collectively refer to changes in the timing, shape, depth and duration as occultation light curve variations (OLVs), which can be analyzed to recover information about the orbits and the planetary properties. In practice, however, the entire light curve should be modeled, which we do further below.}

Measuring the eccentricities can help to break the eccentricity-eccentricity degeneracy \citep{Lithwick2012}, and may also help break the mass-eccentricity degeneracy \citep{Lithwick2012,Deck2015} that affects TTV measurements.  Breaking these degeneracies could lead to more precise measurements of the mass ratios of the planets to the star, as well as better dynamical constraints and estimates of the tidal effects on the planets, such as constraints on the planetary Love numbers  \citep{Mardling2007,Batygin2009}.  The timing offset of secondary eclipses \citep{Charbonneau2005} can impose a constraint on $e_i \cos{\omega_i}$, and thus fewer PPOs may be needed to obtain a unique solution for the planets' eccentricities;  this could be incorporated into the linear analysis.  We note, finally, that if PPOs occur between different pairs of planets, then the above solution may be extended to all planets involved to obtain a simultaneous solution for the eccentricity vectors of all $N$ planets.  However, this subsection ignores the mutual inclinations between planets and between system and observer;  this will not be satisfied in general, and can provide additional constraints upon the orbital geometry, described next.

\subsubsection{Constraining longitudes of nodes}
\label{sec:dynamics:Omegas}
The foregoing analysis assumed edge-on, plane-parallel orbits for the planets; in real planet systems, these assumptions will always be broken to some extent. The inclination of the planets' orbits ($I$) can be obtained from measurement of the transit/eclipse impact parameter(s) \citep{Winn2010}, while the longitudes of nodes ($\Omega$) are more difficult to constrain, even with the measurement of TTVs \citep{Nesvorny2014}, although it can be constrained in some cases in part due to transit duration variations \citep[TDVs;][]{Carter2012}.

In contrast, planet-planet occultations can yield strong constraints on $\Omega$. Assuming circular orbits and identical inclinations, the impact parameter will vary with the longitude of the planet-planet event as
\begin{align}
    b &= \vert a_i \cos \lambda_i \sin \Delta \Omega_{i,j}\vert /(r_i+r_j)\nonumber\\ 
    &= \vert a_j \cos \lambda_j \sin \Delta \Omega_{i,j}\vert /(r_i+r_j),
\end{align}
where $a_i$ and $r_i$ are the semi-major axis and radius of the $i^\mathrm{th}$ planet, respectively. This will yield the absolute value of $\Delta \Omega_{i,j} =\Omega_i-\Omega_j$, the difference between the longitude of nodes of the two planets. The same calculation applies to planet-planet occultations that occur during transit \citep{Pal2012}, but in that case the lever arm is {\it much} smaller, and thus the constraint on $\vert\Delta\Omega\vert$ tends to be poor.

The inclination, $I$, combined with $\vert\Delta \Omega_{i,j}\vert$ will yield a constraint on the full geometry of the system, but with the discrete degeneracy between each pair of planets that undergo multiple occultations due to the absolute value. However, the measurement of $\vert\Delta\Omega\vert$ for more than two planets will yield a unique solution in most cases, thus constraining the full three-dimensional geometry of the planet system, up to an unknown sky angle.

\subsubsection{De-aliasing transit timing}
\label{sec:dynamics:ttvs}
Planet-planet perturbations induce variations in the orbital elements at frequencies that are differences of integer multiples of the orbital frequencies of the planets, $P_{j,k}=(j/P_1-k/P_2)^{-1}$, where $j$ and $k$ are integers, and $P_1$ and $P_2$ are the periods of a pair of adjacent planets \citep{Deck2015}.  Transits are observed every orbital period of each planet, causing aliasing which induces a degeneracy between different $k$ ($j$) values for the inner (outer) planet due to sampling on the period $P_1$ ($P_2$).  This aliasing is the origin of the mass-eccentricity degeneracy \citep{Lithwick2012,Deck2015}.  In contrast, planet-planet occultations will sample the mutual planetary perturbations at a {\it different} orbital phase, which may allow for the measurement of the free orbital eccentricity, as discussed above, but also may allow for the de-aliasing of different components of transit-timing variations.  If sufficient precision can be obtained to measure the variation of the times of PPOs due to forced eccentricity and period variations caused by dynamical interactions, these timing measurements can help break the TTV mass-eccentricity degeneracy. \edited{A similar effect can occur in transiting circumbinary planets, whereby the motion of the stars about the barycenter samples the transit times at irregular intervals, breaking the mass-eccentricity degeneracy as well.} In principle, the full orbital properties of the planets should be modeled to create a photodynamical model \citep{Carter2012}; we describe this approach next.

\subsubsection{N-body code}
\label{sec:dynamics:nbody}
In order to model the general case of eccentric, non-coplanar, massive planets subject to transit timing variations, we use the \texttt{REBOUND} N-body code \citep{ReinLiu2012} to model PPOs in \planetplanet. Starting from an initial state informed by the observational constraints of a multi-planet system, we integrate the orbital positions of all bodies forward in time using \texttt{REBOUND}. We implement both the high order integrator \texttt{IAS15} \citep{ReinSpiegel2015} and the symplectic integrator \texttt{WHFAST} \citep{ReinTamayo2015}. We track the relative sky-projected positions of all bodies, taking Keplerian steps on a finer subgrid to resolve all planet-planet events. At each subgrid timestep, we evaluate the impact parameters between all pairs $\{i,j\}$ of planets, given by
\begin{align}
    \label{eqn:occultation_impact}
    b_{ij} = \frac{\sqrt{(x_i - x_j)^2 + (y_i - y_j)^2}}{r_i + r_j},
\end{align}
where $x_i$ and $y_i$ are the sky-projected Cartesian coordinates of the $i^\mathrm{th}$ planet and $r_i$ is its radius. Occultations between two planets occur when $b_{ij} < 1$. We discuss the implementation of this procedure for the TRAPPIST-1 system in detail in \S\ref{sec:results:dynamics}.

\pagebreak

\subsection{PPO photometry}
\label{sec:photo}

Given the full orbital solution for the system obtained with \texttt{REBOUND}, \planetplanet computes light curves for all transits, secondary eclipses, and planet-planet occultations, as well as phase curves for all planets in the system. Since planet-planet occultations are observable primarily in the mid- and far-infrared, in what follows we model only the thermal emission from planets and neglect reflected light from the star, although the latter may easily be computed with \planetplanet given the symmetry of the problem.

We developed a novel scheme for fast computation of occultation light curves optimized for the case of a body whose thermal emission map is radially symmetric about an \emph{arbitrary} point on its surface. This is the case for a limb-darkened star or limb-darkened/ limb-brightened planet with a thick atmosphere and homogeneous cloud cover, whose surface brightness is to a very good approximation symmetric about the center of its disk. It is also the case for a planet (or moon) with a thin (or nonexistent) atmosphere, whose surface brightness is symmetric about the substellar point but in general lacks radial symmetry in the observer's frame. In the sections below, we describe the geometry of these two limiting cases for a planet's thermal emission map and discuss our integration scheme. At present, we only model planets in these two limits. The presence of spatially variable surface features and/or inhomogeneous clouds break these symmetries, and will be left for future investigation.

\subsubsection{Planets with thick atmospheres}
\label{sec:photo:limbdarkened}
In the limit that a planet's atmosphere is thick, thermal emission from the planet is spatially decoupled from the illumination pattern of the star and the planet appears as a radially symmetric, limb-darkened (or limb-brightened) disk. Barring inhomogeneities due to clouds, this is approximately the case for Venus \citep{Murray1963}, and if one neglects star spots, it is also a good approximation for stars. The case of a spherical body transiting a limb-darkened disk is well studied \citep{MandelAgol2002} and the occultation light curve is analytic under the linear, quadratic, and nonlinear limb darkening laws \citep{Claret2000}. However, the calculation involves multiple evaluations of elliptic integrals or hypergeometric functions, which are expensive to compute. Often it is more expedient to exploit the radial symmetry of the problem to reduce a two-dimensional integral to a one-dimensional integral and solve the latter numerically; this is the approach taken in the nonlinear limb-darkening law case in code provided by \citet{MandelAgol2002}. Recently, \citet{Kreidberg2015} developed a similar method for the \batman transit calculation package. The \batman code discretizes the stellar surface as a series of spherical segments, each with a constant intensity given by an arbitrary radial limb darkening profile. Seen in projection, these are concentric rings. The flux occulted by a transiting planet is then simply the weighted sum over the areas of overlap between the planet and each of the rings, which are expressed in terms of arccosine functions \citep{MandelAgol2002}. These numerical methods allow for the fast computation of light curves under arbitrary limb darkening laws.

In order to treat occultations of limb-darkened/\allowbreak brightened planets (and stars), we adopt a similar integration scheme, discretizing the body's surface with $N$ concentric circles that bound $N+1$ spherical segments equally spaced in $\phi$, the angle between the line of sight and the vector normal to the sphere. Each circular boundary is centered at the origin and has a radius
\begin{align}
    \label{eqn:a}
    a = r_{\scriptscriptstyle \occulted}\sin\phi
\end{align}
where $r_{\scriptscriptstyle \occulted}$ is the radius of the body. This results in an adaptive grid in the radial coordinate, $a$, in which the radial spacing between rings decreases toward the limb, where the change in intensity is fastest. If $\Delta\phi$ is the grid spacing in $\phi$, the spacing in $a$ is
\begin{align}
    \Delta a \approx \Delta\phi\sqrt{1-a^2},
\end{align}
which in the limit $a\rightarrow 1$ is identical to that used in \batman.

We adopt the following generic limb darkening law:
\begin{align}
    \label{eqn:radiance_ld}
    B_\lambda(\phi) = B_\lambda^0 \left[ 1 - \sum_{i=1}^{n} u_i(\lambda) (1 - \cos \phi)^i \right],
\end{align}
where $B_\lambda$ is a spectral radiance (measured in units of power per unit area per unit solid angle per unit wavelength),
$B_\lambda^0$ is the spectral radiance at the center of the disk (see Appendix~\ref{app:wavelength_limbdark}), $\lambda$ is the wavelength, and $u_i$ is the $i^\mathrm{th}$ limb darkening coefficient, which we allow to be an arbitrary function of wavelength. The variation in the intensity of the disk towards the limb can be a strong function of wavelength, even for a perfect blackbody, and it is essential to allow for this when modeling occultations in the mid-infrared. Note that for $u_i(\lambda) = \mathrm{constant}$ and $n = 1$ or $n = 2$, this is equivalent to the standard linear or quadratic limb darkening laws \citep{Claret2000}, respectively.  Also note that this definition is equivalent to the limb-darkening formulation of \citet{Gimnez2006}, although with differently defined coefficients.

In principle, our method is flexible enough to allow for any limb darkening law, including those that cannot be expressed as simple polynomial functions of $\phi$. As both the temperature and the abundance of various species in a planetary atmosphere can change drastically with altitude, light emerging from the limb (where the effective emission layer is higher in the atmosphere) may have a drastically different spectrum than light emerging from the center of the planet disk. Because low-order polynomials are not flexible enough to accurately model this, our model also accepts as input a grid of emission spectra over the planet disk, provided the emission profile is radially symmetric:
\begin{align}
    \label{eqn:radiance_general}
    B_\lambda(\phi) = B_{i,j}(\lambda, \phi),
\end{align}
where $i$ and $j$ are the corresponding indices in the $\lambda$ and $\phi$ grids, respectively, such that
$\lambda_{i-1} < \lambda \leq \lambda_{i}$ and $\phi_{j-1} < \phi \leq \phi_{j}$.

\subsubsection{Eyeball planets}
\label{sec:photo:airless}
In the limit that a planet's atmosphere is thin enough as to have negligible thermal inertia and negligible heat transport, the temperature of a given region on the planet's surface is dictated entirely by radiative equilibrium with the incident stellar flux $F_\star$. If $\phi$ is the angle of the star measured from the zenith as seen from an observer at a point on the planet surface, the temperature at that point is given by
\citep[e.g.,][]{Maurin2012}
\begin{align}
    \label{eqn:tlat}
    T(\phi) = \max\left(T_\mathrm{night}, \left( \frac{F_\star (1 - A)\cos \phi}{\sigma} \right)^\frac{1}{4}\right),
\end{align}
where $A$ is the planetary albedo, $\sigma$ is the Stefan-Boltzmann constant, and $T_\mathrm{night}$ is the temperature of the unilluminated night side, which may be nonzero due to, e.g., a geothermal heat flux. Note that for a planet at full phase, $\phi$ is identical to the angular measure used in \S\ref{sec:photo:limbdarkened}. From here on out, we refer to $\phi$ as the ``zenith angle,'' defined in the range $[0,\pi]$.

The corresponding spectral radiance, $B_\lambda$, is given by Planck's law,
\begin{align}
    \label{eqn:radiance_airless}
    B_\lambda(\lambda, \phi) = \frac{2hc^2}{\lambda^5}\frac{1}{e^\frac{hc}{\lambda k_B T(\phi)} - 1},
\end{align}
where $h$ is Planck's constant, $c$ is the speed of light, $k_B$ is Boltzmann's constant, and $\lambda$ is the wavelength. While the finite angular size of the star as seen from its planets can result in illumination past the day/night terminator ($\phi = \frac{\pi}{2}$), the change in the radiance of the planet is negligible and we do not consider it here.

\begin{figure}[!ht]
\centering
\includegraphics[width=0.39\textwidth]{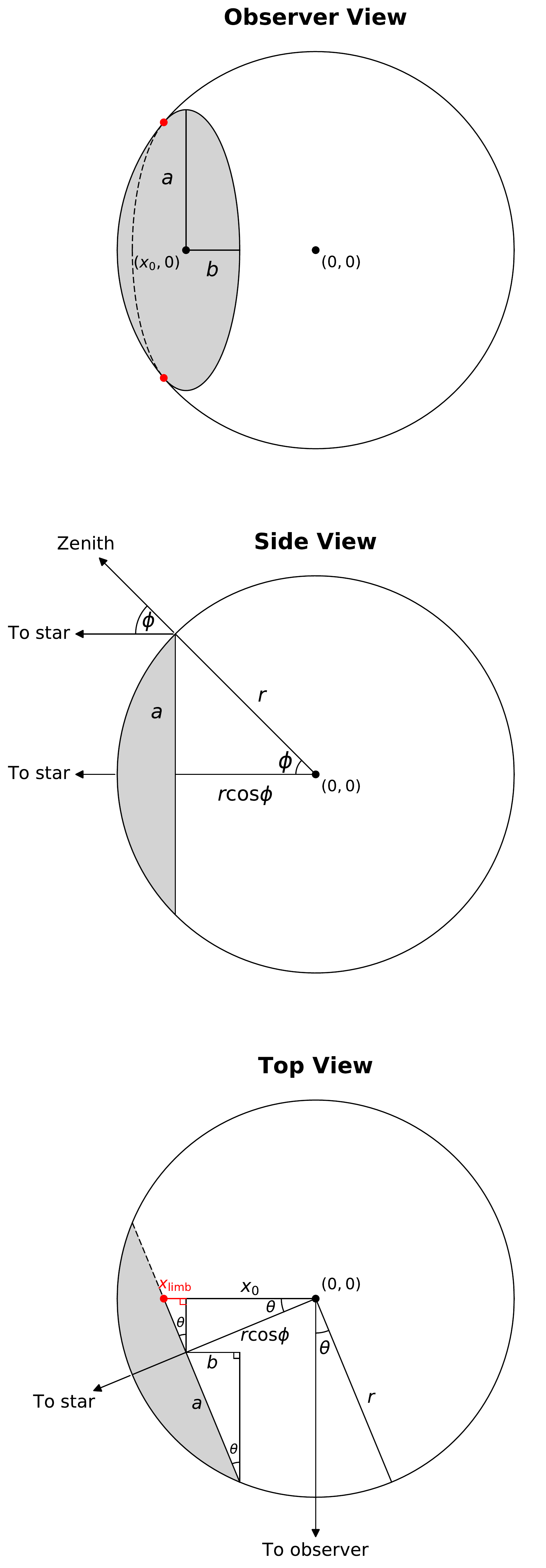}
\caption{Geometry of a region of constant surface brightness on the planet surface, seen from three different vantage points. In general, such a region is a thin spherical segment of radius $a = r\sin{\phi}$, where $r$ is the radius of the body and $\phi$ is the zenith angle. The boundaries of each region are circles, which when projected onto the sky plane become ellipses with semi-major axis $a$ and semi-minor axis $b = a\sin\theta$, where $\theta$ is the phase angle. In this example, we show a spherical cap extending to $\phi = \frac{\pi}{4}$ for a planet at $\theta = \frac{\pi}{8}$. Note that a portion of the elliptical boundary is behind the limb of the planet (dashed lines). See text for details.}
\label{fig:geometry}
\end{figure}

As before, let us discretize the radiance gradient. Regions of constant radiance are once again spherical segments. When the planet is seen at full phase, the curves bounding these regions are concentric circles centered on the sub-stellar point, as in the limb-darkened case. At half phase, these become lines perpendicular to the vector connecting the planet and the star. At intermediate viewing angles, these curves are segments of ellipses. 

Figure~\ref{fig:geometry} shows an example of a region of constant surface brightness extending from the sub-stellar point to a zenith angle $\phi$. This region is a spherical cap, which in the observer's frame is bounded by an ellipse of semi-major and semi-minor axes
\begin{align}
    \label{eqn:a_and_b}
    a &= r_{\scriptscriptstyle \occulted}\sin \phi\nonumber\\
    b &= r_{\scriptscriptstyle \occulted}\sin \phi|\sin\theta|,
\end{align}
respectively, where $r_{\scriptscriptstyle \occulted}$ is the radius of the body and $\theta$ is the phase angle, the angle between the sub-stellar point and the axis perpendicular to the line of sight and parallel to the orbital plane (see Figure~\ref{fig:geometry}). For an orbit at arbitrary eccentricity viewed edge-on and aligned with the $xz$ plane, with the $x$ axis pointing to the right on the sky and the $z$ axis pointing into the sky, this angle is simply the mean longitude of the orbit and is given by
\begin{align}
    \label{eqn:theta0}
    \theta = \mathrm{arctan2}\left(z,x\right),
\end{align}
where $z$ and $x$ are the coordinates of the planet in this frame and $\mathrm{arctan2}$ is the two-argument arctangent function. We discuss how to compute $\theta$ for arbitrary orbital geometries in Appendix~\ref{app:geometry}.

As expected, when $\theta = \pm \frac{\pi}{2}$ (full/new phase), $a = b$ and the bounding region is a circle; when $\theta = 0$ (half phase), $b = 0$ and the bounding region is a line. In general, assuming the planet is centered at the origin, it is straightforward to show that the ellipse is centered at 
\begin{align}
    \label{eqn:x0_and_y0}
    x_0 &= -r_{\scriptscriptstyle \occulted}\cos \phi\cos\theta\nonumber\\
    y_0 &= 0.
\end{align}
Note that we must also account for the fact that the ellipse may not be fully visible to the observer. In the example shown in the figure, the points where the ellipse crosses beyond the limb of the planet are indicated as red dots. It can be shown that the $x$ coordinate of these points is 
\begin{align}
    \label{eqn:xlimb}
    x_\mathrm{limb} &= x_0 - r_{\scriptscriptstyle \occulted} \cos \phi \sin\theta \tan\theta\nonumber\\
                    &= x_0 \sec^2{\theta}.
\end{align}

As we discuss below, this formalism allows us to approximate the occulted planet flux as a sum over the integrals of elliptical segments, which are analytic when one of the axes of the ellipse is parallel to the $x$ axis. This is the case for a planet in an edge-on orbit whose emission is symmetric about the sub-stellar point. However, planets in sufficiently inclined orbits or planets with latitudinal offsets in the location of their peak emission break this symmetry. In Appendix~\ref{app:geometry} we derive the geometry for the general case of a planet in any orbit and with an arbitrary offset in its hotspot, showing that our semi-analytic integration scheme can be straightforwardly adapted to that case. We will see later how this can be used to model phase curves of planets with vigorous winds that shift the position of the hotspot away from the sub-stellar point, provided one replaces Equation~(\ref{eqn:tlat}) with the appropriate temperature map.

\subsubsection{Integration scheme}
\label{sec:photo:integration}
\begin{figure}[!ht]
\centering
\includegraphics[width=0.47\textwidth]{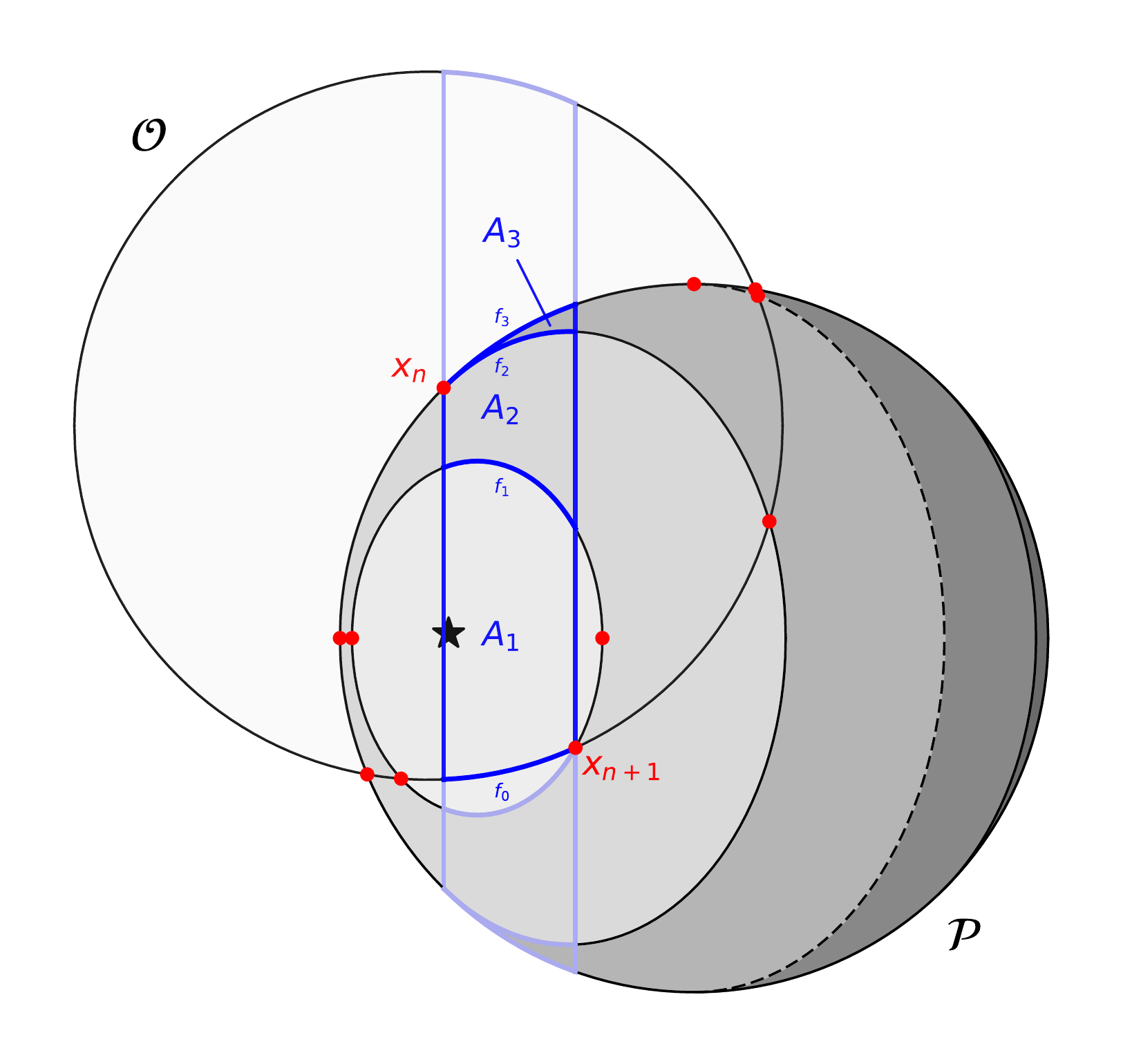}
\caption{An example of the integration scheme for a planet-planet occultation \edited{of an ``eyeball'' planet}. The occultor \occultor is at the top left and the occulted planet \occulted is at the bottom right. The latter is an airless body at a phase angle $\theta = \frac{\pi}{4}$ with a radiance given by Equation~(\ref{eqn:radiance_airless}). This gradient is discretized into regions of constant radiance, shaded accordingly in the figure. The day/night terminator is indicated by the dashed curve, and the sub-stellar point \edited{is indicated by a star}. The flux of \occulted that is occulted by \occultor is computed by identifying all intersection points $x_n$ between curves (red points) and summing the integrals over each of the regions in both \occulted and \occultor that are bounded by adjacent pairs of these points \edited{($x_n$ and $x_{n+1}$ in the figure)}. \edited{The vertical blue lines are the limits of integration.} As the boundary functions $f_j$ of these regions (dark blue curves) are either circles or ellipses, all integrals are analytic. The total occulted flux is then the product of the area $A_j$ of each region and its radiance.}
\label{fig:integration}
\end{figure}

Given the discretized radiance gradients discussed above, we wish to compute the total flux occulted by a body that passes in between the emitting body and the observer. This may be computed as the integral of the occulted planet's radiance evaluated over the region of overlap of two circles. In both the limb-darkened and airless body cases, all curves of constant radiance are ellipses (or circles), so the problem is reduced to a sum over one dimensional integrals of ellipses, which are analytic (see Appendices~\ref{app:circle-circle} and \ref{app:circle-ellipse}). In Appendix~\ref{app:integration} we show how this integration may be fully automated by identifying all relevant points of intersection between the ellipses, sorting them, and evaluating the integrals in the region between all pairs of adjacent points. Figure~\ref{fig:integration} illustrates this integration scheme for a planet-planet occultation.

\subsubsection{Phase curves}
\label{sec:photo:phasecurves}
For an airless planet, the spatially variable radiance map results in a periodic phase curve signal over the course of its orbit. \planetplanet computes phase curves using the same integration scheme described above. In Appendix~\ref{app:geometry} we show sample phase curves of planets in eccentric and inclined orbits computed with \planetplanet.

\subsubsection{Model validation}
\label{sec:photo:validation}
We validate our light curve model against \texttt{pysyzygy}\footnote{\url{https://github.com/rodluger/pysyzygy/}} and \texttt{batman} \citep{Kreidberg2015} (for transit light curves) and against a two-dimensional integration of the radiance map (for planet-planet occultations). \texttt{pysyzygy} is a standard implementation of the \cite{MandelAgol2002} transit model for limb-darkened light curves, while \texttt{batman} is an adaptive algorithm that, similarly to \planetplanet, discretizes the surface brightness gradient into concentric rings and computes light curves semi-analytically with high precision. We find that the transit light curves generated by \planetplanet agree with those of \texttt{pysyzygy} and \texttt{batman} at the $< 1$ ppm level over a range of orbital parameters and limb darkening coefficients. 

We validate our occultation light curves by direct integration of the planet emission profile in the area of intersection between the planet disk and the occultor disk. We discretize the planet disk with a Cartesian grid and compute the radiance at each point from Equation~(\ref{eqn:radiance_airless}), with a temperature given by Equation~(\ref{eqn:tlat}) and zenith angle given by
Equations~(\ref{eqn:zenith_angle1})--(\ref{eqn:zenith_angle3}). We find that our code matches the occultation light curves generated in this fashion to within the error level of the direct integration procedure.

\subsection{\edited{Detectability}}
\label{sec:detect}
\edited{
In order to assess the detectability of PPOs, we compute the expected signal-to-noise ratio (SNR) of occultations modeled with the photodynamical code described above. Since the SNR is instrument- and wavelength-dependent, in our base cases we explore the detectability of PPOs with the James Webb Space Telescope (JWST) Mid-IR Instrument (MIRI) 15 $\mu\mathrm{m}$ filter; in later sections, we assess the detectability of PPOs at different wavelengths and with different instruments. 

Scheduled for launch in \bugfix{early 2019}, JWST will offer an unprecedented view of exoplanetary systems by accessing wavelengths in the range ${\sim}0.6-30$~$\mu$m \citep{Gardner2006}. The mid-IR capability of JWST/MIRI is uniquely suited for secondary eclipse and PPO observations due to the rise in signal contrast towards longer wavelengths. Since the signals of interest are photons emitted from the planet, stellar photons only contribute noise. It is therefore advantageous to observe such occultation events at longer wavelengths where the stellar flux declines along the Rayleigh-Jeans tail and the thermal flux from temperate planets peaks. 

We assess the detectability of PPOs with JWST/MIRI by simulating time-series filter photometry. We consider both shot noise and radiative background noise. Our background noise estimates follow the 6-component gray-body emission spectra from \citet{Glasse2015}, which account for scattered and emitted zodiacal dust and observatory straylight. MIRI has 9 photometric filters spanning ${\sim}5-30$~$\mu$m \citep{Bouchet2015}. We acquired publicly available MIRI filter response curves online.\footnote{\url{http://ircamera.as.arizona.edu/MIRI/pces.htm}} Although we consider all 9 filters, we focus primarily on the 15~$\mu$m filter (F1500W) because we find it to be optimal when considering the rise of both the signal contrast and the background noise towards longer wavelengths. As we show below, the 12.8~$\mu$m filter (F1280W) yields comparable SNR for PPOs in TRAPPIST-1.\footnote{Calculations using all 9 MIRI filters can be performed using our code on \github.}

To construct light curves, we assume consecutive images are captured with the same exposure time and neglect readout and reset time. We interpolate the filter throughput curve $T_{\lambda}$ to the high resolution spectrum grid $F_{\lambda}$ output by the photodynamical model. We then calculate the number of photons detected from the system, 
\begin{align}
\label{eqn:photon_counts}
N_{\text{sys}} = \frac{n A t}{hc} \sum_{\lambda} F_{\lambda} \lambda \Delta \lambda T_{\lambda} ,
\end{align}
where $A$ is the telescope aperture (25 m$^2$ for JWST), $t$ is the exposure time, $n$ is the number of observations (in the case of stacking), $\lambda$ is the wavelength bin, $\Delta \lambda$ is the width of the wavelength bin, and $h$ and $c$ are Planck's constant and the speed of light, respectively. The calculation for the background photons $N_{\text{back}}$ is analogous.

Given Poisson errors, the SNR of an individual measurement $j$ is
\begin{align}
\label{eqn:snr_indiv}
    \text{SNR}_j = \frac{N_{0,j} - N_{\text{sys},j}}{ \sqrt{N_{\text{sys},j} + N_{\text{back},j}}},
\end{align}
where we have defined the ``signal'' as the difference between the number of photons one would detect if no occultation event occurred ($N_{0}$) and the number of photons actually observed ($N_{\text{sys}}$).

The SNR of an occultation (or a set of multiple occultations) is the quadrature sum of the SNR on each individual measurement taken during the event(s):
\begin{align}
\label{eqn:snr_occult}
    \text{SNR} = \sqrt{ \sum_j \frac{( N_{0,j} - N_{\text{sys},j} )^2}{ N_{\text{sys},j} + N_{\text{back},j}} }.
\end{align}
In the case that the occultation model has a single degree of freedom (such as the time of the occultation), the SNR is equivalent to the significance of the detection under the model; i.e., an $\mathrm{SNR} = 5$ feature is a 5$\sigma$ detection of an occultation. In general, however, if the orbital parameters of the two bodies involved in the occultation are not well constrained, the significance of the detection will be lower than the SNR.
%
%
%
%
%
%
%
Moreover, the equation above implicitly assumes that the continuum ($N_0$) is known precisely; in general, stellar and instrumental variability will introduce uncertainty on this value, which will increase the noise and decrease the SNR. We therefore caution that the SNR estimates presented in this paper are optimistic.

We note, finally, that our SNR metric can be scaled to telescopes with different areas and throughputs:
\begin{align}
    \mathrm{SNR'} \approx \sqrt{\left(\frac{A}{25\ \mathrm{m^2}}\right)\left(\frac{\left<T_\lambda\right>}{0.3}\right)}\ \mathrm{SNR}
\end{align}
where $A$ is the collecting area of the telescope and $\left<T_\lambda\right>$ is the effective throughput in the ${\sim}13.5-16.5\mu\mathrm{m}$ bandpass. Note that the SNR is not linear in other parameters, such as the wavelength or the telescope background level. We therefore provide code on \github to compute the SNR for arbitrary telescope specifications.
}

\section{Application to TRAPPIST-1}
\label{sec:results}
In this section we present our results for the TRAPPIST-1 system, but our methodology is general and can be used to model planet-planet occultations in any nearby multi-planetary system. \edited{In our photometric calculations, we assume an optimistic albedo of 0 for all planets unless otherwise stated. We consider the two atmospheric limits outlined above: airless (``eyeball'') planets, which we assume have a fixed night side temperature $T_\mathrm{night} = 40$ K, and bodies with thick atmospheres and uniform radiance, which we model in the limb-darkened limit with $u_i(\lambda) = 0$. These assumptions can be easily changed or relaxed in \planetplanet. The stellar luminosity is sampled from $L_\star = 0.000524 \pm 0.000034\ \mathrm{L_\odot}$ \citep{Gillon2017} and the effective temperature is computed from this value via the radius, $R_\star = 0.121 \pm 0.003\ \mathrm{R_\odot}$ \citep{BurgasserMamajek2017}.}

\subsection{Dynamics}
\label{sec:results:dynamics}
The frequency of occurrence and the dynamical properties of planet-planet occultations (including, for instance, their distribution in orbital phase, their durations, and their impact parameters) are extremely sensitive to the full three-dimensional architecture of a planetary system. Transit photometry and TTV analyses have thus far constrained many of the orbital parameters of the seven planets in TRAPPIST-1. In our analyses below, we compute the statistics of planet-planet occultations by sampling the posterior distributions reported in \cite{Gillon2017} and \cite{Luger2017}. At present, the eccentricities of the planets are constrained with only upper limits ($\la 0.01$). For these, we use estimates based on a migration and tidal evolution model \citep{Luger2017}, drawing the longitude of pericenter from uniform distributions in the range (0, 2$\pi$]. Table~\ref{tab:sysparams} shows the distributions assumed for the orbital parameters of each of the planets.

Currently, the largest source of uncertainty relevant to PPOs are the longitudes of ascending nodes ($\Omega$) of the planets, which are completely unconstrained. For an edge-on system like TRAPPIST-1, $\Omega$ is the angle of rotation of the orbital plane on the sky relative to some reference direction. In general, transit light curves are completely insensitive to the value of $\Omega$; however, the relative value $\Delta\Omega$ for each pair of planets controls their sky-projected separation, such that planets with large $\Delta\Omega$ may never occult each other away from the disk of the star. In the following section, we describe a Monte Carlo technique we developed to place constraints on $\Delta\Omega$ for the TRAPPIST-1 planets.

\subsubsection{Coplanarity of TRAPPIST-1}
\label{sec:results:dynamics:coplanarity}
\begin{figure}[!ht]
\centering
\includegraphics[width=0.47\textwidth]{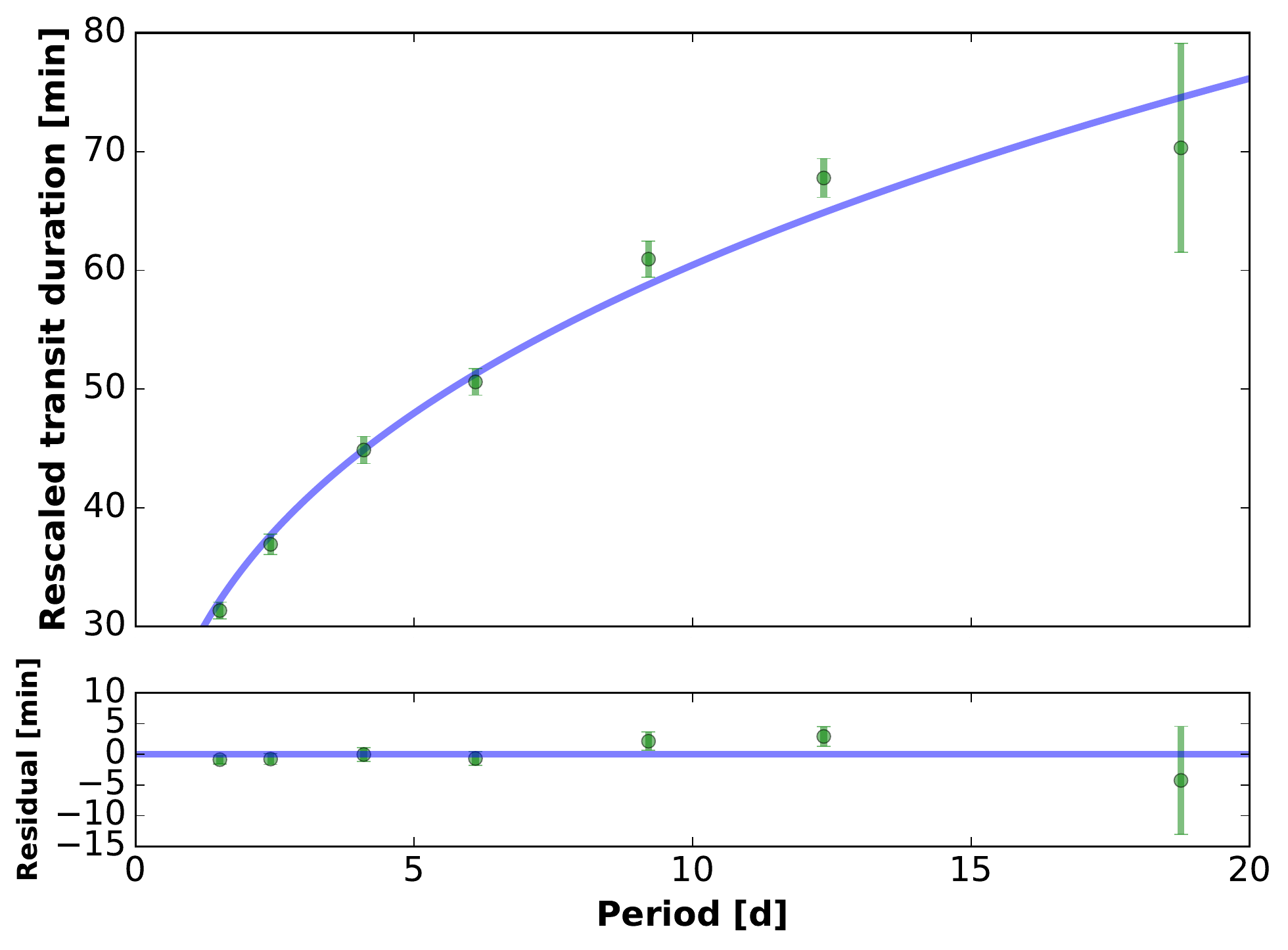}
\caption{Rescaled transit duration, $T^\prime$, versus orbital period.  The median impact parameter, $b$, and radius ratio, $p=r_{\scriptscriptstyle \occulted}/R_\star$, is used to compute $T^\prime=T/((1+p)^2-b^2)^{1/2}$.  A curve with $T^\prime \propto P^{1/3}$ is overplotted assuming a stellar density of $\rho_\star=51 \rho_\odot$.}
\label{fig:duration}
\end{figure}

For a perfectly coplanar, circular planetary system, the planets should have $T^\prime = T/\sqrt{(1+p)^2-b^2} \propto P^{1/3}$, where $b$ is the impact parameter, $p=r_{\scriptscriptstyle \occulted}/R_\star$ is the planet/star radius ratio, $a$ is the semi-major axis, $P$ is the period, and $T$ is the transit duration from first contact to last. Figure~\ref{fig:duration} shows the value of $T^\prime$ for each of the TRAPPIST-1 planets alongside this relation. The transit durations satisfy the above relation surprisingly well, suggesting a very coplanar system. 
 
The coplanarity of the system can be used to place strong prior bounds on the values of $\Delta\Omega$ for each of the planets. To quantify this coplanarity, we generated random planetary systems with planets in different orbital planes and simulated observations of each system, assigning a probability to each based on how well the observed transit durations matched the simulated durations.

\begin{figure}[!t]
\centering
\includegraphics[width=0.47\textwidth]{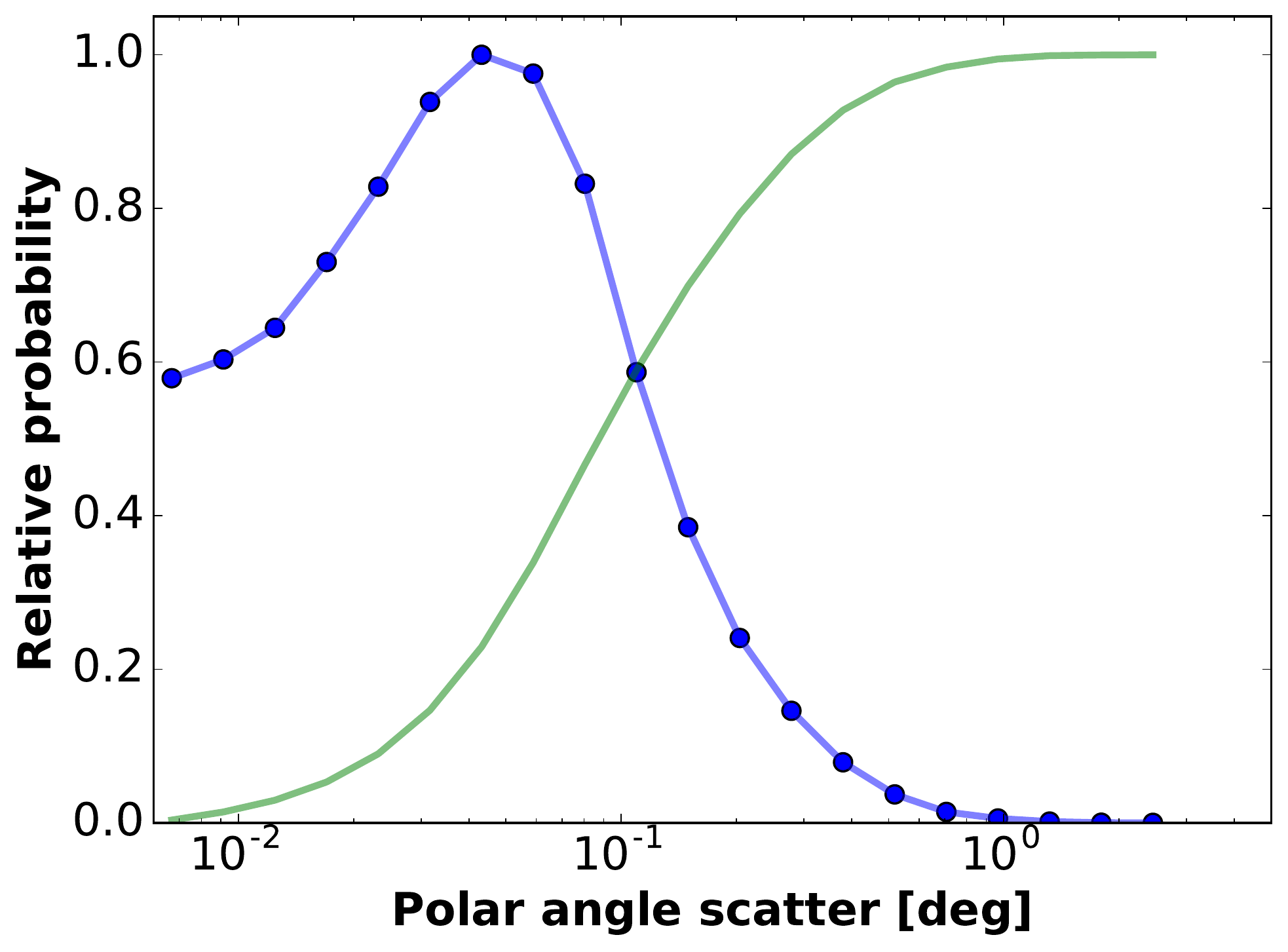}
\caption{Probability of polar angle scatter, $\sigma_\vartheta$, marginalized over $\rho_\star$ and assuming circular orbits for the planets (blue dots/line).  Cumulative probability distribution shown in green.}
\label{fig:coplanarity}
\end{figure}

We randomized orbits with periods drawn from the observed values of the seven planets and zero eccentricity. We justify the assumption of zero eccentricity since for the small ($\la 0.01$) values estimated from the tidal/migration simulations \citep{Luger2017}, the transit duration should only be affected by $<$1\%, which is comparable to the uncertainty on the transit durations of the outer five planets. The inner two planets have smaller duration uncertainty, but are expected to have even smaller eccentricity due to tidal circularization. 
We drew the angular momentum vectors of the planets in each simulated system from a Gaussian in polar angle, $\vartheta$, with an assumed width, $\sigma_\vartheta$, relative to the system angular momentum axis. We drew the azimuthal angle of the angular momentum vector uniformly, and finally drew an observer from a uniform location within 2$^\circ$ of edge-on (since outside this region one or more of the planets are not seen to transit). We made a grid of ($\sigma_\vartheta$, $\rho_\star$) values, where $\rho_\star$ is the stellar density, and computed the expected transit duration for each planet, drawing the planet radius ratios from the values observed by \cite{Gillon2017} and \cite{Luger2017}. Note that we did not fit the observed impact parameters, as these are less well constrained from the observed transits and are thus strongly correlated with one another and with the transit durations. The transit durations, on the other hand, are not correlated amongst the planets as they are well constrained by the data. For each grid point of ($\sigma_\vartheta,\rho_\star)$, we ran $10^6$ simulations of planet system plus observer, computing the probability of the transit durations compared to the durations measured by \cite{Gillon2017};  if one or more simulated planets does not transit, we set the probability to zero.  We summed up the probabilities over all simulations for each grid point, assigning the summed probability to each grid point.  This procedure yields a joint probability distribution estimate of the stellar density and the coplanarity of the planets. \edited{Although this is not as efficient as the analytic approach of \citet{Brakensiek2016}, we expect it to give good constraints on the system parameters given our assumptions.}

\begin{figure}[!t]
\centering
\includegraphics[width=0.47\textwidth]{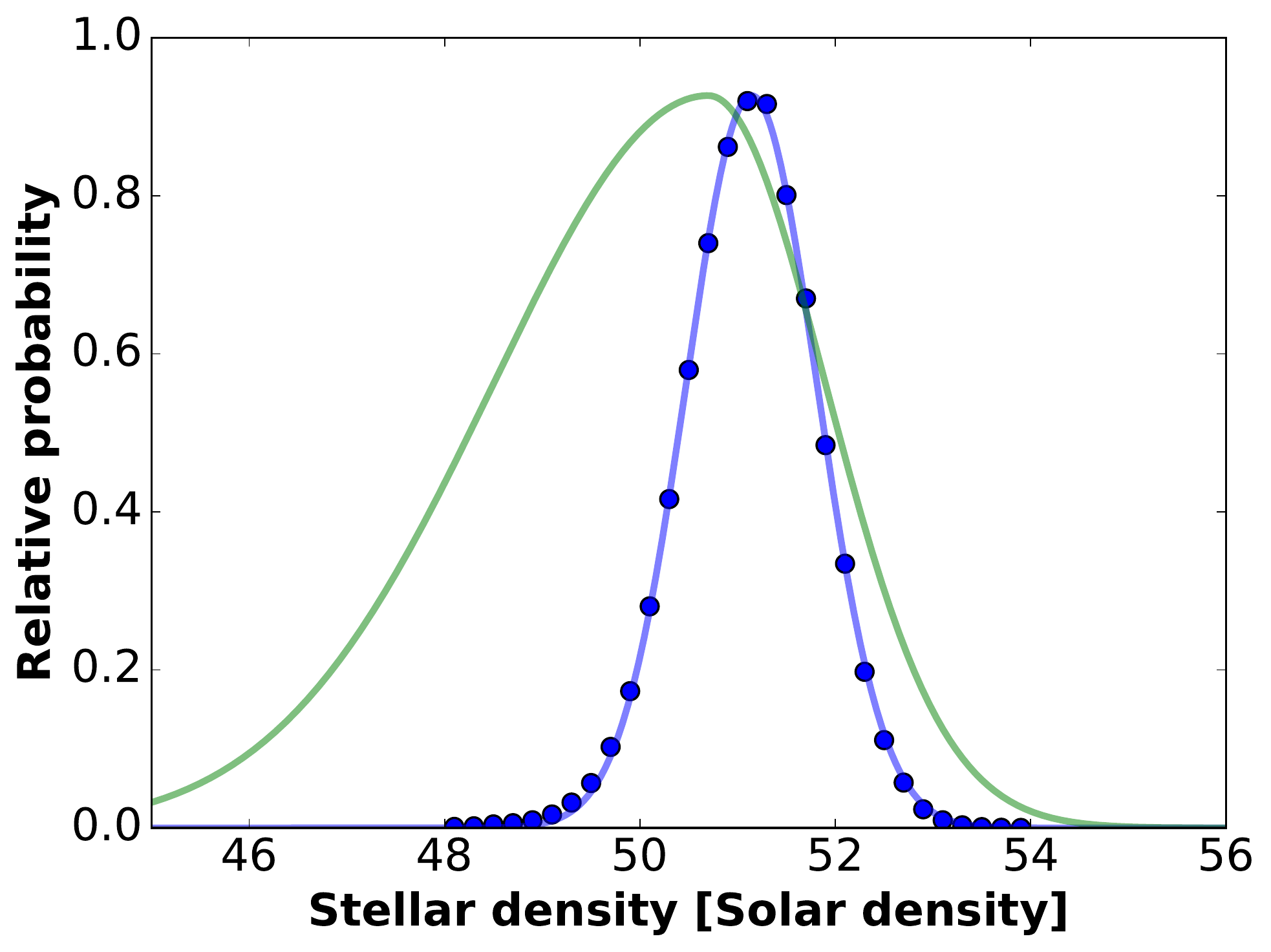}
\caption{Density of the star, $\rho_\star$, in units of the Solar density, $\rho_\odot$, marginalized over $\sigma_\vartheta$.  Best-fit Gaussian (blue curve), and double-sided Gaussian (green) with uncertainties from \cite{Gillon2017}.\nudge}
\label{fig:stellar_density}
\end{figure}

Figure \ref{fig:coplanarity} shows the inferred value of $\sigma_\vartheta$, which indicates that this system is {\it extremely} coplanar, with $\sigma_\vartheta < 0.3^\circ$ at 90\% confidence.  Figure \ref{fig:stellar_density} shows the inferred stellar density of $\rho_\star/\rho_\odot=51.14\pm 0.67 $, which has a narrower distribution than that computed in \cite{Gillon2017}.  The more precise value of the stellar density results from the assumption that the planets' angular momentum vectors are drawn from a single distribution rather than allowing the inclinations of each planet to vary independently.  We have tried relaxing the circularity constraint, and found a consistent result, as expected.

\subsubsection{Sample integration of TRAPPIST-1}
\label{sec:results:dynamics:sample}
\begin{figure*}[p!]
\centering
\includegraphics[width=\textwidth]{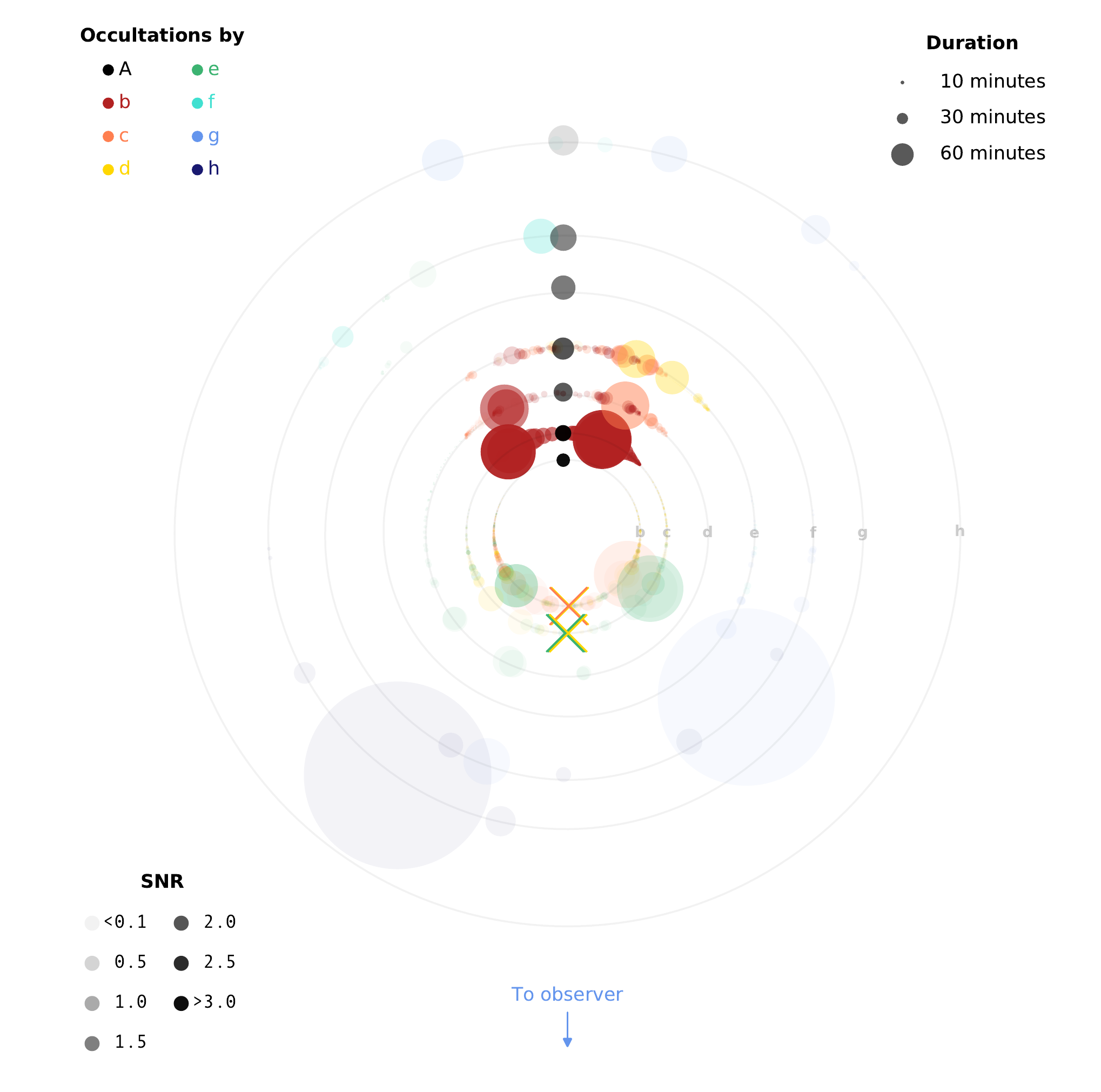}
\caption{Three years of planet-planet occultations in TRAPPIST-1, for a single random draw from the orbital parameter distributions in Table~\ref{tab:sysparams}. The system is seen from above, with the observer located towards the bottom of the plot and planets orbiting counter-clockwise. The initial orbital outlines of each of the seven planets are shown in grey, and each occultation is indicated as a colored circle placed at the location of the occulted planet at the time of the event. Circle colors correspond to different occultors: black is an occultation by the star (i.e., secondary eclipse), red is an occultation by b, and so forth (see legend at top left). Circle sizes are proportional to the event duration (legend at top right), and the \edited{opacity is proportional to the SNR of the occultation as seen by JWST/MIRI at 15 $\mu\mathrm{m}$} (legend at lower left). The ``X''s indicate mutual transits, or planet-planet occultations occurring on the face of the star.}
\label{fig:scatter}
\end{figure*}

At present, the uncertainties on the orbital parameters of the TRAPPIST-1 planets are too large to permit a deterministic prediction of the time of planet-planet occultations. Nevertheless, we may compute the frequency of occultations and their distribution in phase space by drawing planet properties from their respective distributions (Table~\ref{tab:sysparams}). We assume a stellar mass $M_\star = 0.0802 \pm 0.0073\ \mathrm{M_\odot}$ \citep{Gillon2017} and radius $R_\star = 0.121 \pm 0.003\ \mathrm{R_\odot}$ \citep{BurgasserMamajek2017}. Planet radii are self-consistently computed from the stellar radius and the transit depths in Table~\ref{tab:sysparams}. We draw the longitudes of ascending node of each planet from a Gaussian with standard deviation $\sigma_\Omega$, which for simplicity we take to be the same as $\sigma_\vartheta$, whose distribution was derived in \S\ref{sec:results:dynamics:coplanarity}. This is technically incorrect, as the polar angle, $\vartheta$, of the angular momentum vector of a planet has both an inclination component and a longitude of ascending node component, requiring $\sigma_\Omega \leq \sigma_\vartheta$.  The $\sigma_\Omega$ distribution does not have an analytic expression resulting from our dynamical simulations, but our choice is conservative in the sense that the mutual inclinations of the planets are slightly \emph{smaller} than we assume here. In practice, however, since the system is so coplanar, this choice does not significantly affect our conclusions. Finally, we also neglect covariances between the orbital parameters of the different planets, except for the inclinations, which are are significantly correlated. We draw the set of inclinations from the full posterior distributions of \citet{Gillon2017}.

For a given set of orbital parameters, we use \texttt{planetplanet} to integrate the orbits forward in time with the \texttt{REBOUND} N-body code in order to capture dynamical interactions among the planets. \edited{We use the 15$^\mathrm{th}$ order \texttt{IAS15} integrator} with a timestep of one hour, which we find leads to a maximum error in the sky-projected separation of two planets of less than one percent the planet radii over one (Earth) year. We oversample each timestep on a finer 10 second grid, over which we take Keplerian steps. We use Equation~(\ref{eqn:occultation_impact}) to determine when occultations occur.

Figure~\ref{fig:scatter} shows the results of a sample 3-year integration of the TRAPPIST-1 system. System parameters were drawn from the distributions in Table~\ref{tab:sysparams}. The figure shows a top-down view of the system, with the observer located towards the bottom and planets orbiting counter-clockwise. Each planet-planet occultation event is indicated as a circle, whose color, size, and transparency indicate the occultor, the duration of the event, and the \edited{signal-to-noise ratio (SNR) of the occultation, respectively; refer to the legends for details. To compute SNRs, we assume zero albedo ``eyeball'' planets observed in the JWST/MIRI 15 $\mu\mathrm{m}$ filter (see \S\ref{sec:detect})}. The ``X''s indicate mutual transits, which are planet-planet occultations on the disk of the star (see \S\ref{sec:results:mutual}).

For this particular set of system parameters, \edited{1,499} planet-planet occultations occur among the TRAPPIST-1 planets over the course of 3 years, averaging to about \edited{1.4} per (Earth) day. We find that this is a typical rate when marginalizing over the uncertainties on the orbital parameters. Nevertheless, it is evident from the figure that the vast majority of occultations are very short-lived ($\ll 10$ minutes), which poses challenges to their detectability. Longer, higher SNR occultations are less frequent, but still occur several tens of times per year for certain planet pairs (see \S\ref{sec:results:dynamics:stats}). \edited{Occultations of b and c (circles along the two innermost orbital tracks) are the most common and occur over a wide range of orbital phases.}
Their duration is a strong function of phase: occultations \edited{of b} on the near side of the star (mean longitude $\lambda_b \approx -90^\circ$) last upwards of 30 minutes, while those on the far side of the star ($\lambda_b \approx 90^\circ$) last on the order of tens of seconds and are not visible in the figure. This is because the former case happens when the sky-projected velocity vectors of b and \edited{its occultor (usually c)} are aligned, resulting in the smallest relative velocity among the two planets, while the latter case happens when the occultor is on the near side of the star and b is on the far side, such that the two planets are moving in opposite directions on the sky and thus have large relative velocities. \bugfix{Conversely, the longest occultations of c occur on the far side of the star, sometimes exceeding a few hours (see below). Occultations of the other planets span a wide range of durations. Most notably, an occultation of f by g and one of g by h last ${\sim}8$ hours due to the longer orbital periods of the outer planets.}

\begin{figure}[!hbt]
\centering
\includegraphics[width=0.47\textwidth]{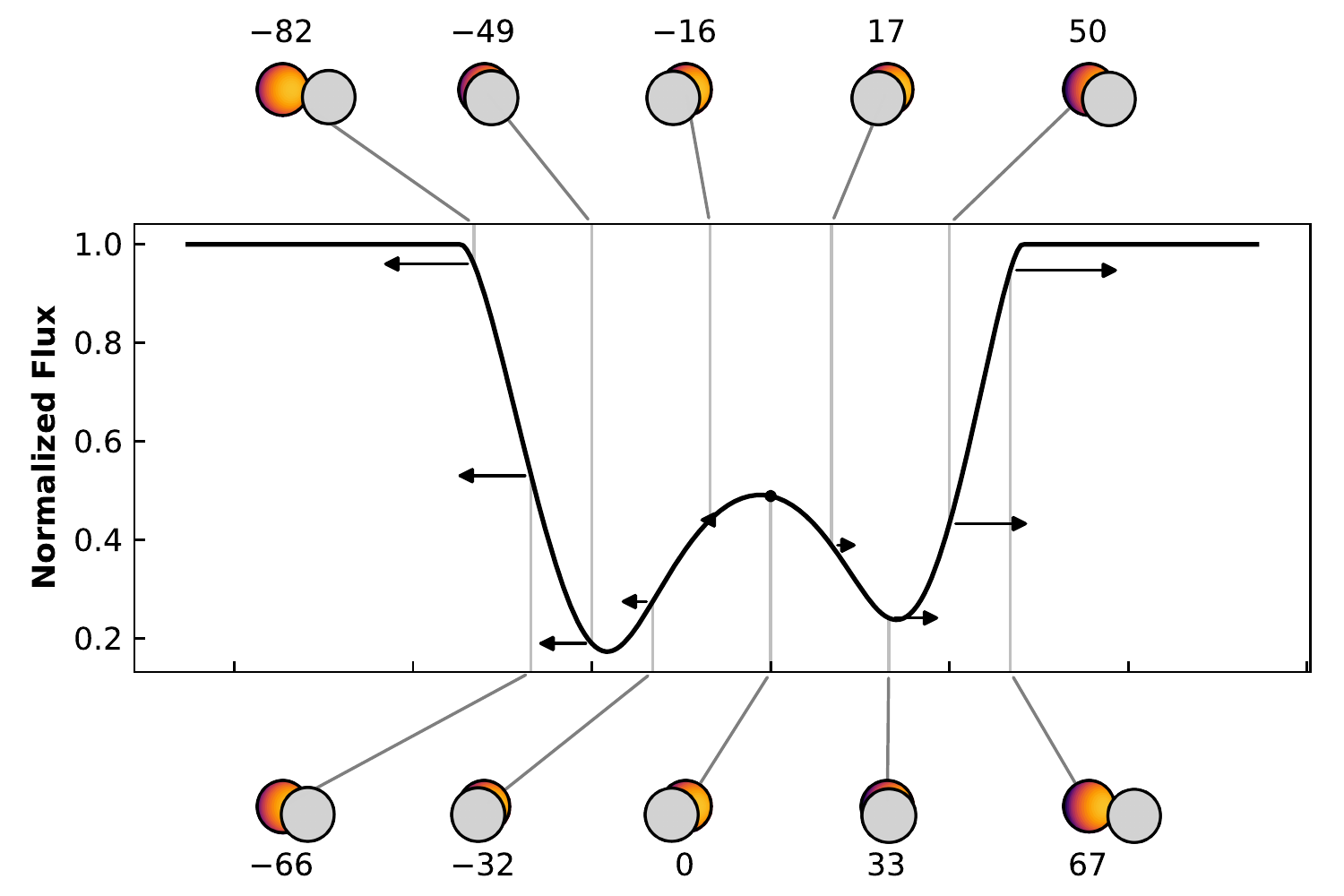}
\caption{\edited{A PPO doublet in TRAPPIST-1. This is an occultation of planet c (colored disks) by planet b (gray disks) when c is close to full phase. The time in minutes is indicated above or below each image, and an arrow indicates the relative velocity vector of the two planets. The ``W'' light curve shape is due to the fact that the relative velocity changes sign at $t = 0$. This occultation lasts about 160 minutes. An animation of this PPO can be viewed with the code on \github.}\nudge}
\label{fig:doublet}
\end{figure}

\bugfix{The highest SNR PPOs are occultations of c by b occurring on the far side of the star. A few stand out, in particular, as they are much longer than the others and approach $\mathrm{SNR}{\sim}3$. These are repeated occultations, in which the relative velocity of the two planets changes sign one or more times over the course of the event, effectively extending the duration of the occultation. The highest SNR one is an occultation by b just before new phase (large red circle to the right of secondary eclipse). It begins when b has just passed quadrature ($\lambda_b \approx 0^\circ$), moving primarily away from the observer, while c is closer in phase to secondary eclipse, moving faster on the sky plane. The occultation is prolonged due to the fact that as b passes quadrature it speeds up on the sky and catches up to c. Eventually, as b approaches secondary eclipse, its sky velocity exceeds that of c and c emerges from the occultation. For clarity, we refer to an occultation that is prolonged due to a change of sign in the planets' relative velocity as a ``doublet.'' This applies to events that appear as a single, continuous occultation or to two occultations separated in time but happening in the same orbit; see Figure~\ref{fig:doublet}. In some cases (see \S\ref{sec:results:photo:jwst}) there may be \emph{two} successive changes of sign in the relative velocity, which we refer to as ``triplets.''}

\bugfix{Doublets can be seen for other planets as well, in particular for occultations of b and c by e (light green circles on either side of new phase) and occultations of d by b and c (light red and orange circles on either side of full phase). These are less detectable than doublets of planet c, but are also quite common. In general, doublets in which c is occulted by b are likely to be the most detectable PPOs in TRAPPIST-1 because of their high SNR.} Given semi-major axes $a_b$ and $a_c$ of b and c, respectively, these necessarily occur when c is on the far side of the star with mean longitude
\begin{align}
    \label{eqn:delta_theta}
    \frac{\pi}{2} - \sin^{-1}\left(\frac{a_b}{a_c}\right) &\leq \lambda_c \leq \frac{\pi}{2} + \sin^{-1}\left(\frac{a_b}{a_c}\right)\nonumber\\
    43^\circ &\lesssim \lambda_c \lesssim 137^\circ.
\end{align}
These are therefore occultations of primarily the day side of c. \edited{While occultations by b typically last between 30 and 60 minutes, doublets like the ones seen in the figure can last over 2 hours.} Moreover, since b is slightly larger than c, these are often full or near-full occultations, although the impact parameter of the occultation is sensitive to the longitudes of ascending nodes of the two planets. Figure~\ref{fig:doublet} shows an example of such a doublet.

\subsubsection{PPO statistics for TRAPPIST-1}
\label{sec:results:dynamics:stats}
The results presented in the previous section were for a single draw from the distributions of orbital parameters allowed by observations of the system up to the present time. Planet-planet occultations are aperiodic, and therefore different realizations of the system lead to very distinct occultation maps like the one shown in Figure~\ref{fig:scatter}. It is therefore instructive to consider the statistics of PPOs in TRAPPIST-1 when marginalizing over the uncertainties on the orbital parameters, so that we may quantify the expected frequency of PPOs given all the information known at present. We therefore ran an ensemble of 50,000 1-year integrations of TRAPPIST-1, each time drawing orbital parameters from their respective distributions and integrating the system forward in time as before. \edited{We performed two such simulations, one for each atmospheric limit (uniform and ``eyeball'' planets) so that we may compute statistics on the expected SNR of the occultations.}

\begin{figure}[!t]
\centering
\includegraphics[width=0.47\textwidth]{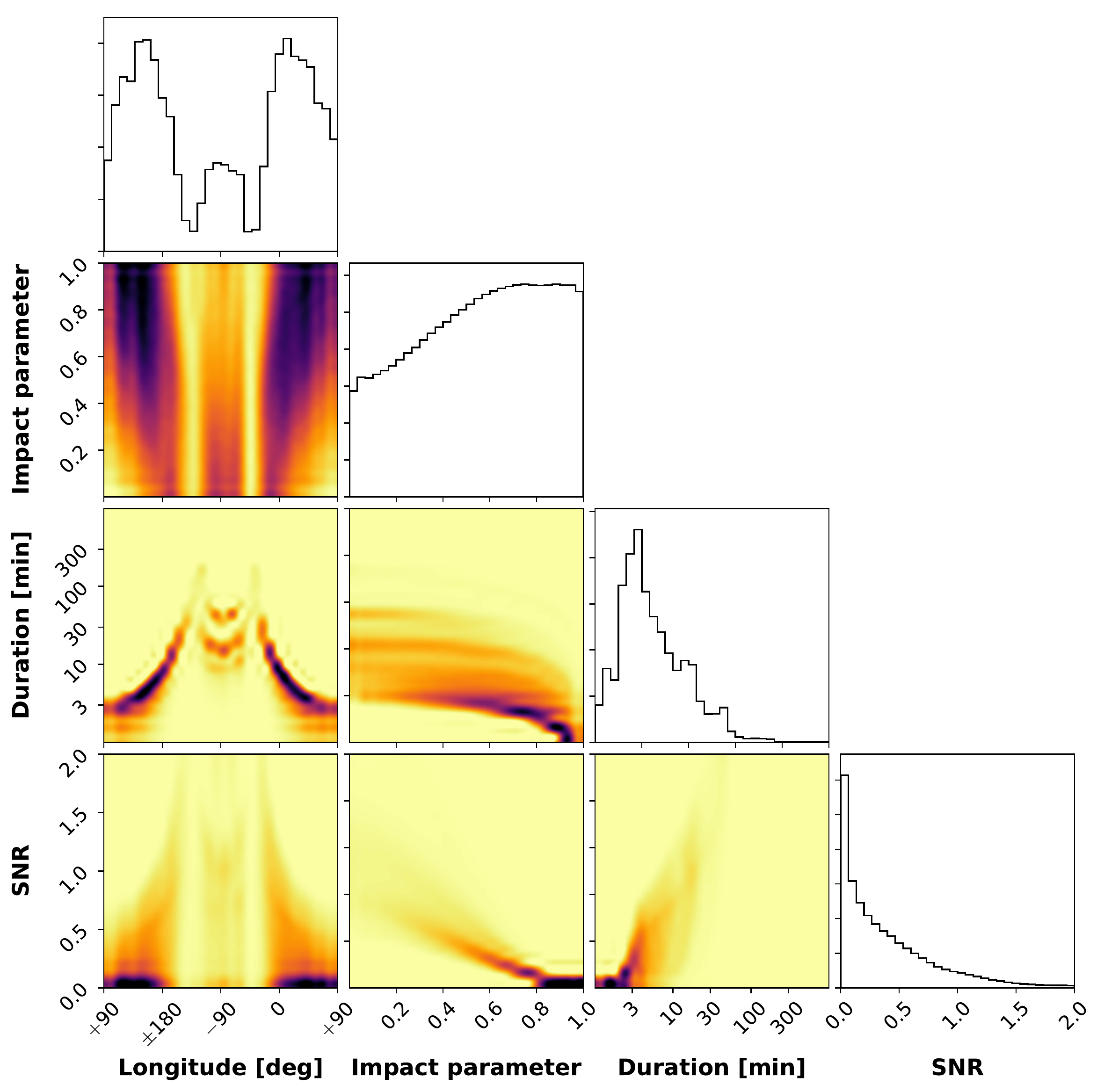}
\caption{\edited{Posterior distributions for the mean longitude, impact parameter, duration, and JWST/MIRI 15 $\mu\mathrm{m}$ signal-to-noise ratio (SNR) of all occultations of TRAPPIST-1b, assuming a thick atmosphere with uniform radiance. These were plotted using the \texttt{corner} package \citep{ForemanMackey2016}. Panels at the top of each column are the marginalized distributions; other panels are the joint posteriors, where the darkness is linearly proportional to the probability density. These distributions are marginalized over the uncertainties on the orbital parameters of all bodies in the system. Occultations of b are piled up in phase near quadrature and transit. Grazing, low SNR occultations are most common, with a long tail to occultations exceeding \bugfix{SNR$\sim$1.5}; these are the most detectable ones, occurring primarily near quadrature. Occultations lasting about 5 minutes are typical, but a small fraction can exceed one hour.}}
\label{fig:b_corner}
\end{figure}

Figure~\ref{fig:b_corner} shows the posterior distributions for the mean longitude $\lambda$, impact parameter $b$, duration $\Delta t$, \edited{and MIRI 15 $\mu\mathrm{m}$ SNR} of all planet-planet occultations of TRAPPIST-1b assuming uniform radiance. These include occultations by all planets, though occultations by c are the most common. PPOs occurring behind the star (i.e., during secondary eclipse) are not included in the figure, as they are not observable. The plots at the top of each column are the fully marginalized distributions. These show that occultations of b occur primarily near quadrature ($\lambda_b \approx 0^\circ$ or $180^\circ$), and somewhat less frequently near transit ($\lambda_b \approx -90^\circ$). The duration peaks around 5 minutes, with a long tail toward long ($> 1$ hour) occultations. \edited{The SNR of the occultations peaks at very low values: the vast majority of PPOs of b are either very short and/or grazing events that are not individually detectable with JWST. However, the SNR distribution also shows a long tail extending past \bugfix{SNR$\sim$1.5}. As we show below, these occultations occur, on average, about ten times per (Earth) year and may be jointly detectable.}

Below the marginalized distributions, we plot the joint posterior distributions for each combination of the three parameters. \edited{There is a strong correlation between longitude and duration:} the shortest events are those where b is on the far side of the star near secondary eclipse, when the occultor is necessarily on the near side of the star moving in the opposite direction. Conversely, for $\lambda_b \approx -150^\circ$ or $-30^\circ$, the long ``doublet'' occultations of b by c discussed in \S\ref{sec:results:dynamics:sample} are possible, resulting in events that can exceed two hours. \edited{The SNR is also strongly correlated with the duration: longer events typically have the highest SNR. These primarily occur very close to quadrature or in the vicinity of transits of b.}

\begin{figure}[!t]
\centering
\includegraphics[width=0.47\textwidth]{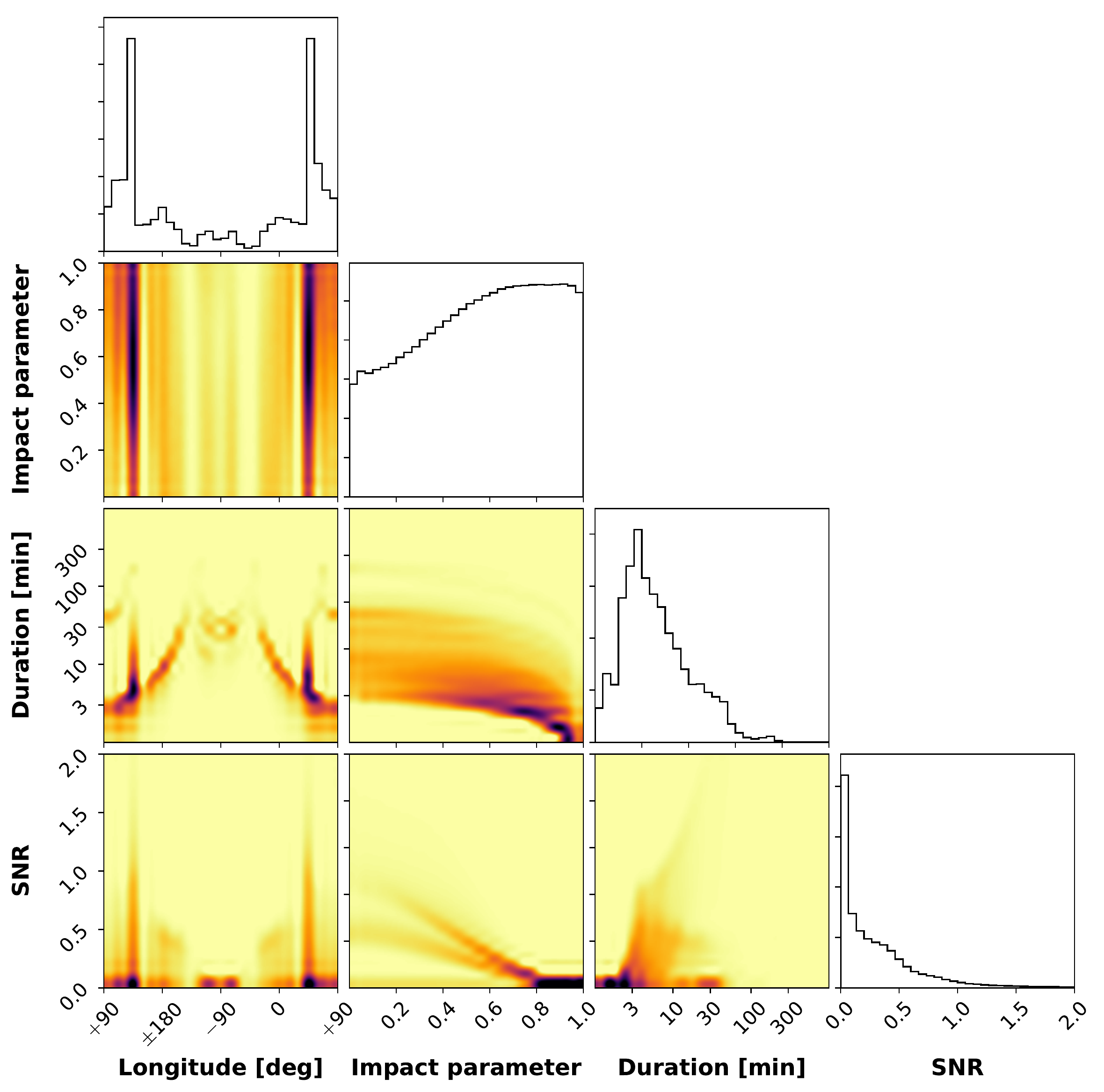}
\caption{\edited{Same as Figure~\ref{fig:b_corner}, but for occultations of TRAPPIST-1c and assuming an ``eyeball'' planet with a dark night side. The distributions are qualitatively similar to those of b, although now the vast majority of occultations are by b, occurring halfway between quadrature and secondary eclipse, in a narrow region near $\lambda_c \approx 43^\circ$ and $\lambda_c \approx 137^\circ$ (see Equation~\ref{eqn:delta_theta}).}}
\label{fig:c_corner}
\end{figure}

Figure~\ref{fig:c_corner} shows the same distributions, but for occultations of TRAPPIST-1c \edited{and this time assuming an ``eyeball'' planet}. The distributions of impact parameter, duration \edited{and SNR} are quite similar to those of b, but the \edited{longitude} posterior now peaks at $\lambda_c \approx 43^\circ$ or $137^\circ$, as expected from Equation~(\ref{eqn:delta_theta}). These are primarily occultations by b when its sky-projected velocity is close to zero, resulting in a much higher occultation probability. Even though our calculations marginalize over the uncertainties on all orbital parameters, the two peaks in the histogram are \emph{extremely} narrow, with FWHM $\sim$ 2$^\circ$, as this width is due primarily to the uncertainty on the semi-major axes of the two planets. \edited{The peaks in the longitude posterior coincide with the peaks in the SNR, which also extend past \bugfix{SNR$\sim$1.5}. As with occultations of b, these may be detectable by JWST if several are observed.}

\edited{Occultation statistics of all TRAPPIST-1 planets, in both atmospheric regimes, can easily be plotted with the tools made available on \github. Plots for planets d--h are qualitatively similar to those shown here, although the SNR of the occultations is significantly lower.}

\begin{figure*}[!p]
\centering
\includegraphics[width=0.47\textwidth]{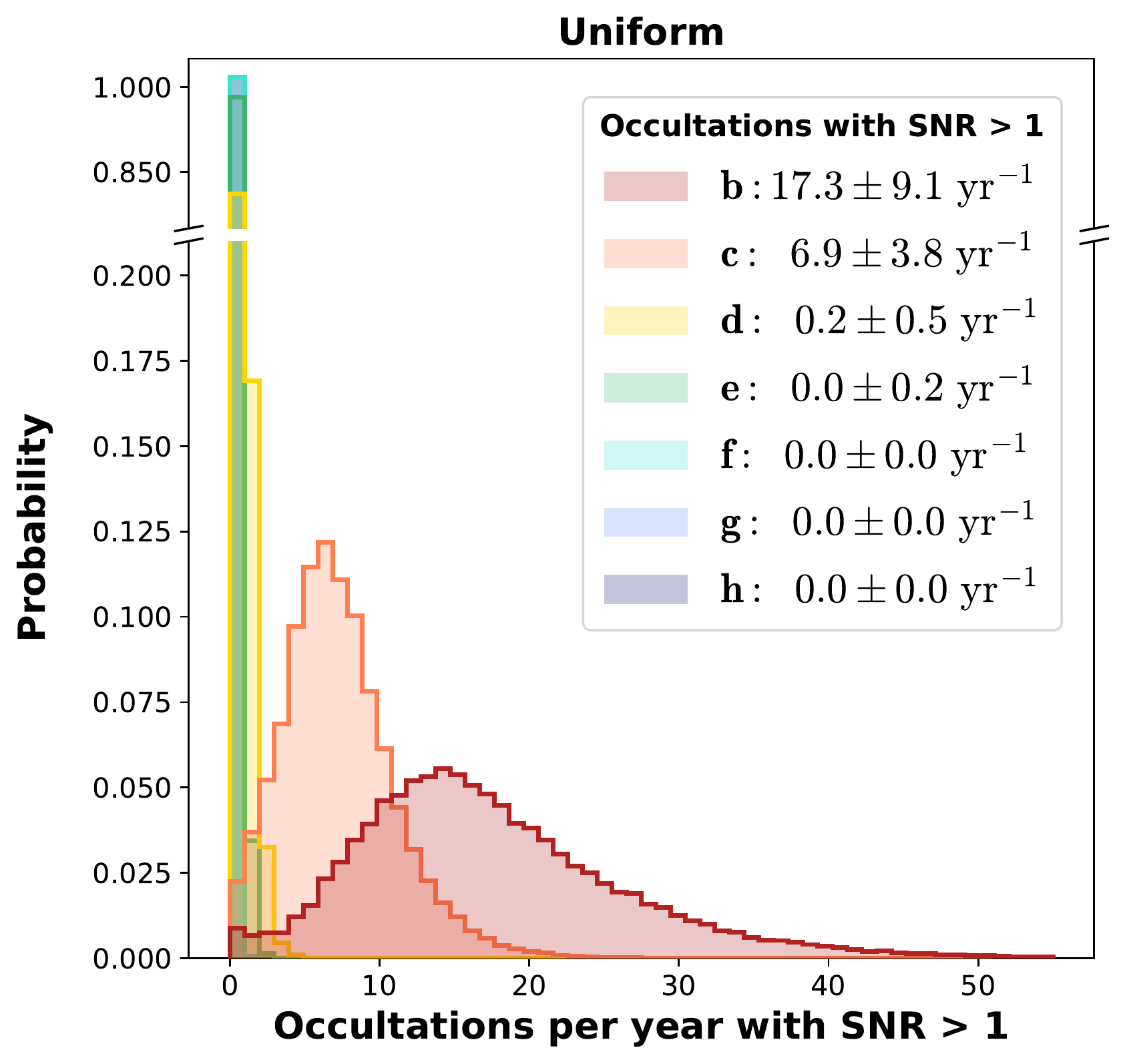}
\includegraphics[width=0.47\textwidth]{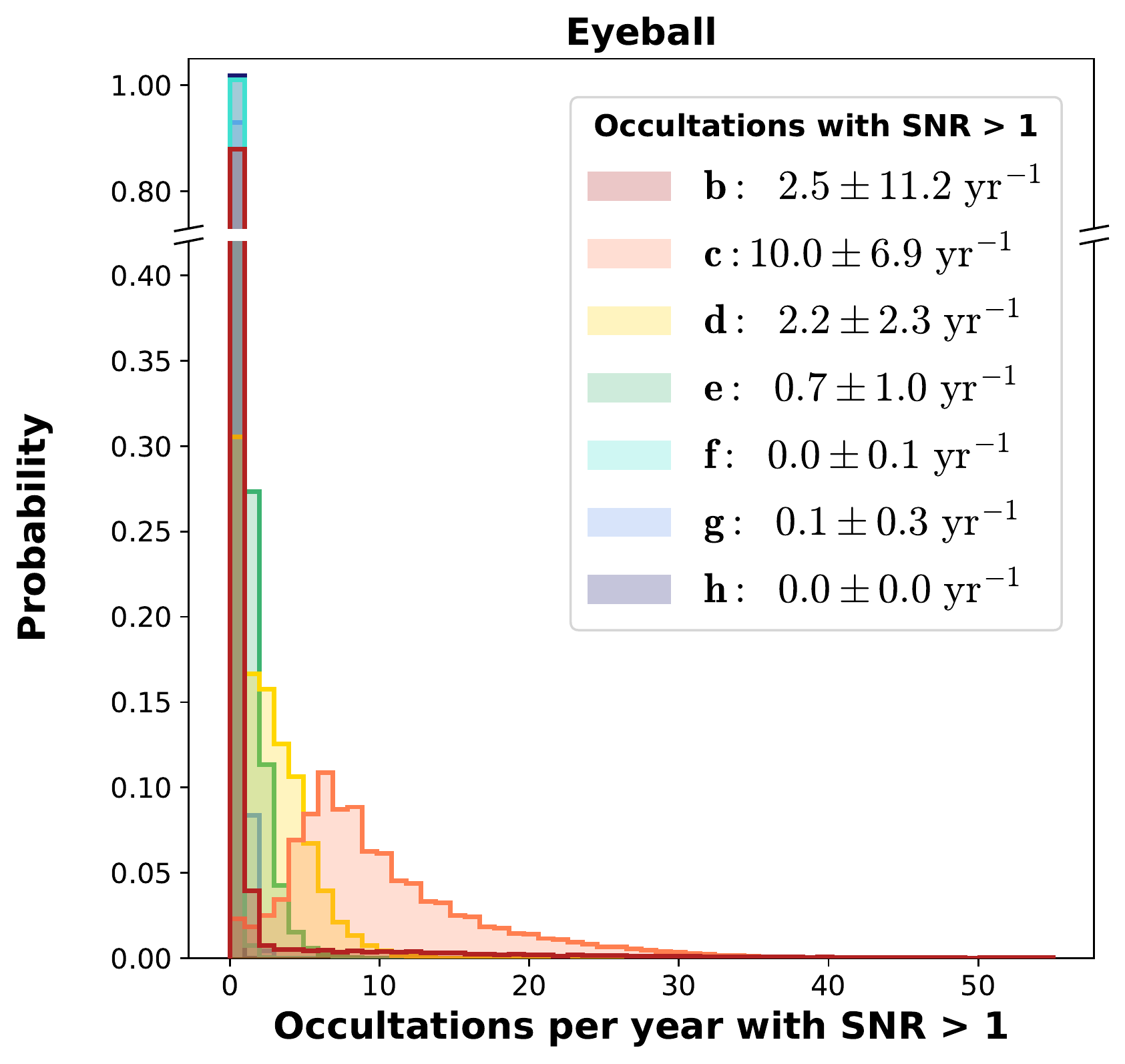}
\caption{\edited{Frequency of planet-planet occultations in TRAPPIST-1 for each of the planets, marginalized over all system parameters. Only occultations with \bugfix{SNR$>$1} in the 15 $\mu\mathrm{m}$ JWST/MIRI filter are included. The legend shows the expectation value and the standard deviation for the occultation frequency of each planet. (Left) SNR computed assuming uniform planets with no day/night temperature contrast (thick atmosphere limit). (Right) SNR computed assuming ``eyeball'' planets with 40 K night sides. \bugfix{Depending on the atmospheric regime of the TRAPPIST-1 planets, there may be ${\sim}10-20$ PPOs of planets b and c with SNR$>$1 each year.}}}
\label{fig:hist}
\end{figure*}

\begin{figure*}[!p]
\centering
\includegraphics[width=0.47\textwidth]{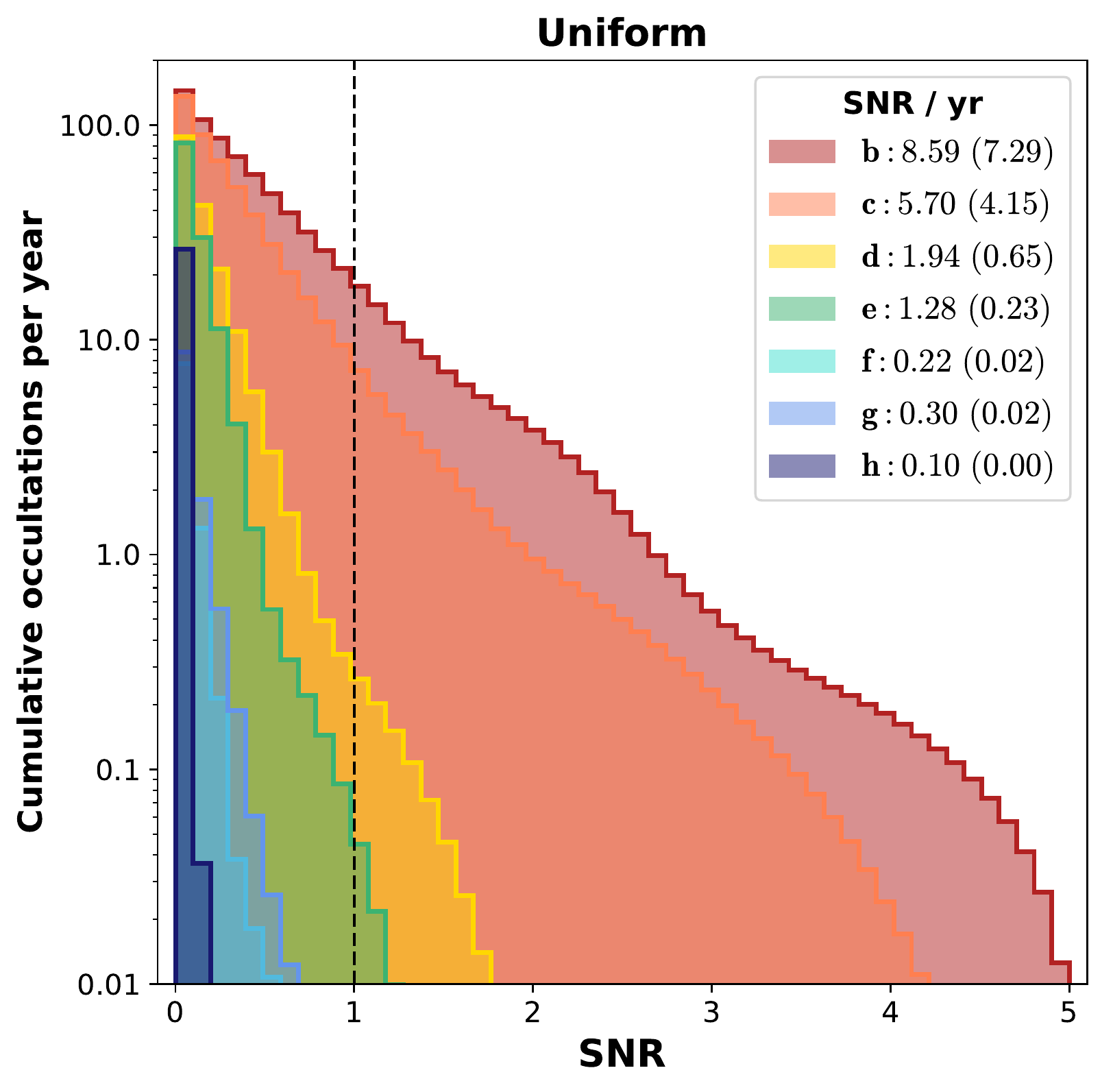}
\includegraphics[width=0.47\textwidth]{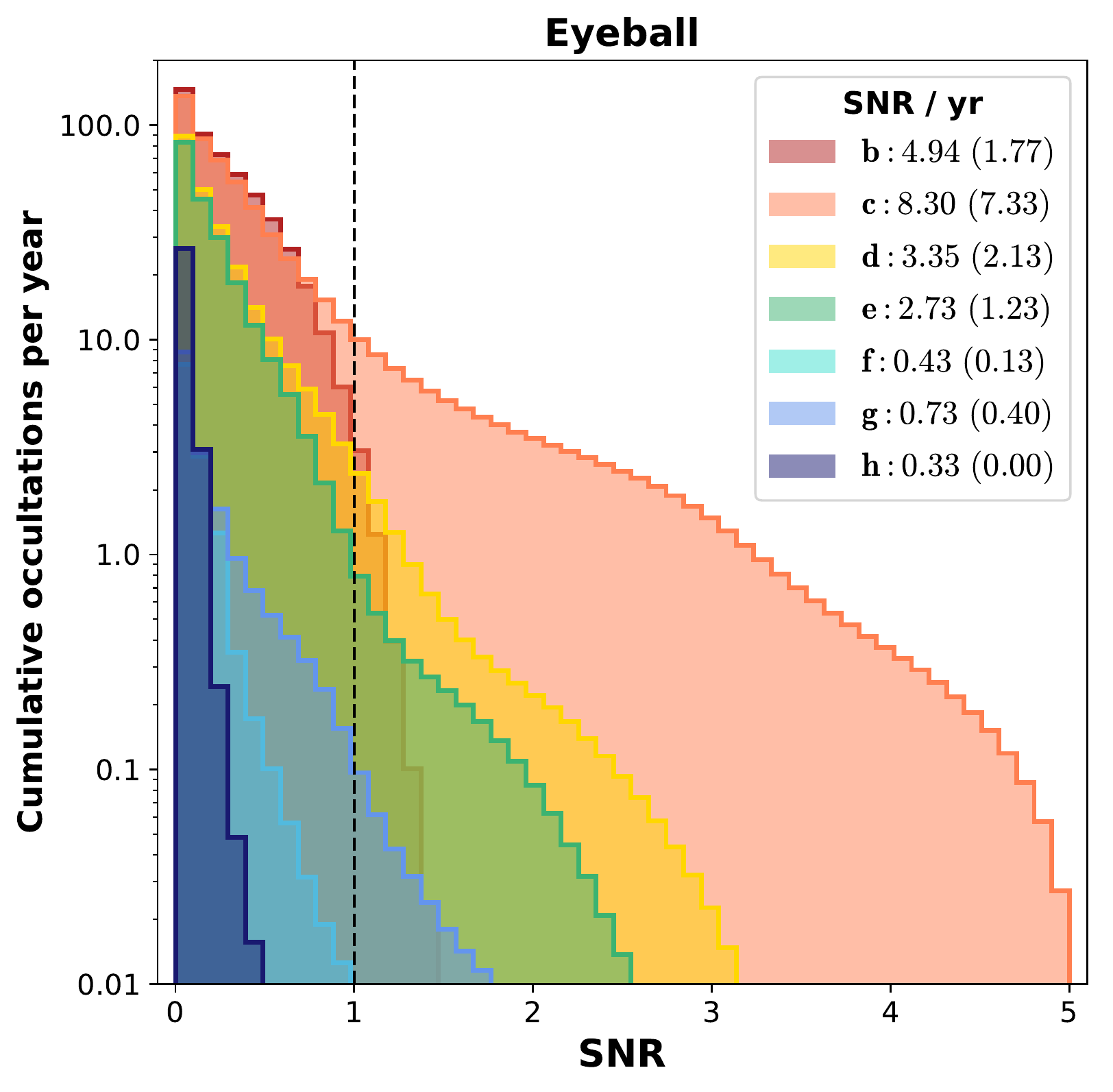}
\caption{\edited{Cumulative distributions of the expected number of occultations per year above a certain SNR for each of the seven TRAPPIST-1 planets, under two atmospheric limits: uniform radiance (left) and ``eyeball'' planets (right). For reference, the dashed vertical line highlights the expected number of occultations with \bugfix{SNR$>$1}, which is the mean of the distributions shown in Figure~\ref{fig:hist}. Very low SNR occultations are the most common, but the distributions for planets b and c have long tails to SNR exceeding \bugfix{5}. 
The legend indicates the expected total SNR per year for each of the planets, which is the quadrature sum of the SNR of all occultations. In parenthesis is the expected SNR when accounting only for occultations with individual \bugfix{SNR$>$1}. For TRAPPIST-1b and c, high SNR occultations contribute most of the total signal. Observations with JWST/MIRI at 15 $\mu\mathrm{m}$ can thus detect PPOs in TRAPPIST-1 with \bugfix{SNR$\gtrsim$7} by observing \bugfix{${\sim}10-20$} PPOs in one year.
}}
\label{fig:snr_hist}
\end{figure*}

\edited{In Figure~\ref{fig:hist}, we marginalize over all system parameters and plot histograms of the number of ``potentially observable'' occultations of each of the planets expected over the course of one (Earth) year under the two atmospheric regimes: uniform (left) and ``eyeball'' (right). We define ``potentially observable'' occultations as those with \bugfix{$\mathrm{SNR} > 1$}, for which less than \bugfix{25} must be observed for a $\mathrm{SNR} = 5$ joint detection. The mean and standard deviation from a Gaussian fit to these histograms are shown in the legends. Given our current knowledge of the orbital parameters of TRAPPIST-1, we expect there to be \bugfix{$17.3 \pm 9.1$ and $6.9 \pm 3.8$} potentially observable occultations per year of TRAPPIST-1b and TRAPPIST-1c, respectively, if they are uniform emitters. If these planets have dark night sides and the day side emission dominates, occultations of planet c are more common, at \bugfix{$10.0 \pm 6.9$ per year (versus $2.5 \pm 11.2$ per year for b)}. Note the extremely high variance in these estimates, originating in the uncertainty in the orbital parameters. In particular, if TRAPPIST-1b has a dark night side, there is an \bugfix{86\%} probability that it will have no \bugfix{$\mathrm{SNR}>1$} occultations in a given year. This is because occultations of the day side of TRAPPIST-1b can only occur by planets on the opposite side of the star and are therefore extremely short, low SNR events. If TRAPPIST-1b has a bright night side, on the other hand, the probability that there will be no significant occultations is less than \bugfix{1\%}, as night side occultations by c are common for nearly all configurations of the system allowed by the prior. The probability of no significant occultations of TRAPPIST-1c is \bugfix{about 2\%} in both cases. Note also that the distributions for both planets have long tails extending to upwards of \bugfix{30} occultations per year, particularly in the ``eyeball'' limit. Finally, high SNR occultations of the other planets are rare and are likely not detectable with JWST. The exception to this is TRAPPIST-1d, which could undergo \bugfix{a few} potentially detectable occultations per year if it has a strong day/night temperature contrast.

In Figure~\ref{fig:snr_hist}, we plot the number of occultations per year with SNR above a certain threshold for all TRAPPIST-1 planets, again in both atmospheric regimes. These show that a uniform radiance TRAPPIST-1b undergoes, on average, PPOs with $\mathrm{SNR}>1$ about \bugfix{20} times per year, PPOs with $\mathrm{SNR}>2$ about \bugfix{4 times per year}, and so forth. An ``eyeball'' TRAPPIST-1c experiences $\mathrm{SNR}>1$ PPOs about \bugfix{10} times per year, and $\mathrm{SNR}>2$ PPOs \bugfix{also about 4 times per year}. In the plot legends, we indicate the total average SNR of all occultations of each of the planets in one year; this is equal to the quadrature sum of all events involving a given planet. Depending on the atmospheric properties of the planets, if one were to observe all occultations of b or c in a given year, these could be jointly detected with \bugfix{$\mathrm{SNR}{\gtrsim}8$}. In parentheses, we indicate the total SNR if only ``potentially detectable'' \bugfix{($\mathrm{SNR}>1$)} occultations are observed. Interestingly, the total SNR does not decrease significantly, as most of the signal originates in the highest SNR occultations. Thus, if one were to observe the \bugfix{${\sim}20$} highest SNR occultations of TRAPPIST-1b or the \bugfix{${\sim}10$} highest SNR occultations of TRAPPIST-1c, a \bugfix{$\mathrm{SNR}>7$} detection of PPOs would still be possible in a single year.

A few words of caution are in order at this point. As we mentioned in \S\ref{sec:detect}, the SNR is only equivalent to the significance of the detection (in standard deviations) when the occultation model has a single degree of freedom. Given our current (lack of) knowledge of many of the orbital parameters of the TRAPPIST-1 planets, the large number of degrees of freedom in our model would likely preclude a robust detection of PPOs with JWST by jointly modeling events that are not individually detectable in the data. However, the precision on the masses, inclinations and eccentricity vectors for the TRAPPIST-1 planets is expected to increase substantially in the next few years as TTV measurements and eventually secondary eclipse measurements reduce the uncertainties on these values. This will in turn greatly reduce the uncertainty on the timing and properties of individual PPOs, increasing the statistical significance of a detection involving multiple events. We return to this point in \S\ref{sec:discussion:bestcase}.

\bugfix{Alternatively, Figure~\ref{fig:snr_hist} suggests that events with $\mathrm{SNR}{\gtrsim}4$ may occur once every few years, on average. While rare, these could be potentially individually detectable with JWST. These are long-lasting PPO doublets that are in general only possible if the mutual inclination of TRAPPIST-1b and c is small, leading to a full or near-full occultation that lasts several hours.}

Finally, we note that the histograms in Figure~\ref{fig:snr_hist} correspond to the \emph{mean} occultation rate when considering the present-day uncertainties on the orbital parameters. The variance in these estimates is high and the values are likely to change as more orbital information becomes available. However, even if the orbital parameters of all planets are known exactly, the aperiodicity of PPOs will still lead to yearly variations in the occultation rate, albeit with much smaller variance than that shown in Figure~\ref{fig:hist}.
}

\subsection{Photometry}
\label{sec:results:photo}

We used \planetplanet to produce synthetic light curves of TRAPPIST-1 over a range of wavelengths, modeling all transits, secondary eclipses, planet-planet occultations, and phase curves. Stellar variability is not modeled. \edited{All planet and system parameters are drawn from the same distributions as in the previous sections. Below we present sample system light curves and discuss their features.}

\subsubsection{Sample light curve}
\label{sec:results:photo:sample_lc}

\begin{turnpage}
\leavevmode
\begin{figure*}[!p]
\begin{center}
\leavevmode
\includegraphics[width=1.3\textwidth]{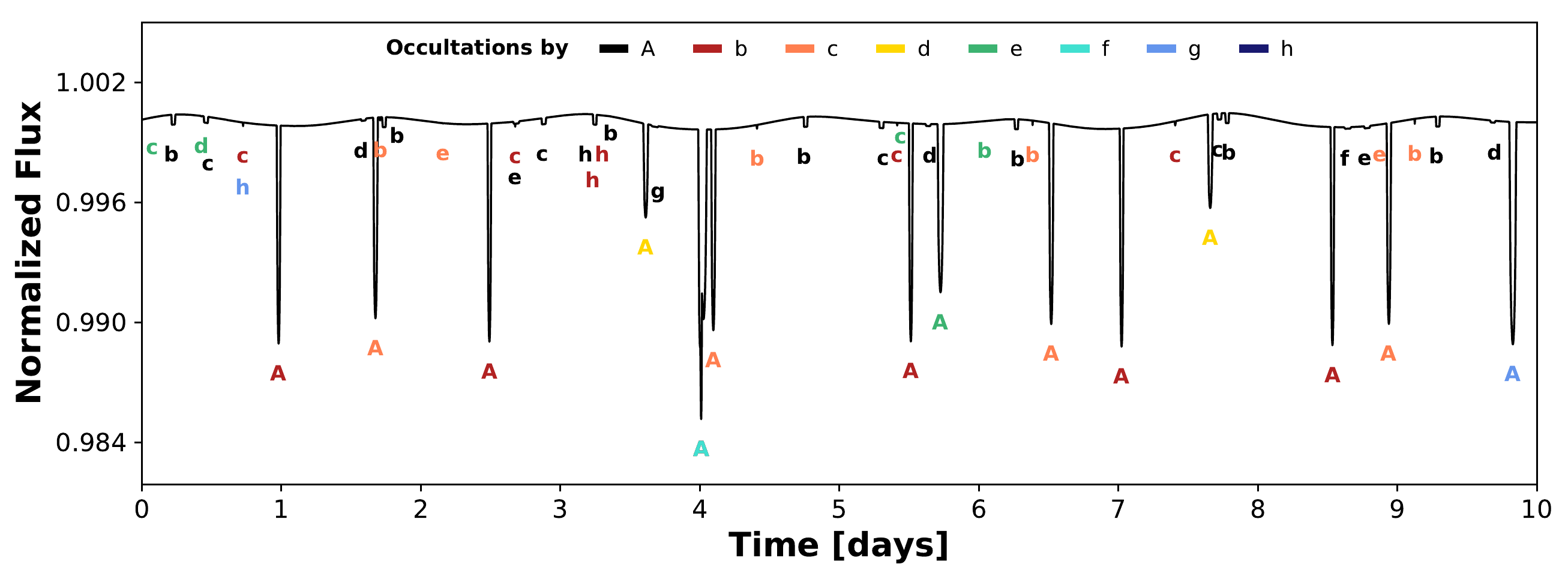}
\caption{A light curve of TRAPPIST-1 over ten days at 15 $\mu$m, generated from a random draw from the orbital parameter distributions (Table~\ref{tab:sysparams}). All transits, secondary eclipses, and planet-planet occultations are labeled with the occulted body name and colored according to the occultor (see legend at the top). The planets are assumed to have thin atmospheres, giving rise to a bright day side and a dark night side. This results in distinct phase curves, which are visible in the figure; the overall signal is dominated by the phase curves of b and c. Stellar variability is not modeled. Note that a simultaneous (but not mutual) transit of b and f occurs at $t = 4$ days.}
\label{fig:lightcurve}
\end{center}
\end{figure*}
\end{turnpage}

Figure~\ref{fig:lightcurve} shows the full light curve of the TRAPPIST-1 system over the course of ten days for a random realization. The light curve is computed at a wavelength of 15 $\mu$m and is normalized to a mean of unity. Planets are assumed to be airless ``eyeballs'', giving rise to the prominent phase curve modulation. All events are labeled with the name of the occulted body and colored according to the occultor. Transits by nearly all planets (labeled ``A'') are evident, with depths on the order of 1\%. Secondary eclipses (black labels) are also visible for most planets, with depths ranging from 0.05\% (500 ppm) for b to 0.0015\% (15 ppm) for h. These are also the magnitudes of the phase curve amplitudes; the overall phase curve is dominated by that of b and c. All other labeled events are planet-planet occultations (4 of b, 6 of c, 1 of d, 2 of e, and 3 of h). The occultation of c by b at $t =$ 7.4 days has a depth of 200 ppm, and that of b by c at $t =$ 9.1 days has a depth of 250 ppm, which is within a factor of 2 of the depth of their secondary eclipses.

\subsubsection{Sample occultation}
\label{sec:results:photo:sample_ppo}
\begin{figure}[!ht]
\centering
\includegraphics[width=0.47\textwidth]{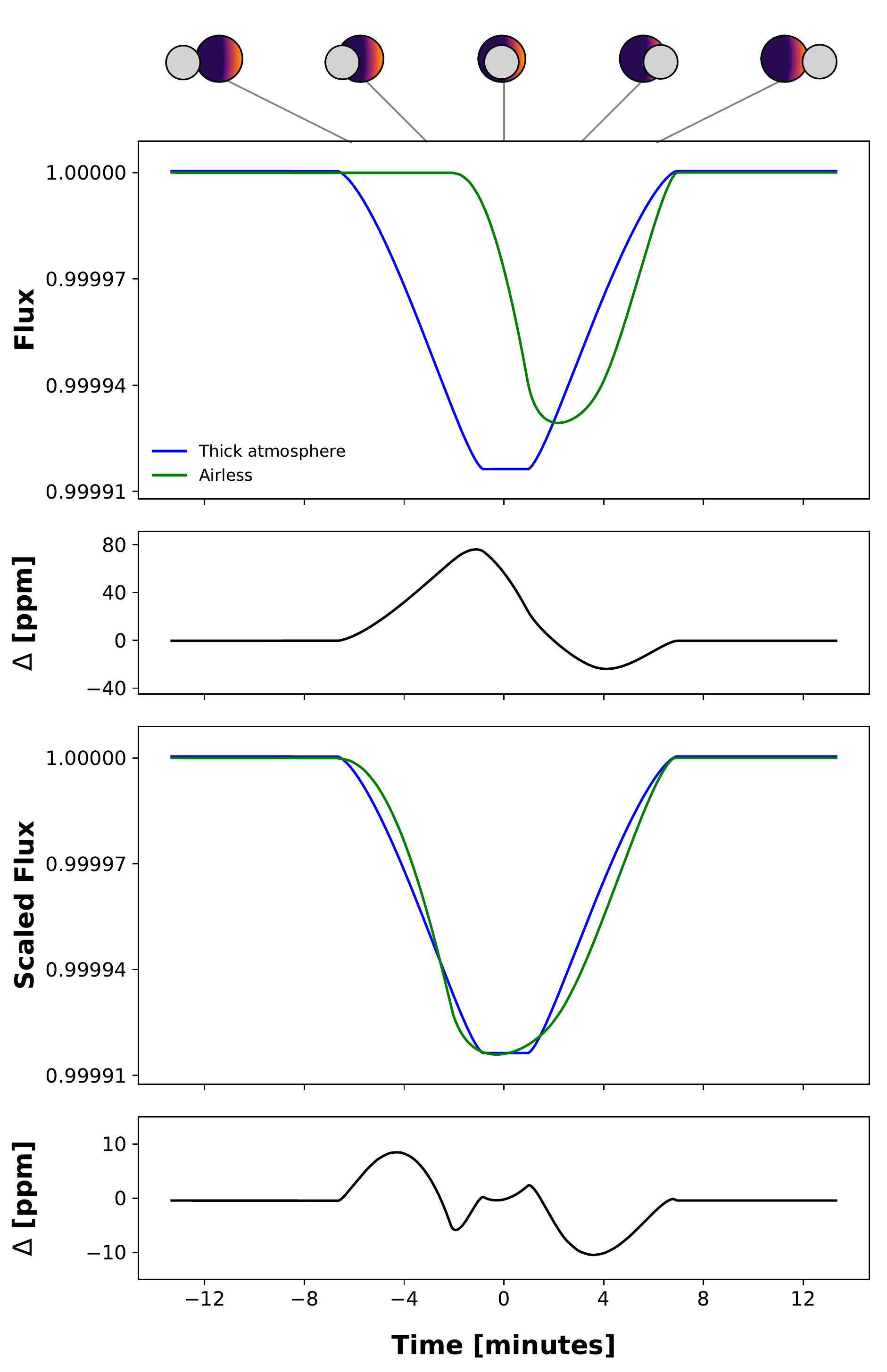}
\caption{An occultation of TRAPPIST-1c by TRAPPIST-1d at 15 $\mu$m as c approaches quadrature, for two different atmospheric regimes: the thick atmosphere limit (blue curves) and the airless body limit (green curves). In the former case, the planet disk is radially symmetric, and the light curve is symmetric about the midpoint of the occultation ($t = 0$). In the latter case, the stark day/night temperature contrast leads to an asymmetry in the light curve and a shift in the time of flux minimum. The light curves at 15 $\mu$m are shown in the top panel. Below it, we plot the difference of the two curves, showing that the day/night temperature contrast corresponds to a $\sim$80 ppm signal. However, if the time of occultation, the albedo of the planet, and the duration of the event are not known \emph{a priori}, discrimination between an airless planet and one with a thick atmosphere must be made based on the shape alone. To this end, in the third panel, we shift and scale the green curve so that the timing, duration, and depth coincide with those of the blue curve. The residuals are shown in the bottom panel: the curves are different at the $\sim$10 ppm level.}
\label{fig:contrast}
\end{figure}

Figure~\ref{fig:contrast} shows a sample occultation of TRAPPIST-1c by TRAPPIST-1d. We chose this occultation in particular because it highlights the ability of PPOs to probe a planet's day/night temperature contrast and even generate crude surface maps in the mid-infrared. This occultation occurs shortly after c has passed quadrature, such that about half its day side and half its night side are visible. TRAPPIST-1d is on the near side of the star, close to halfway between quadrature and transit, and the occultation occurs as it overtakes c on the sky. At occultation center, d is fully within the disk of c, but because d is smaller, only about half of the disk of c is occulted. The illustration at the top of the figure shows the progression of the occultation: d is the grey disk, \edited{and c is colored with an intensity proportional to the surface radiance in the ``eyeball'' limit.} The occultation proceeds such that the night side is occulted first, followed by the day side. The event lasts a total of 15 minutes.

The top panel of the figure shows the occultation light curve normalized to a stellar continuum of unity. The light curve is plotted for two different regimes of TRAPPIST-1c: a thick atmosphere with uniform radiance (blue; see \S\ref{sec:photo:limbdarkened}) and a thin/negligible atmosphere with an ``eyeball'' radiance map (green; see \S\ref{sec:photo:airless}). The former case leads to a light curve that is symmetric about the occultation midpoint, reflecting the assumed radial symmetry of the planet disk. The latter case leads to a distinctly asymmetric light curve, whose time of minimum flux is shifted by $\sim$2.5 minutes. In the airless regime, occultation of the cold ($T = 40$ K) night side leads to a negligible change in flux; ingress only occurs when TRAPPIST-1d begins to occult the bright day side of c and flux minimum occurs when the region in the vicinity of the substellar point is occulted. The occultation depth in the airless case is thus smaller by about 15\% compared to the thick atmosphere case.

The next panel shows the difference between the two light curves in parts per million (ppm) of the total signal. If the timing of the occultation is precisely known \emph{a priori}, these residuals suggest that one can discriminate between the airless case and the thick atmosphere case at the $\sim$80 ppm level, assuming the observation cadence is short enough and sources of correlated noise such as stellar variability can be properly removed. 


However, in the case that the time of occultation is not precisely known---due to uncertainty in the orbital parameters of the two planets---there will be a degeneracy between the day/night temperature contrast of the planet and the orbital parameters of the occulted/occultor pair, as a change in the latter can easily cause a similar shift in the occultation time. Moreover, if the albedo of the planet is unknown \emph{a priori}, the depth of the occultation is similarly not sufficient to distinguish between an airless body and one with a thick atmosphere. Finally, the duration of the event is in general degenerate with the impact parameter: a shorter event can be explained by either a large day/night contrast or a grazing occultation. Therefore, absent constraints on the orbital parameters and the albedo of the planet, distinguishing between the two atmospheric limits must be done based on the shape of the occultation alone. To this end, in the third panel we plot the same light curves, but having shifted and scaled the light curve corresponding to the airless case so that its duration, time of minimum, and depth match those of the thick atmosphere case. We again plot the difference between the two below, which has a maximum of $\sim$10 ppm during both ingress and egress. This signal is unlikely to be detectable with JWST (see \S\ref{sec:results:photo:jwst}), but constraints on the orbital parameters (from TTVs and secondary eclipses) and on the albedo (from secondary eclipses) can break several of the degeneracies mentioned above, allowing one to constrain the day/night temperature contrast based on the timing, depth, and/or duration of the occultations. We return to this point in \S\ref{sec:discussion:bestcase}.


\subsection{Observability with JWST}
\label{sec:results:photo:jwst}

\begin{figure*}[!p]
\centering
\includegraphics[width=\textwidth]{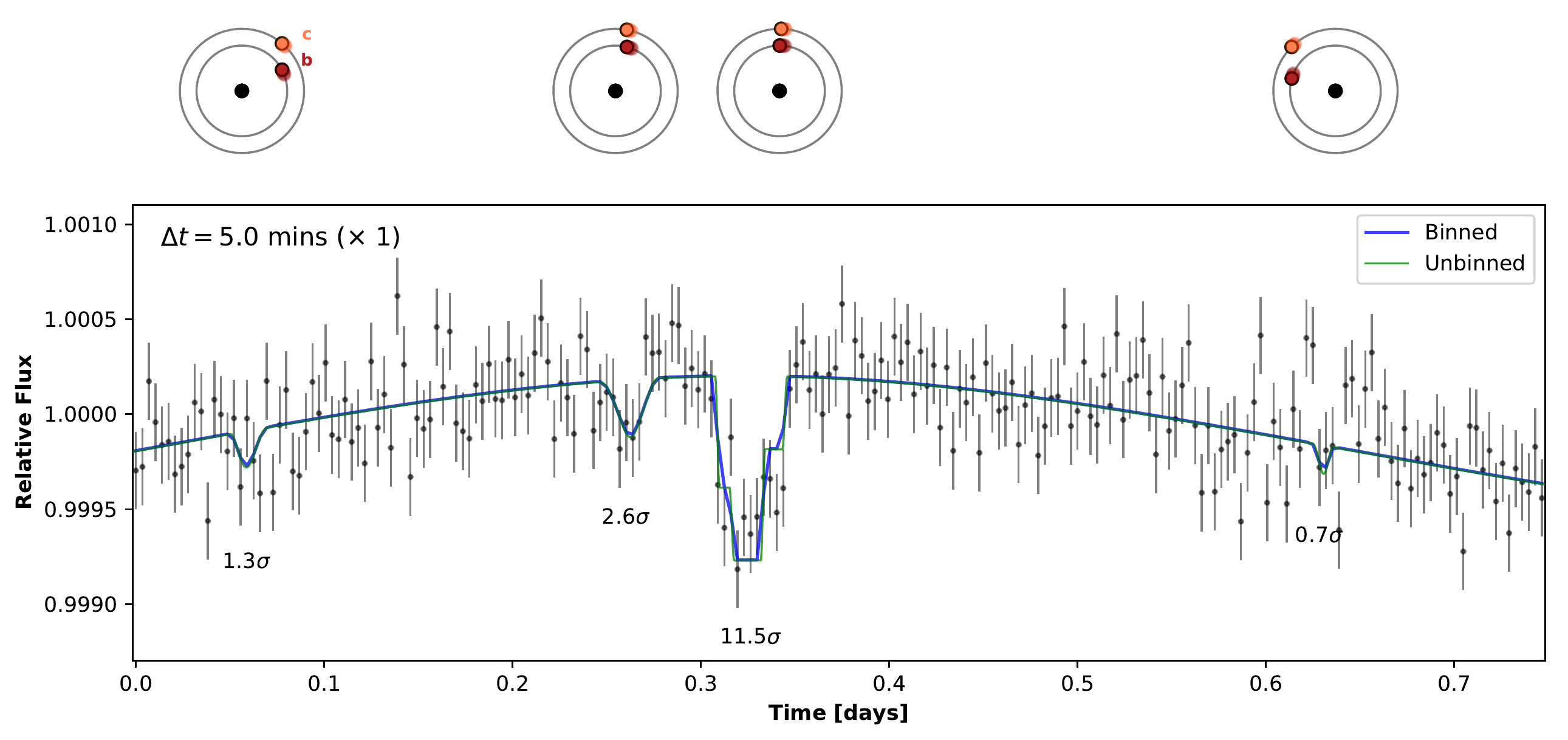}
\caption{A simulated triple occultation of TRAPPIST-1c by TRAPPIST1-b seen with JWST/MIRI at $15\mu$m with \edited{5 minute} exposures over 17 hours. The top panel shows the orbital positions of b and c during each of the events, seen from above the orbital plane; the observer is towards the bottom. The lower panel shows the full light curve (green), the light curve binned to the exposure time (blue), and the simulated observations with 1$\sigma$ error bars (black). From left to right, c overtakes b and is occulted; b overtakes c and occults it again; b and c are successively occulted by the star; c overtakes b a final time and is occulted. While the simultaneous eclipse of b and c is detectable above the noise (\edited{SNR$\sim$11.5}) in a single observation, the occultations of c by b are \edited{only marginally above the noise}. The deepest one, occurring at $t = 0.27$ days \edited{($\lambda_c = 81^\circ$, $\Delta t = 45$ minutes)}, has \edited{SNR$\sim$2.6}. Several of these must be \edited{modeled jointly} in order to permit detections of PPOs in TRAPPIST-1 with JWST; \edited{see Figure~\ref{fig:stacked_bc}}.}
\label{fig:triple_bc}
\end{figure*}

\begin{figure*}[!p]
\centering
\includegraphics[width=\textwidth]{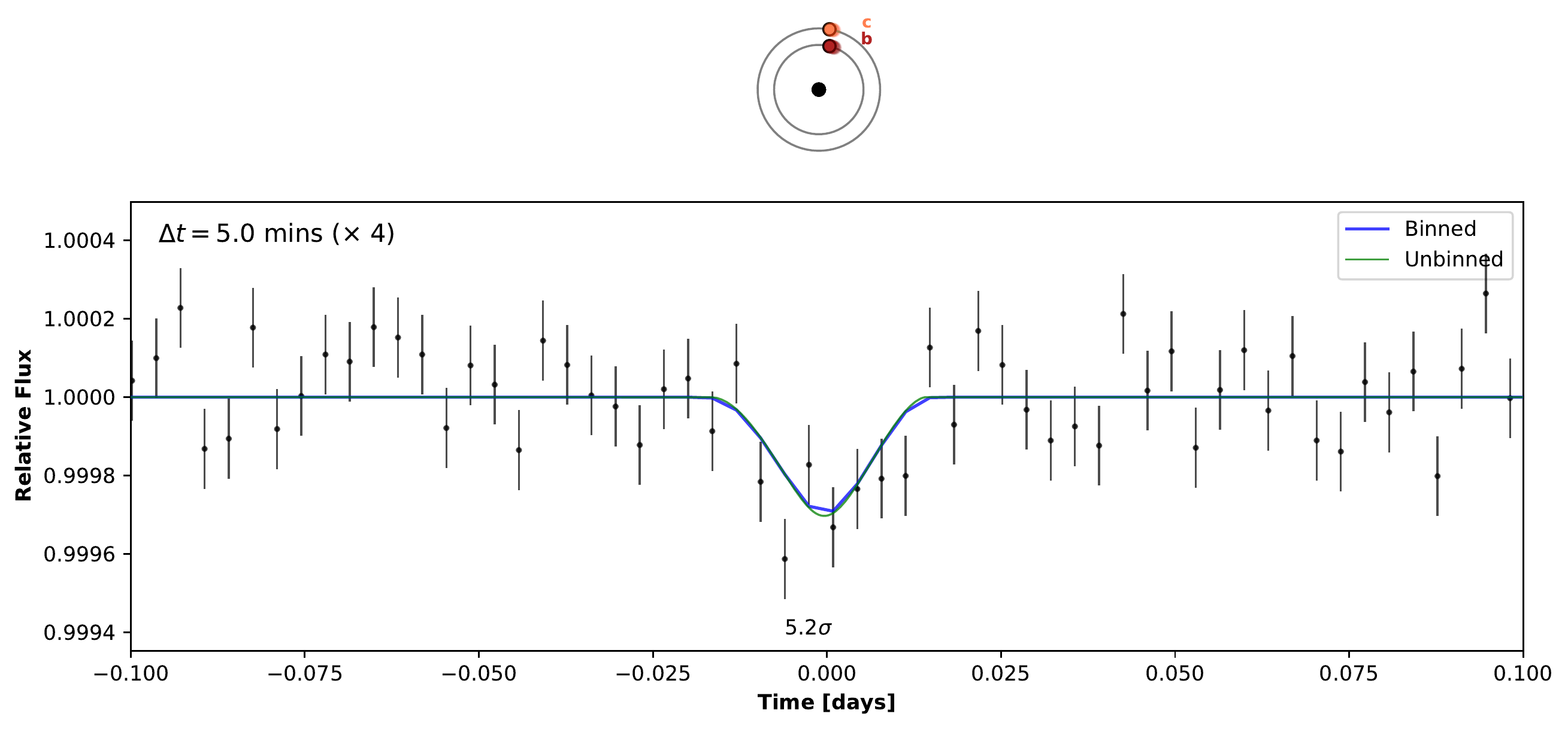}
\caption{Similar to Figure~\ref{fig:triple_bc}, but showing \edited{four} stacked observations of an occultation of TRAPPIST-1c by TRAPPIST-1b seen with JWST/MIRI at 15 $\mu$m with 5 minute exposures. The phase curves have been removed. The occultation is detectable above the noise with \edited{SNR$\sim$5.2.}}
\label{fig:stacked_bc}
\end{figure*}

\begin{figure}[!ht]
\centering
\includegraphics[width=0.47\textwidth]{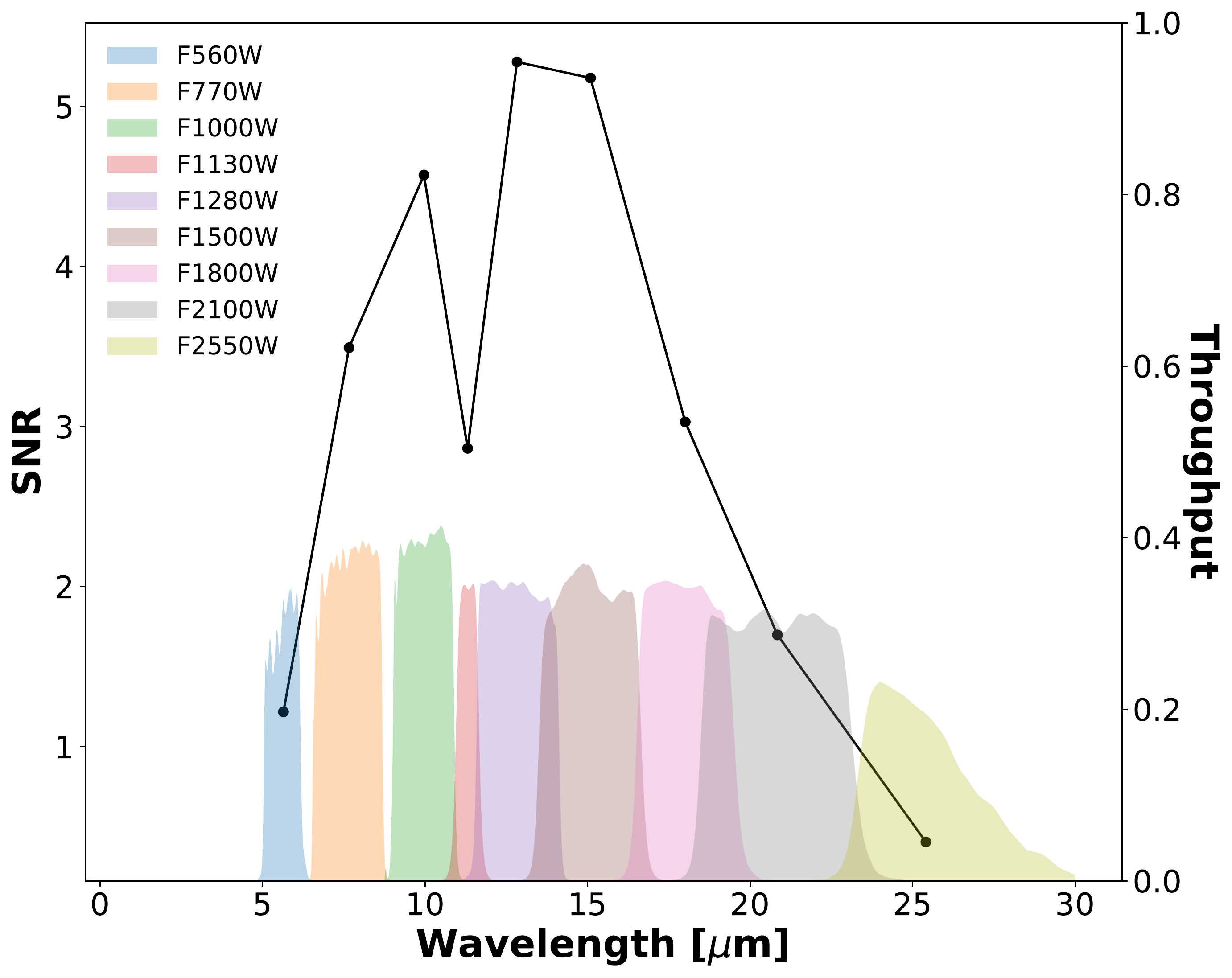}
\caption{Expected signal-to-noise ratio (SNR; black curve) for \edited{four} stacked occultations of TRAPPIST-1c by TRAPPIST-1b observed in each of the nine JWST/MIRI photometric filter bands. Filter throughput curves are shown in color and are plotted on the right $y$-axis.\nudge\nudge}
\label{fig:stacked_bc_all_filters}
\end{figure}

\edited{In this section we use the methodology discussed in \S\ref{sec:detect} to produce simulated observations of PPOs in TRAPPIST-1 with JWST.} Figure~\ref{fig:triple_bc} shows a simulated \edited{PPO triplet} of TRAPPIST-1c by TRAPPIST-1b observed with MIRI at 15 $\mu$m with \edited{5 minute} exposure times. \edited{As we discussed before, triplets} occur because of retrograde motion: when b is at quadrature and c is approaching secondary eclipse, the sky-projected velocity of b is smaller than that of c. But as b approaches secondary eclipse, it speeds up on the sky and overtakes c. On the opposite side of the star, b slows down once more and c overtakes it. A \edited{simultaneous} secondary eclipse of both planets occurs at $t = 0.32$ days. The low frequency modulation of the light curve is due to the combined phase curves of b and c, which are assumed to be \edited{airless bodies with zero albedo}. For reference, diagrams of the orbital positions of the two bodies are shown above each of the occultations.

In green we show the high resolution noise-free light curve, and in blue we show the light curve binned to the exposure time; the latter is what we would observe in the limit of infinite SNR. Black dots are the simulated noised observations. \edited{The first ($\mathrm{SNR}{\sim}1.3$) and last ($\mathrm{SNR}{\sim}0.7$) occultations of c are not statistically significant. At a SNR of ${\sim}2.6$, the occultation just before secondary eclipse is marginally significant but not robustly detectable.} The \edited{simultaneous} secondary eclipse, on the other hand, is a detectable \edited{SNR${\sim}11.5$} event. \edited{Unless TRAPPIST-1c has a higher surface temperature and observations are made in an atmospheric window, \bugfix{most} individual PPOs of TRAPPIST-1c will likely be just below the detectability threshold of JWST. As we discuss in \S\ref{sec:discussion:bestcase}, this likely also applies to  occultations of planet b.}

However, given the high frequency of occultations between TRAPPIST-1b and c (Figure~\ref{fig:hist}), the observation of multiple PPOs with JWST could lead to a statistically significant detection. In Figure~\ref{fig:stacked_bc} we plot the result of \edited{``stacking'' four} observations of the occultation at $t = 0.27$ days in Figure~\ref{fig:triple_bc}, this time at a 5 minute cadence. The phase curves and the secondary eclipses have been subtracted out. In this case, stacking \edited{four} observations leads to a detectable, \edited{SNR${\sim}5.2$} PPO. \edited{From Figure~\ref{fig:snr_hist}, occultations \bugfix{with SNR at least as high as} this one are expected to occur about \bugfix{three times per year} in TRAPPIST-1, but could occur more (or less) frequently due to the high variance.}

Figure~\ref{fig:stacked_bc_all_filters} shows the SNR on \edited{four} stacked occultations of planet c by planet b \edited{like the one in Figure~\ref{fig:stacked_bc}} in all nine MIRI filters ($5\ \mu\mathrm{m} \lesssim \lambda \lesssim 26\ \mu\mathrm{m}$). The F1280W (12.8 $\mu$m) and F1500W (15 $\mu$m) filters are approximately equal in their ability to optimally detect this particular PPO event given our assumption here that TRAPPIST-1c is an airless body. However, the presence of an atmosphere will affect the wavelength-dependent thermal emission from the planet and may change the optimal filter band. In particular, if TRAPPIST-1c has an atmosphere with CO$_2$ then we might expect relatively strong CO$_2$ absorption at 15 $\mu$m, as it is present in the spectra of Venus, Earth, and Mars. CO$_2$ absorption will make the planet darker at these wavelengths and therefore harder to detect in the F1500W filter. It is worth noting that secondary eclipse observations in both F1280W and F1500W may offer a simple test for a CO$_2$-bearing atmosphere by measuring the slope between the two eclipse depths. A positive slope may be fit by a featureless blackbody while a flat line or negative slope could indicate CO$_2$ absorption.

\edited{Finally,} a word of caution is in order. As PPOs are not periodic or even strictly repeatable events, it is not in general possible to simply stack multiple ones to improve the SNR of the detection, as we did above. Even for PPOs occurring at the same phase between the same two planets, slight differences in the impact parameter and the relative velocities of the two planets are likely to lead to deconstructive interference in the stacked signal. This is particularly problematic when the orbital parameters of the two planets are not well constrained (as is currently the case), since the timing of the occultation---and whether or not one occurs at all---is at best uncertain.

Instead, given a series of $n$ observations of the system, one must jointly model all of them with a full photodynamical model to derive posterior probabilities of the quantities of interest, such as the effective temperature of the occulted planet or its day/night temperature contrast. This allows one to marginalize over the unknown orbital parameters---and over the uncertainty of whether or not an occultation occurs at all---instead of risking the contamination of real occultations with noise when stacking. Figure~\ref{fig:stacked_bc} is therefore merely illustrative of the fact that if the orbital parameters are well constrained, on the order of \edited{four $2.6\sigma$} occultations of TRAPPIST-1c are necessary for a robust detection. We return to this point in \S\ref{sec:discussion:joint}.

\subsection{Observability with OST}
\label{sec:results:photo:ost}
The Origins Space Telescope \citep[OST;][]{Cooray2017} is a NASA mission concept currently under study for the 2020 Astronomy and Astrophysics Decadal Survey. OST would be an actively-cooled, mid- to far-IR, large aperture (8--16 m) space telescope, and successor to JWST. OST's access to the far-IR may make it an ideal observatory for thermal emission studies of nearby exoplanets, including PPO observations. 

Similar to our estimates for JWST, we assess the observability of PPOs in the TRAPPIST-1 system with OST. Given the active cooling of OST, we neglect background noise and assume photon-limited observations. We simulate time-series photometry for a 12 meter OST primary mirror and investigate filter configurations. 

\begin{figure}[!t]
\centering
\includegraphics[width=0.47\textwidth]{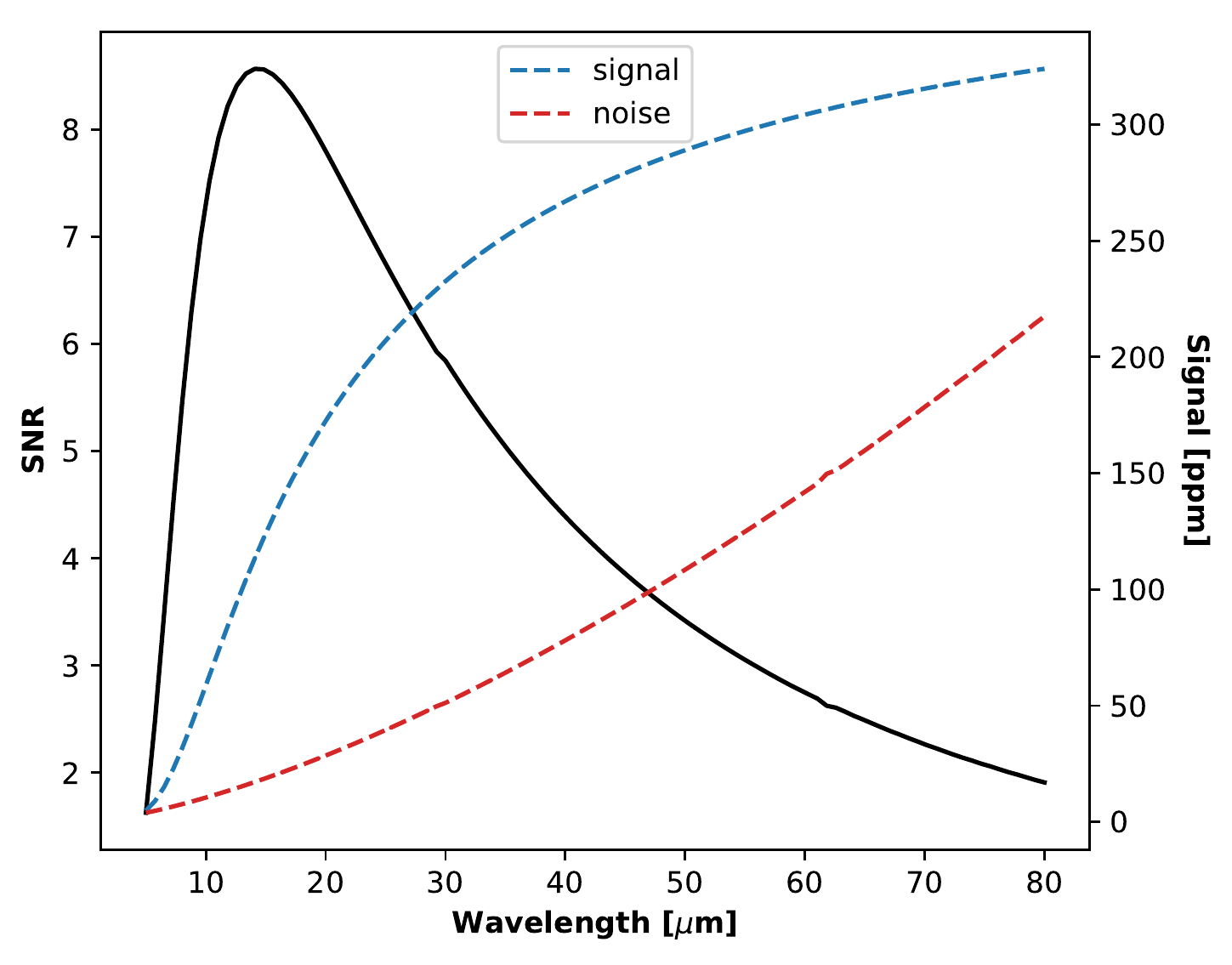}
\caption{\edited{Expected signal-to-noise (SNR; black curve) for an occultation of TRAPPIST-1c by TRAPPIST-1b if observed with the Origins Space Telescope (OST). Dashed blue and red curves are plotted on the right $y$-axis and show PPO signal and noise terms (see Equation \ref{eqn:snr_occult}), respectively, in units of parts-per-million (ppm) of total photons observed from the system during the occultation. OST calculations assume 30\% throughput in a 5 $\mu$m wide filter centered at each wavelength.\nudge\nudge}}
\label{fig:snr_bc_ost}
\end{figure}

In Figure~\ref{fig:snr_bc_ost} we plot the signal, noise, and SNR for an occultation of TRAPPIST-1c by TRAPPIST-1b (the same one shown in Figure~\ref{fig:stacked_bc}). For our photometry estimates we assume a tophat filter with 30\% throughput in a 5 $\mu$m-wide filter centered at a range of wavelengths between 5 and 80 $\mu$m. Both the signal and noise terms (numerator and denominator in Equation~\ref{eqn:snr_occult}, respectively) are monotonically increasing functions of wavelength, but with second derivatives of opposite sign. Interestingly, the SNR curve reveals that the optimal wavelength for the detection of this particular PPO event is ${\sim}15$ $\mu$m, similar to the conclusion we reached with JWST. However, because of the absence of background noise and the larger mirror size, PPOs are detectable at much higher SNR with OST. 
%

\begin{figure*}[!ht]
\centering
\includegraphics[width=\textwidth]{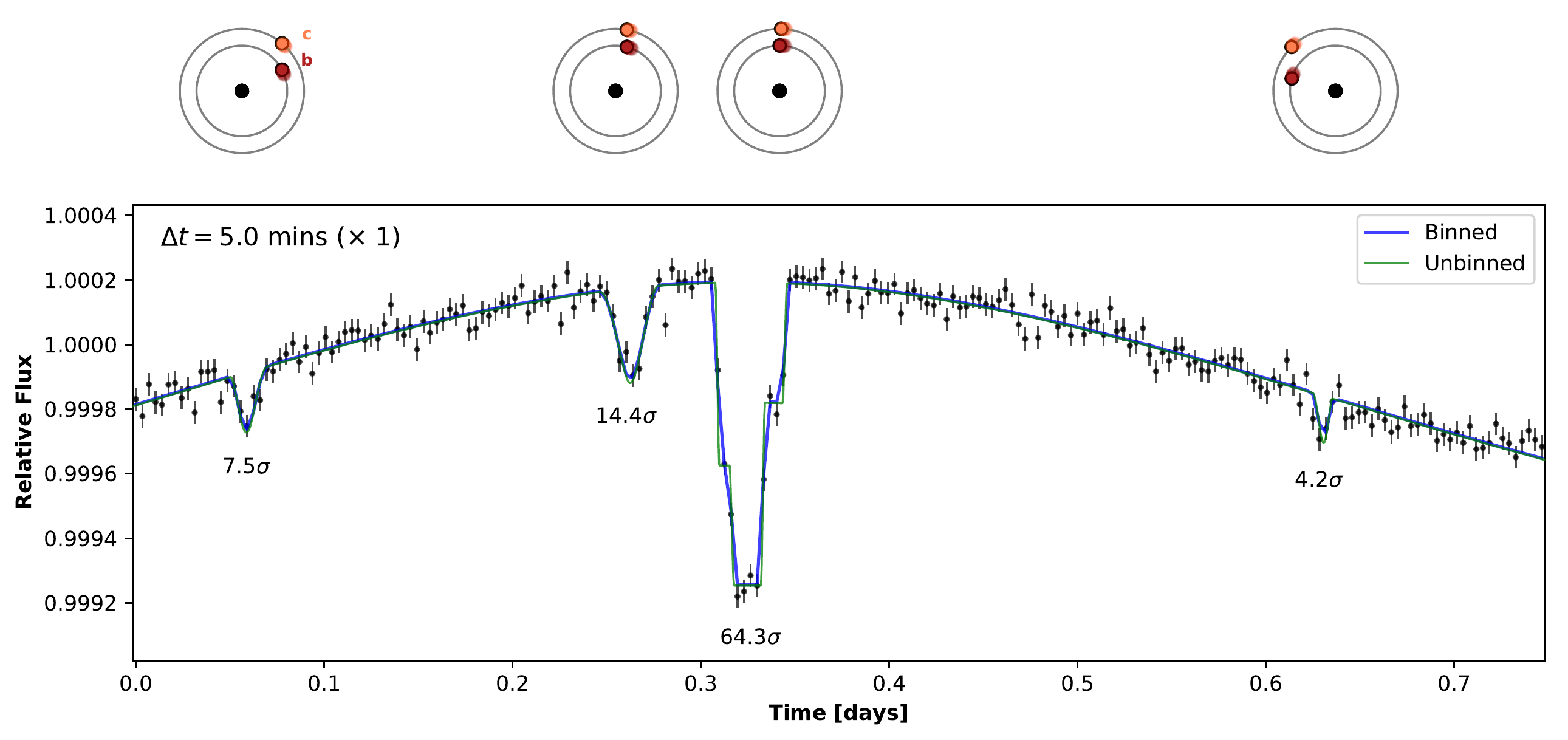}
\caption{Similar to Figure~\ref{fig:triple_bc}, but for a single pointing of the Origins Space Telescope, assuming a \edited{13.5m} diameter mirror and an observation in a filter between $10-30$ $\mu$m with a throughput of 30\%. The exposure time is 5 minutes. All three planet-planet occultations of TRAPPIST-1c are detectable above the noise. The deepest one has \edited{SNR$\sim$14.4.}\nudge}
\label{fig:ost}
\end{figure*}

In Figure~\ref{fig:ost} we plot the same triple occultation as in Figure~\ref{fig:triple_bc}, but this time observed with OST in a broad mid-IR filter between 10 and 30 $\mu$m with a 30\% throughput. All three occultations of c by b are statistically significant events and are visible by eye in the data, particularly the \edited{SNR$\sim$14.4} event just prior to the double secondary eclipse. 

\subsection{Observability with Spitzer}
\label{sec:results:photo:spitzer}
Finally, we briefly explore the observability of PPOs in the TRAPPIST-1 system with the \emph{Spitzer} Space Telescope. \emph{Spitzer} has a small (85 cm diameter) primary mirror and is limited to photometry at 3.6~$\mu$m and 4.5~$\mu$m due to the loss of coolant, both of which are detrimental to PPO detectability. We find that a statistically significant (\edited{SNR $\sim$ 5}) detection of the occultation of TRAPPIST-1c by TRAPPIST-1b considered in the previous sections would require the observation of on the order of $10^4$ PPOs at 4.5~$\mu$m. However, if the brightness temperature of TRAPPIST-1c at 4.5~$\mu$m were \edited{$\sim$850 K ($\sim$500 K hotter than the equilibrium value at zero albedo)}, a \edited{SNR $\sim$ 5} detection could be made with \edited{approximately 10} stacked observations. This could in principle be possible if TRAPPIST-1c had a thick, greenhouse-heated atmosphere with low opacity at 4.5~$\mu$m; the same is true for TRAPPIST-1b. We therefore conclude that PPO observations with \emph{Spitzer} are generally infeasible, except under vigorous greenhouse heating and the presence of transparent atmospheric windows coincident with the \emph{Spitzer} filters.

\section{Discussion}
\label{sec:discussion}

Here we discuss qualitatively some aspects of PPO detection, including planet mapping (\S \ref{sec:discussion:mapping}), comparison with phase-curve measurements (\S\ref{sec:discussion:phasecurves}), optimistic prospects for detection (\S\ref{sec:discussion:bestcase}), and potential pitfalls, including degeneracies and systematics (\S\ref{sec:discussion:degeneracies}).  We argue that photodynamical analysis with our code offers the best potential for detection of multiple PPO events (\S\ref{sec:discussion:joint}), which can account for mutual events (\S\ref{sec:results:mutual}), as well as offset hot spots (\S\ref{sec:discussion:hotspotoffset}) and tidal heating (\S\ref{sec:discussion:tidalheating}).  We end with a discussion of prior related work (\S\ref{sec:discussion:priorwork}) and the potential for application to other systems (\S\ref{sec:discussion:kepler444}), including non-transiting systems or white dwarfs.

\subsection{PPO Mapping}
\label{sec:discussion:mapping}


Planet-planet occultations offer an opportunity to map the disk of the occulted planet \citep{Ragozzine2010}. High SNR, high cadence mid-IR observations of PPO light curves will, in principle, reveal the brightness temperature map, including any heterogeneities, along circular arcs having the radius of the occultor. While PPO mapping is similar to secondary eclipse mapping \citep[e.g.][]{Williams2006, Rauscher2007, Agol2010, Majeau2012, deWit2012}, the events will differ in orbital phase, duration, depth, and impact parameter. In principle, different events allow one to probe the surface map of the planet along different circular arcs, permitting the deconvolution of the signal into separate latitudinal and longitudinal maps. Moreover, occultations at different phases could allow one to map the entire surface of the planet---not just the day side, as with secondary eclipse mapping.

The procedure outlined above may likely not be possible with JWST, given the low SNR of individual PPO events using the nominal noise model (\S\ref{sec:results:photo:jwst}); future generation telescopes such as OST (\S\ref{sec:results:photo:ost}) may be required. However, there are several diagnostics in PPO events arising from the spatial distribution of thermal emission that could be possible with JWST:
\begin{enumerate}
\item A larger day/night contrast will cause deeper events at full phase and shallower events at new phase.
\item A larger day/night contrast will cause a shift in the timing of events (see Figure~\ref{fig:contrast}).
\item A phase offset of the peak emission due to recirculation may also cause a shift in timing (see \S\ref{sec:discussion:hotspotoffset}).
\item A latitudinal offset of the peak emission may cause a depth variation (see \S\ref{sec:discussion:hotspotoffset}).
\end{enumerate}

In principle, these \edited{OLVs (\S\ref{sec:dynamics:eccentricities})} could be degenerate with the orbital dynamics of the system. For example, if the eccentricity is such that the occultations at new phase have a smaller impact parameter, this will cause shallower depths. This may be degenerate with a larger day/night temperature contrast, but the shorter duration and the timing shift could perhaps break this potential degeneracy. Moreover, the detection of multiple PPOs will help constrain the orbital parameters and eventually allow precise determination of the impact parameter and timing of individual events, which \edited{may also help to} break these degeneracies.

In \S\ref{sec:results:photo:sample_ppo} we argued how the signature of a strong day/night temperature contrast corresponds to a ${\sim}80$ ppm signal if the timing of the PPO is known \emph{a priori}. Incidentally, this is slightly above the noise level in the simulated stacked observation with JWST shown in Figure~\ref{fig:stacked_bc}, suggesting that \bugfix{$\sim$4} PPOs of c are needed to detect the day/night asymmetry with SNR $>$ 1. A statistically significant detection \bugfix{($\mathrm{SNR}{\gtrsim}5$)} would require on the order of 100 PPOs with JWST, which could in principle be achievable over the mission's lifetime. However, targeted observations of occultations of c at new phase would likely be a more efficient way of constraining the day/night asymmetry, as the (non)detection of PPOs could place strong (upper) limits on its night side emission. Alternatively, just a few PPOs seen with OST could lead to a statistically significant detection.

\subsection{PPOs versus phase curves}
\label{sec:discussion:phasecurves}
Thermal phase curves are another method used to map short-period planets by measuring the day/night temperature contrast of an exoplanet from its flux as a function of that planet's orbital phase about the star \citep{Knutson2007,Selsis2011,Maurin2012}. However, these observations will be challenging to acquire with JWST for several reasons, both practical and fundamental. The practical limits are that phase curves require a significant observing time investment, since the phase curve must be mapped over a significant fraction of the orbit. The high demand for JWST will likely make it difficult to obtain observing time for such long observations. Additionally, over longer timescales, instrumental systematic variations can be significant. For example, in the case of 55 Cancri e, the systematic variations were several hundred times larger than the claimed phase amplitude precision measured with {\it Spitzer} \citep{Demory2016}.

The fundamental limits are that multiple phase curves can produce complex photometric behavior due to the different amplitudes and frequencies of each planet \citep{Kane2013}, which when added to stellar variability of a periodic or random nature, may make it difficult to uniquely recover longitudinal maps of any of the planets. In contrast, PPO events are much shorter in duration, and can be scheduled in advance (within the dynamical uncertainties), using a much smaller fraction of time in which the star has less chance to vary, and systematic variations will likely be less severe.

Finally, phase curves of transiting planets are \edited{primarily sensitive to the longitudinal surface brightness of the planet \citep{Cowan2008,Cowan2009}.
While latitudinal information can be extracted for planets with significant obliquity \citep{Cowan2013,Cowan2017}, phase curves tend to act as a low-pass filter, which makes the recovery of unique surface maps difficult. PPOs, on the other hand, probe both longitudinal and latitudinal brightness, similar to secondary eclipse mapping.} However, unlike secondary eclipse mapping, which can only occur at full phase and probe the day side, PPOs can occur at multiple phases allowing the disk to be mapped for different illumination fractions.

\subsection{Best case scenarios}
\label{sec:discussion:bestcase}
In our detectability discussion above, we focused on occultations of TRAPPIST-1c by TRAPPIST-1b. In principle, occultations of \emph{b by c} should be deeper events, given that b is exposed to $\sim$2 times the irradiation of c. However, day side occultations of b by c are only possible when c is on the opposite side of the star, closest to the observer, in which case the relative velocity of the two planets is at a maximum. These occultations are thus extremely short-lived and will not be detectable. In contrast, long occultations of the night side of b by c \emph{are} possible, when both planets are on the near side of the star, and could be detectable if b has a bright night side due to strong atmospheric recirculation. By the same token, if b has a cold night side, the nondetection of these occultations could place strong constraints on its day/night temperature contrast. Given the higher irradiation of b relative to c, fewer occultations \edited{may} need to be observed to obtain statistically significant results.

If the TRAPPIST-1 planets are airless bodies, occultations of c by b like the ones shown in Figure~\ref{fig:triple_bc} will have the \edited{highest SNR}. If, on the other hand, TRAPPIST-1b and c have thick atmospheres, they may be analogous to Venus, whose extreme greenhouse forcing results in a surface that is much hotter than the planet's equilibrium temperature. In general, the surface of such a planet will be radiatively decoupled from the effective emitting layer, which is at a much lower temperature, and PPOs may be difficult to observe. This is an issue particularly for a CO$_2$-dominated atmosphere, which has strong absorption bands near 15 $\mu$m. Nevertheless, if the orbital parameters are well constrained, non-detections of PPOs at 15 $\mu$m could in principle be used to infer the presence of CO$_2$ or other strongly absorbing atmospheric species.

Alternatively, observations in atmospheric windows---such as the 2.4 $\mu$m \emph{K} band window in Venus' atmosphere \citep{Arney2014}---could allow one to probe to a much deeper (and hotter) layer, significantly enhancing the detectability of PPOs. \edited{Moreover, atmospheric dynamics of planets close to the inner edge of the habitable zone could result in a thick cloud layer that suppresses outgoing infrared radiation on the day side, effectively increasing night side emission \citep{Yang2013}. This could potentially increase the detectability of PPOs of planets b and c close to new phase. We leave investigation of these and other atmospheric effects to future work.}

Whatever the atmospheric regime of TRAPPIST-1c, there are two particular strategies for choosing when to observe occultations by TRAPPIST-1b. First, the \edited{PPO triplet} shown in Figure~\ref{fig:triple_bc} is perhaps the best observational scenario, as three occultations of TRAPPIST-1c plus a secondary eclipse of both TRAPPIST-1c and b occur over the span of about 15 hours, yielding a large amount of information for a single pointing of JWST, \edited{with a total $\mathrm{SNR}{\sim}3.0$}. \edited{Triplets} are relatively common, and happen on both sides of the star; on the near side, it is TRAPPIST-1b which is occulted, but the orbital geometry is otherwise identical by symmetry. Once the orbital parameters of b and c are better constrained, these events can be predicted using our photodynamical code.

Second, from Figure~\ref{fig:c_corner} it is clear that occultations of TRAPPIST-1c are most common when the planet is at a mean longitude $\lambda_c \approx 43^\circ$ or $137^\circ$. As we discussed previously, the peak in the histogram has FWHM $\sim$ 2$^\circ$, corresponding to a time window of just under 20 minutes. Thus, even if the eccentricities and longitudes of ascending node of TRAPPIST-1b and c are not well known, the likelihood of observing an occultation of c is maximum for an observation centered at this mean longitude lasting 20 (or more) minutes. In order for the occultation to occur, TRAPPIST-1b must be near quadrature, which can easily be predicted in advance. Alternatively, and by symmetry, when TRAPPIST-1c is at a mean longitude $\lambda_c \approx 43^\circ$ or $137^\circ$ and b is near quadrature, occultations of b are most common (note the peaks near $0^\circ$ and $180^\circ$ in Figure~\ref{fig:b_corner}).

Finally, we note that occultations of the other planets \bugfix{will be difficult to observe} with JWST, but may be detectable with OST. Nevertheless, occultations of b and c \emph{by} the outer planets are potentially observable and can be long-lived, especially occultations of c by d. \bugfix{These can be used to constrain the orbital properties of the other planets in the system.}

\subsection{Degeneracies and other issues}
\label{sec:discussion:degeneracies}
There exist several degeneracies in PPO light curves that merit special attention, particularly for low SNR observations. We already discussed how a timing offset due to a day/night temperature contrast could be degenerate with the orbital parameters (\S\ref{sec:discussion:mapping}). The planetary albedo is also degenerate with the impact parameter of the occultation, since both can change the depth of the event; in principle, at high SNR the duration can be used to resolve this.

If the orbital parameters are \emph{very} poorly constrained, it may not be possible to tell with certainty which pairs of planets are occulting. This could be an issue if simultaneous or near-simultaneous occultations occurred between two different sets of planets. This is likely not going to be the case for TRAPPIST-1, since only occultations of b and c are at present potentially detectable and the orbital parameters are fairly well constrained. In principle, however, this could lead to interesting degeneracies between the orbital and surface properties across multiple planets in the system.

It is important to remember, however, that transits occur about 6\% of the time in TRAPPIST-1 (as do secondary eclipses). Moreover, the star flares at a rate of about 0.26 day$^{-1}$ \citep{Luger2017}. It is therefore quite likely that some PPOs will occur during one of these events, which will negatively impact their detectability.

Another degeneracy that may arise concerns the potential heterogeneity of the surface. Spatial albedo variations due to clouds or geography could lead to asymmetries in the occultation light curve that could in principle be mistaken for a day/night temperature contrast. \edited{Moreover, temporal variability due to weather could also negatively impact the joint modeling procedure necessary to robustly detect PPOs with JWST.} We defer a proper treatment of heterogenous surfaces to a future paper.

Finally, we note that in our detectability calculations we have neglected stellar variability, even though brightness modulations due to spots are clearly visible in the \emph{K2} light curve of the system \citep{Luger2017}. However, while this may complicate phase curve observations, it is unlikely to significantly impact the detection of PPOs, given the rotation period of the star \citep[3.3 days;][]{Luger2017} is two orders of magnitude greater than the typical duration of a PPO. Moreover, spot contrasts and granulation noise are both much weaker in the mid-IR. The same is true for flares, whose emission peaks at shorter wavelengths. Nevertheless, unmodelled stellar variability could in principle decrease the SNR of PPO events and increase the number of observations needed to achieve a detection.

\subsection{Joint modeling}
\label{sec:discussion:joint}
\bugfix{Since it will be challenging to robustly detect individual PPOs with JWST}, joint modeling of multiple events must be performed to obtain constraints on the orbital and atmospheric properties of the planets. As \planetplanet is a full photodynamical model, it may be used in a Markov Chain Monte Carlo (MCMC) simulation to derive posterior probabilities of the quantities of interest. We described in \S\ref{sec:results:photo:jwst} why this approach is preferable to ``stacking'': while multiple occultations of a given planet result in different light curves, all of these light curves are functions of the same set of model parameters $\Theta$. Instead of stacking to infer the occultation depth (which will always be different), joint modeling can be performed to marginalize over all model parameters and constrain, for instance, the day/night temperature contrast (which is the same across all occultations, if the planet has weak temporal variability). Given a dataset of measured fluxes $D$, the probability of a given value of the day/night temperature contrast $\Delta T$ is
\begin{align}
    p(\Delta T | D) = \int p(\Delta T, \Theta | D) d\Theta
\end{align}
where $\Theta$ is the set of all other parameters in the model (eccentricities, albedos, etc.), over which we have marginalized. 

MCMC provides an efficient and robust way of computing the expression above. More generally, it yields the posterior probability distributions of all parameters conditioned on the data and all prior information about the system, such as the current constraints on the orbital parameters (Table~\ref{tab:sysparams}).
We provide sample code for performing full photodynamical modeling of TRAPPIST-1 with MCMC on the project \github page. For more details on MCMC and its implementation, the reader is referred to \cite{Mackay2003} and \cite{ForemanMackey2013}.

\subsection{Mutual transits}
\label{sec:results:mutual}

The extreme coplanarity of the TRAPPIST-1 system suggests that mutual transits should also occur in the system. These are planet-planet occultations occurring on the face of the star, which have been studied for \emph{Kepler} systems \citep[e.g.,][]{Hirano2012,Masuda2013,Masuda2014} and for moon-planet occultations \citep[e.g.,][]{Kipping2011,Pal2012}. The advantage of mutual transits over PPOs is that since the star is being occulted, the signal of the mutual event is much stronger, often a significant fraction of the transit depth. Typical mutual transits result in a brightening close to transit center, due to the fact that the total area occulted on the star is less than the sum of the areas of the two planets (see, for example, Figure 3 in \citealt{Pal2012}). However, for a coplanar system like TRAPPIST-1, mutual transits are much more infrequent than planet-planet occultations. This is because the region of overlap of the orbital paths of any two planets on the face of the star is much smaller than the total region of overlap (see Figure~\ref{fig:orbits}). \edited{Nevertheless, when marginalizing over the uncertainties on all orbital parameters, we find that mutual transits between all seven planets should occur at an average rate of 27 times per year, or once every ${\sim}14$ days. Mutual transits involving planets b, c, d, and e are the most common, occuring at average rates of 16, 13, 10, and 12 times per year, respectively.}

\begin{figure}[!t]
\centering
\includegraphics[width=0.47\textwidth]{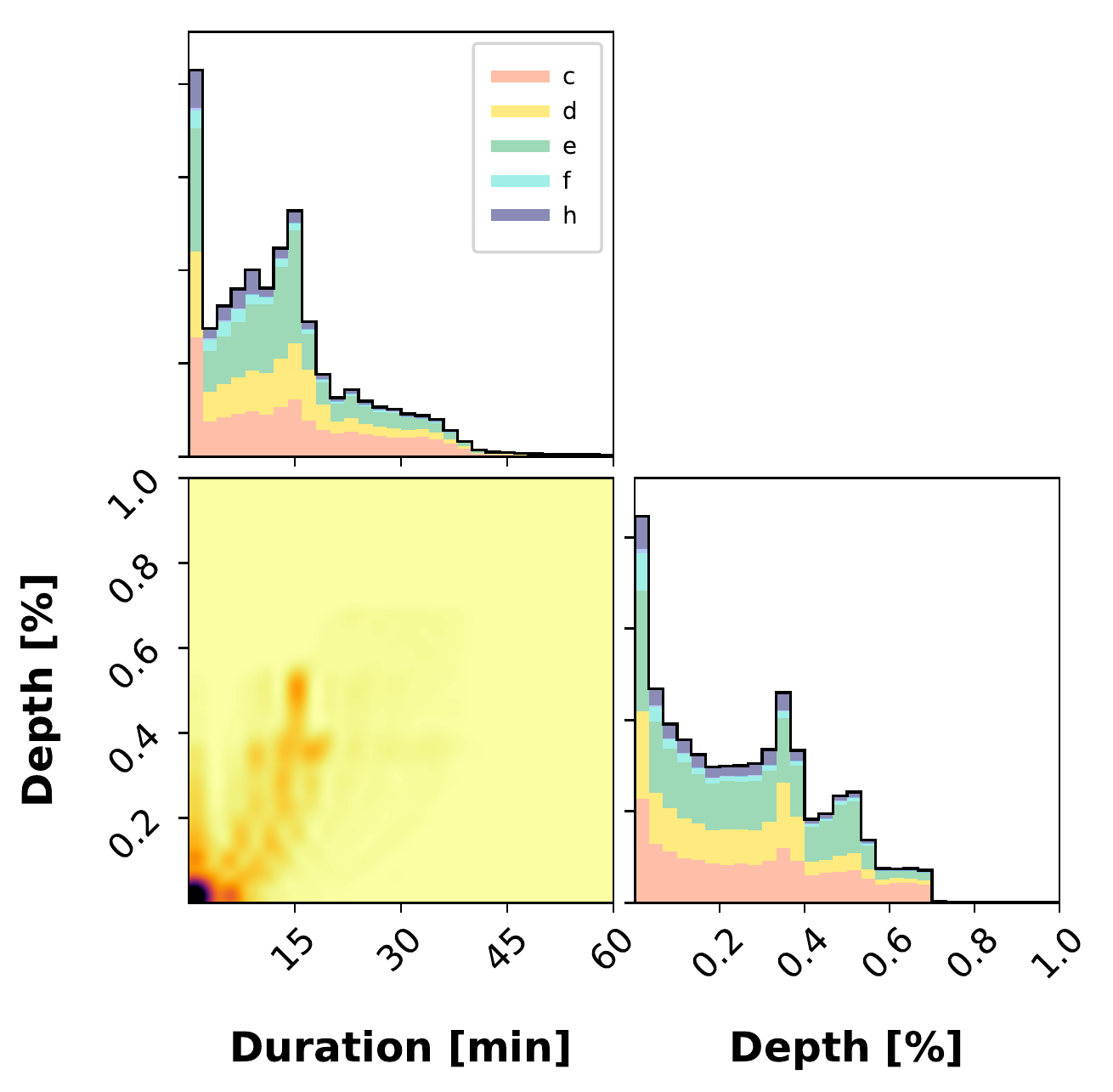}
\caption{\edited{Posterior distributions of the duration (in minutes) and fractional depth (in percent) of mutual transits involving TRAPPIST-1b. The panel at the bottom left is the joint distribution, where darker colors indicate higher probability density. The other two panels are the marginal distributions. The black lines are histograms of all mutual transits involving planet b; the colored distributions are stacked histograms of mutual transits involving b and each of the other planets.}}
\label{fig:mutual_transit}
\end{figure}

\edited{In Figure~\ref{fig:mutual_transit} we show the posterior distributions of the duration and depth of mutual transits involving TRAPPIST-1b for 30,000 1-year integrations of the system (as before, plots for the other planets can be easily generated with the code on \github). The duration is measured as the total time b is occulted by another planet while in transit across the star, and the depth is the maximum fractional brightening of the transit light curve during the event measured as a percentage of the stellar flux. Marginal distributions of the duration and depth of mutual transits by each of the other six planets are shown as the colored stacked histograms; their sum is the distribution for all mutual transits involving planet b, shown in black. The panel at the bottom left is the joint distribution. As with PPOs, many mutual transits are grazing, short-lived events; however, the majority of these events last upwards of 10 minutes and cause a brightening of the transit light curve at the $\gtrsim 0.1\%$ level, which should be easily detectable with JWST. The deepest events, which are mutual transits primarily by c (though also by d and e), have depths $\gtrsim 0.5\%$ and should be readily detectable by \emph{Spitzer}. A search for these events in current and future \emph{Spitzer} data may help place strong constraints on the impact parameters (and hence the inclinations) of the TRAPPIST-1 planets and thus aid in the search for PPOs in the coming years.}

\begin{figure}[!t]
\centering
\includegraphics[width=0.47\textwidth]{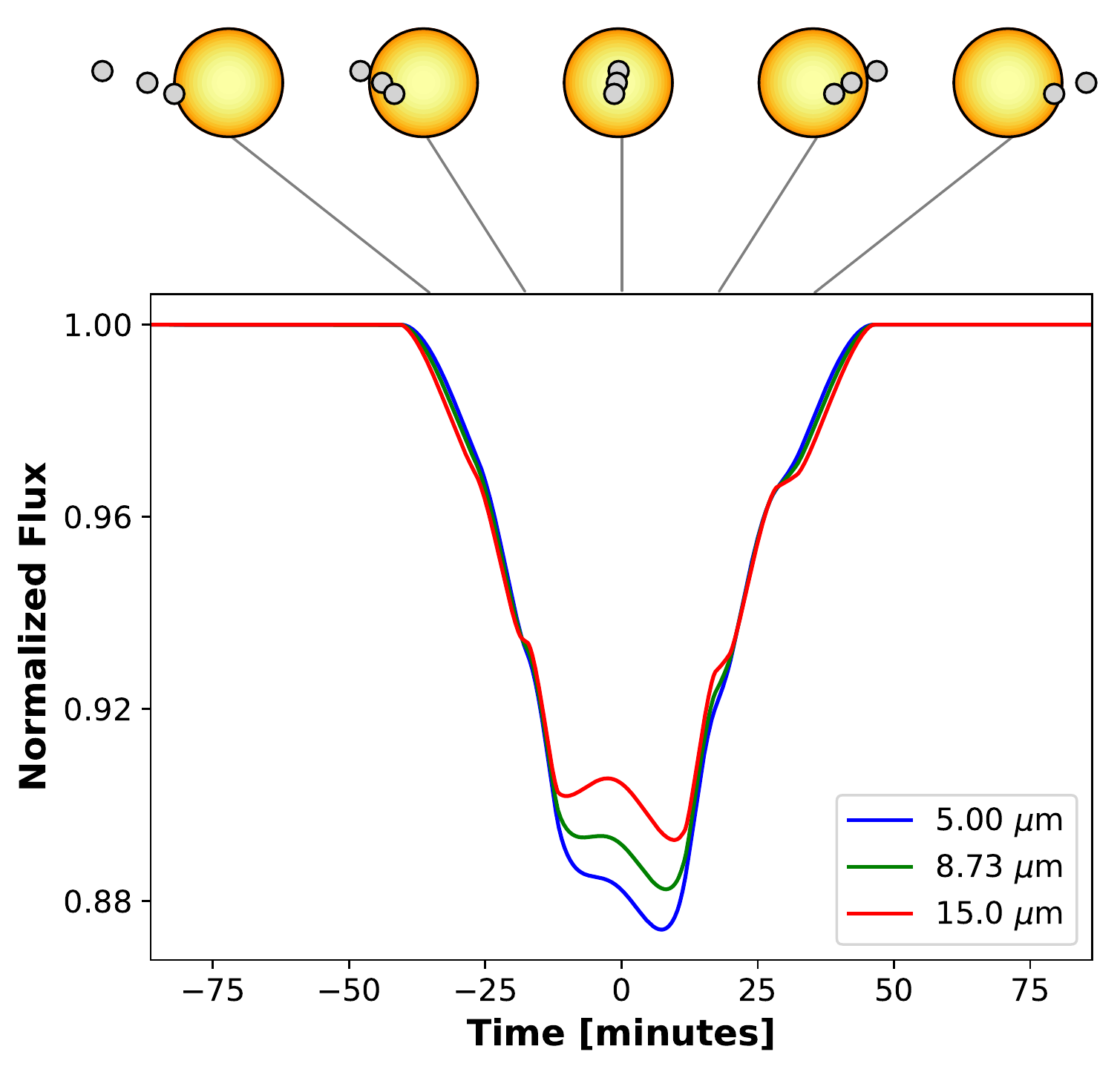}
\caption{An example of a triple mutual transit across a star with an arbitrary limb darkening profile in a hypothetical system. The integration scheme in \planetplanet is completely general and allows one to easily calculate occultation light curves for arbitrary numbers of overlapping bodies. In this example, we use a simple linear limb darkening law with a limb-to-center contrast that decreases linearly with wavelength, resulting in changes to the transit depth and shape at different wavelengths.\nudge}
\label{fig:triple}
\end{figure}

We note, \edited{finally}, that our generalized integration scheme  (Appendix~\ref{app:integration}) allows one to easily compute \edited{mutual transit and} occultation light curves for an arbitrary number of overlapping bodies. Figure~\ref{fig:triple} shows a hypothetical double mutual transit computed with \planetplanet, in which three planets occult each other as they transit their star. The brightening due to the mutual transits is evident near transit center.

\subsection{Hotspot offsets}
\label{sec:discussion:hotspotoffset}
Secondary eclipse measurements are capable of measuring latitudinal and longitudinal offsets in the position of peak emission of a planet due to atmospheric recirculation \citep{Majeau2012,deWit2012}; planet-planet occultations are no different. The symmetry of the emission about the hotspot suggests we can use the same integration scheme discussed in \S\ref{sec:photo:integration} to model light curves for planets with shifted hotspots. \planetplanet allows users to specify the latitudinal ($\Phi$) and longitudinal ($\Lambda$) shifts when computing light curves. In Appendix~\ref{app:geometry}, we derive the geometry of the general case of an ``eyeball'' planet with an arbitrary hotspot offset, and discuss how we adapt the \planetplanet integration scheme to handle these bodies in the same semi-analytic fashion. Note that, in general, the temperature distribution of an ``eyeball'' planet with an offset hotspot will not follow Equation~(\ref{eqn:tlat}), since that expression assumes the surface is in pure radiative equilibrium with the stellar flux. To this end, \planetplanet allows users to input custom functions for $T(\phi)$.

\subsection{Tidal heating}
\label{sec:discussion:tidalheating}
Our signal-to-noise considerations make the assumption that the planet temperatures are governed by the absorbed flux on each planet which is re-radiated in the thermal infrared.  Another possibility is that planets may be tidally heated to the point that their thermal emission may be dominated by tidal heat flux \citep{Selsis2013,Bolmont2013}. However, on a short timescale this dissipation can cause the rotation of a planet to be synchronized \citep{Jackson2008a} and the eccentricity to be damped, causing tidal forces to be diminished unless additional forcing maintains the planet in an asynchronous or eccentric state \citep{Jackson2008b,Leconte2015,Arras2010}.  If the eccentricity is maintained, tidal heating of short-period, eccentric exoplanets can compete with heating by the host star, even for modest eccentricities \citep{Driscoll2015}.  

Using Equation (2) from \citet{Driscoll2015}, the ratio of the
tidal heating flux to the equilibrium flux is given by
\begin{equation}
\left(\frac{T_{tid}}{T_{eq}}\right)^4 = 1344 \pi^3 \frac{\vert {\rm  Im}(k_2)\vert}{1-A} \frac{M_\star}{L_\star} \frac{e^2 r_{\scriptscriptstyle \occulted}^3}{aP^3},
\end{equation}
where $A$ is the albedo, $T_{tid}$ is the effective temperature resulting from tidal flux, $T_{eq}$ is the effective temperature resulting from absorbed flux, $\mathrm{Im}(k_2)$ is the imaginary component of the tidal Love number \footnote{\edited{This approximately relates to the tidal quality factor via the expression $Q\approx -\mathrm{Re}(k_2)/\mathrm{Im}(k_2)$ \citep{Driscoll2015}.}}, $M_\star$ and $L_\star$ are the mass and luminosity of the star, and $a$, $e$, $P$, and $r_{\scriptscriptstyle \occulted}$ are the semi-major axis, eccentricity, orbital period, and radius of the planet, respectively. For TRAPPIST-1b, we find that
\begin{equation}
\left(\frac{T_{tid}}{T_{eq}}\right)^4 = 2.9 \times 10^6 \vert {\rm Im}(k_2)\vert \frac{e^2} {1-A}.
\end{equation}
For $\vert {\rm Im}(k_2)\vert = 3 \times 10^{-3}$ and $A \ll 1$, these fluxes are equal for $e \approx 0.011$. However, this is about a factor of $\sim$10 larger than the eccentricity predicted by tidal evolution of the system \citep{Luger2017}. Nevertheless, if the eccentricity were this large, the tidal flux should give a higher signal-to-noise on PPO measurements. If strong enough, this could perhaps enable a constraint on $\vert \mathrm{Im}(k_2)\vert$ if $A$ and $e$ are known.

\subsection{Comparison to other work}
\label{sec:discussion:priorwork}
Previous studies have developed similar integration schemes to the one employed in \planetplanet for computing various types of light curves. The \texttt{LUNA} code \citep{Kipping2011} is a photodynamical algorithm for computing mutual transits of planet--moon systems across limb-darkened stars. \texttt{LUNA} computes mutual light curves by considering 34 different cases for the relative positions of the three disks (the star, the planet, and the moon) and deriving analytic expressions for the occulted stellar flux for each one. While extremely fast, \texttt{LUNA} accounts only for occultations of the stellar flux, and not for occultation of the flux from the planet and/or moon, and can therefore not be easily extended to the general case of planet-planet occultations.

\texttt{LUNA} is also limited to the three-body case and to the limb darkening laws that can be modeled analytically with the \citet{MandelAgol2002} algorithm. Seeking to relax these assumptions and generalize the computation of mutual transits, \citet{Pal2012} developed an algorithm to analytically compute light curves for arbitrary numbers of overlapping bodies transiting stars with various limb darkening profiles. This algorithm relies on a clever implementation of Green's theorem, which allows one to analytically evaluate the surface integral of the stellar flux occulted by an arbitrary shape via computation of the line integral along the occultor boundary. In principle, the method is general enough that it can be extended to non-radially symmetric surface maps, as is the case for airless bodies or even bodies with arbitrary heterogeneous surface features. However, the algorithm requires finding (and integrating) a function whose exterior derivative is the surface brightness map, which can become arbitrarily complex for all but the simplest cases. Application of this algorithm to the airless body case would require finding the exterior derivative of Equation~(\ref{eqn:radiance_airless}), with the temperature profile given by Equation~(\ref{eqn:tlat}) and the zenith angle given by Equations~(\ref{eqn:zenith_angle1})--(\ref{eqn:zenith_angle3}), which is intractable.

As we mentioned earlier, the \batman code \citep{Kreidberg2015} discretizes the radial brightness gradient of the star to compute fast transit light curves. \planetplanet extends the \batman algorithm to brightness gradients that are symmetric about \emph{any} point on the surface of the body, not just the center of the disk. Because in the general case the curves of constant radiance are ellipses, \planetplanet light curves are slower to evaluate (as they require solving a quartic equation to determine the points of intersection with the occultor; see Appendix~\ref{app:circle-ellipse}). We therefore adopt the \batman algorithm when modeling single-planet transits and occultations of planets whose emission is radially symmetric, making \planetplanet fast for wavelength-dependent transit calculations.

\subsection{Kepler-444 and other Systems}
\label{sec:discussion:kepler444}
While we focused our discussion on TRAPPIST-1, planet-planet occultations likely occur in other compact, multi-planet systems and may be detectable in those. TRAPPIST-1 is the closest known edge-on, compact multi-planet system, which enhances its detectability, but because of the small sizes of its planets and their modest irradiation, there could exist other systems for which PPO detection is more favorable.

At 36 pc (three times more distant that TRAPPIST-1), Kepler-444 is a K dwarf hosting five edge-on, extremely coplanar sub-Earths with periods less than 10 days \citep{Campante2015,Mills2017}. The Kepler-444 planets receive 50--100 times higher irradiation than the TRAPPIST-1 planets and, with higher equilibrium temperatures and stronger thermal emission, could in principle produce detectable PPOs. However, despite the fact that the Kepler-444 planets emit more flux than the TRAPPIST-1 planets, the larger luminosity of the star and greater distance to the system strongly decrease the occultation signal and increase the noise on PPO observations. We find that the $7.7$ $\mu$m JWST/MIRI filter (F770W) would be the optimal filter for secondary eclipse and PPO observations of Kepler-444b, but on the order of 100 events would be required to build up a SNR$ {\sim} 3$, which makes their detection infeasible.  On the other hand, only ${\sim}15$ secondary eclipses/occultations of Kepler-444b would be required for a SNR$ {\sim} 3$ detection using our nominal OST setup from  \S\ref{sec:results:photo:ost} with the same F770W filter.

Alternatively, surveys such as TESS and PLATO may soon discover nearby systems of multiply-transiting super-Earths and mini-Neptunes, which could have stronger PPO signals than TRAPPIST-1. In particular, PPO searches should focus on late-M or brown dwarf multi-planet systems with large, short-period planets that are equally distant or closer than TRAPPIST-1.

\subsection{Other applications}
In principle, PPOs could be used to search for non-transiting planets. This was pointed out by \citet{Ragozzine2010}, with the caveat that the rare and aperiodic nature of these events would make detections based on PPOs alone extremely challenging. \edited{In practice, it may be easier to use PPOs to find additional non-transiting planets in multi-planet systems.} \edited{Alternatively}, there may exist systems of extremely coplanar planets that are inclined just enough so that they do not transit. Around main sequence stars, even small system inclinations will in general greatly reduce the frequency of PPOs, since the minimum projected separation of coplanar planets increases steeply with inclination. Compact planetary systems orbiting white dwarfs or other compact objects, on the other hand, can still be extremely close to edge-on and not transit. If planetary systems orbiting these stars are common, PPOs may be a way to detect and characterize them.

This last point merits further discussion. Not only may PPOs be ideally suited to detecting planets around white dwarfs, but they should also be \emph{easier} to detect for these systems than for main-sequence stars (should they exist). In the Rayleigh-Jeans limit, the depth of a PPO scales as $T_p r_{\scriptscriptstyle \occulted}^2/(L_\star^{1/4} R_\star^{3/2})$, where $T_p$ and $r_{\scriptscriptstyle \occulted}$ are the temperature and radius of the planet and $L_\star$ and $R_\star$ are the luminosity and radius of the star, respectively. For fixed planet properties and stellar luminosity, a planet orbiting a smaller star undergoes deeper occultations. In fact, if TRAPPIST-1 were a white dwarf with the same luminosity, the depth of PPO events in the system would be $\sim$30 times larger because of the relative faintness of the star in the infrared.


Finally, we only explored PPO detectability from space-based telescopes. 
It may be possible in the future to consider the observation of PPOs with ground-based telescopes.  This is challenging due to the small PPO depth, limited visibility from any given site, high sky-brightness in the mid-infrared, and limited sky transmittance.  However, future extremely large telescopes (ELTs) will have a smaller PSF, decreasing the contribution from sky brightness to the noise, and a large collecting area, leading to smaller photon noise.  In practice, though,  variable sky brightness and variable seeing will likely limit the precision of ground-based ELTs, just as it limits the precision of transit and secondary eclipse measurements.  The transmittance in the 20~$\mu$m $Q$ band requires high and/or dry sites for observation, as well as limited cloud emissivity, all leading to a small probability of success.  Nevertheless, the possibility of ground-based observations of PPOs may be worth examining in the future. 


\section{Conclusions and future directions}
\label{sec:conclusions}
We have developed a formalism to predict, model, and extract orbital and atmospheric information from planet-planet occultations (PPOs) in extrasolar multi-planet systems. These events occur when a planet occults another planet in the same system, blocking thermal or reflected light from the occulted body and producing a photometric signal analagous to (and of similar depth to) a secondary eclipse. While PPOs are in general rare events, for extremely coplanar, edge-on, and compact systems such as TRAPPIST-1 they are frequent and occur preferentially at certain orbital phases, which may in some cases permit advanced scheduling of observations.

PPOs are a powerful tool to assess the day/night temperature contrast of exoplanets. Unlike secondary eclipse, which always probes a planet's day side emission, PPOs can occur at any phase and thus can yield information about the entire surface of the planet, enabling crude two-dimensional surface maps that are potentially not degenerate in latitude. Moreover, PPOs yield strong constraints on the three-dimensional architecture of exoplanetary systems. Observations of multiple events can yield the full eccentricity vectors of pairs of planets, breaking both the eccentricity-eccentricity and eccentricity-mass degeneracy inherent in transit timing variation (TTV) analyses. Because they occur off the face of the star, PPOs also constrain the mutual inclinations of pairs of planets precisely.

At this time, the mutual inclinations of the TRAPPIST-1 planets are unconstrained. We developed a Monte Carlo framework to assess their values given current data on the system, finding that the scatter in their mutual inclinations is less than 0.3$^\circ$ with 90\% confidence, suggesting the TRAPPIST-1 system is \emph{extremely} coplanar. We marginalize over the uncertainty on the mutual inclinations and all other orbital parameters to determine the frequency of PPOs in TRAPPIST-1. We find that occultations among the TRAPPIST-1 planets occur, on average, about \edited{1.4 times} per (Earth) day.

In order to model PPO light curves, we developed the open-source photodynamical code \texttt{planetplanet}, which uses an $N$-body solver to calculate the sky-projected separations of all pairs of planets and computes wavelength-dependent light curves for occultation events. We model planet thermal emission in two limits: the thick atmosphere limit, in which the planet is assumed to have symmetric emission about the center of its disk, and the thin atmosphere \edited{(``eyeball'')} limit, in which its thermal emission is proportional to the stellar irradiation profile. We developed a novel integration scheme that takes advantage of the elliptical symmetry of the problem to compute fast PPO light curves and use it to also model transits, secondary eclipses, phase curves, and mutual transits.

We modeled the detectability of PPOs with future space-based telescopes. We find that observations with JWST/MIRI at 12.8 and/or 15 $\mu$m are best suited to study PPOs in TRAPPIST-1. \bugfix{Although most events will be below the level of the noise, our simulations show that on the order of 20 occultations of b should have $\mathrm{SNR}>1$ in a given year if it is a uniform emitter. Occultations of c are most detectable if it has a strong day/night temperature contrast, in which case on the order of 10 $\mathrm{SNR}>1$ occultations are expected per year. Joint modeling of these occultations could lead to a $\mathrm{SNR}{\sim}7-8$ detection in one year. Individually detectable (SNR${\gtrsim}4$) occultations are rare but may occur once every few years for TRAPPIST-1b and c thanks to retrograde motion. These ``doublet'' PPOs occur when the relative velocity of TRAPPIST-1b and c changes sign during an occultation, extending the duration of the event to a few hours.}

For a future far-infrared surveyor such as the Origins Space Telescope (OST), we find that single PPOs are detectable in TRAPPIST-1 at \edited{${\gtrsim}10\sigma$}. Future observations of these events may therefore paint the most complete picture yet of the three-dimensional architecture \edited{and densities} of an exoplanet system and of the surfaces of its planets.

PPOs likely occur in other compact multi-planet systems, such as Kepler-444; TESS and PLATO may reveal other nearby systems that can be explored with JWST using this technique. Finally, we suggest that PPOs may be a way to detect near-edge-on systems of planets around white dwarfs and other compact objects, given the low geometric probability of transit in these systems.

All code used to generate the figures in this paper is open source and available at \url{https://github.com/rodluger/planetplanet}. The photodynamical code \planetplanet is also installable with the \texttt{pip} package manager. \bugfix{In addition to PPO light curves, \planetplanet can be used to model transits, secondary eclipses, phase curves, mutual transits, exomoons, and more.} Complete documentation is available at the url listed above.

\acknowledgments

This work was supported by the NASA Astrobiology Institute's Virtual Planetary Laboratory under Cooperative Agreement number NNA13AA93A. This work made use of the advanced computational, storage, and networking infrastructure provided by the Hyak supercomputer system at the University of Washington. \edited{We would like to thank Darin Ragozzine for his comments and suggestions, which greatly helped improve the paper.} We would also like to thank Daniel Foreman-Mackey, Dan Fabrycky, David Fleming, and Michael Gillon for useful discussions. E.A. acknowledges support from NASA grant NNX13AF62G and NSF grant AST-1615315.  Simulations in this paper made use of the \texttt{REBOUND} code which can be downloaded freely at \url{http://github.com/hannorein/rebound}. \edited{The version of \planetplanet used to generate the figures in this paper has been archived at \url{http://dx.doi.org/10.5281/zenodo.997391}.}

\clearpage
\bibliography{bib}\clearpage

\begin{thebibliography}{76}
\expandafter\ifx\csname natexlab\endcsname\relax\def\natexlab#1{#1}\fi

\bibitem[{{Agol} {et~al.}(2010){Agol}, {Cowan}, {Knutson}, {Deming}, {Steffen},
  {Henry}, \& {Charbonneau}}]{Agol2010}
{Agol}, E., {Cowan}, N.~B., {Knutson}, H.~A., {Deming}, D., {Steffen}, J.~H.,
  {Henry}, G.~W., \& {Charbonneau}, D. 2010, \apj, 721, 1861

\bibitem[{{Albers}(1979)}]{Albers1979}
{Albers}, S.~C. 1979, \skytel, 57

\bibitem[{{Arney} {et~al.}(2014){Arney}, {Meadows}, {Crisp}, {Schmidt},
  {Bailey}, \& {Robinson}}]{Arney2014}
{Arney}, G., {Meadows}, V., {Crisp}, D., {Schmidt}, S.~J., {Bailey}, J., \&
  {Robinson}, T. 2014, Journal of Geophysical Research (Planets), 119, 1860

\bibitem[{Arras \& Socrates(2010)}]{Arras2010}
Arras, P. \& Socrates, A. 2010, The Astrophysical Journal, 714, 1

\bibitem[{{Barstow} \& {Irwin}(2016)}]{Barstow2016}
{Barstow}, J.~K. \& {Irwin}, P.~G.~J. 2016, Monthly Notices of the Royal
  Astronomical Society, 461, L92

\bibitem[{{Batygin} {et~al.}(2009){Batygin}, {Bodenheimer}, \&
  {Laughlin}}]{Batygin2009}
{Batygin}, K., {Bodenheimer}, P., \& {Laughlin}, G. 2009, \apjl, 704, L49

\bibitem[{{Bolmont} {et~al.}(2013){Bolmont}, {Selsis}, {Raymond}, {Leconte},
  {Hersant}, {Maurin}, \& {Pericaud}}]{Bolmont2013}
{Bolmont}, E., {Selsis}, F., {Raymond}, S.~N., {Leconte}, J., {Hersant}, F.,
  {Maurin}, A.-S., \& {Pericaud}, J. 2013, \aap, 556, A17

\bibitem[{{Borucki} {et~al.}(2011){Borucki}, {Koch}, {Basri}, {Batalha},
  {Brown}, {Bryson}, {Caldwell}, {Christensen-Dalsgaard}, {Cochran}, {DeVore},
  {Dunham}, {Gautier}, {Geary}, {Gilliland}, {Gould}, {Howell}, {Jenkins},
  {Latham}, {Lissauer}, {Marcy}, {Rowe}, {Sasselov}, {Boss}, {Charbonneau},
  {Ciardi}, {Doyle}, {Dupree}, {Ford}, {Fortney}, {Holman}, {Seager},
  {Steffen}, {Tarter}, {Welsh}, {Allen}, {Buchhave}, {Christiansen}, {Clarke},
  {Das}, {D{\'e}sert}, {Endl}, {Fabrycky}, {Fressin}, {Haas}, {Horch},
  {Howard}, {Isaacson}, {Kjeldsen}, {Kolodziejczak}, {Kulesa}, {Li}, {Lucas},
  {Machalek}, {McCarthy}, {MacQueen}, {Meibom}, {Miquel}, {Prsa}, {Quinn},
  {Quintana}, {Ragozzine}, {Sherry}, {Shporer}, {Tenenbaum}, {Torres},
  {Twicken}, {Van Cleve}, {Walkowicz}, {Witteborn}, \& {Still}}]{Borucki2011}
{Borucki}, W.~J., {Koch}, D.~G., {Basri}, G., {Batalha}, N., {Brown}, T.~M.,
  {Bryson}, S.~T., {Caldwell}, D., {Christensen-Dalsgaard}, J., {Cochran},
  W.~D., {DeVore}, E., {Dunham}, E.~W., {Gautier}, III, T.~N., {Geary}, J.~C.,
  {Gilliland}, R., {Gould}, A., {Howell}, S.~B., {Jenkins}, J.~M., {Latham},
  D.~W., {Lissauer}, J.~J., {Marcy}, G.~W., {Rowe}, J., {Sasselov}, D., {Boss},
  A., {Charbonneau}, D., {Ciardi}, D., {Doyle}, L., {Dupree}, A.~K., {Ford},
  E.~B., {Fortney}, J., {Holman}, M.~J., {Seager}, S., {Steffen}, J.~H.,
  {Tarter}, J., {Welsh}, W.~F., {Allen}, C., {Buchhave}, L.~A., {Christiansen},
  J.~L., {Clarke}, B.~D., {Das}, S., {D{\'e}sert}, J.-M., {Endl}, M.,
  {Fabrycky}, D., {Fressin}, F., {Haas}, M., {Horch}, E., {Howard}, A.,
  {Isaacson}, H., {Kjeldsen}, H., {Kolodziejczak}, J., {Kulesa}, C., {Li}, J.,
  {Lucas}, P.~W., {Machalek}, P., {McCarthy}, D., {MacQueen}, P., {Meibom}, S.,
  {Miquel}, T., {Prsa}, A., {Quinn}, S.~N., {Quintana}, E.~V., {Ragozzine}, D.,
  {Sherry}, W., {Shporer}, A., {Tenenbaum}, P., {Torres}, G., {Twicken}, J.~D.,
  {Van Cleve}, J., {Walkowicz}, L., {Witteborn}, F.~C., \& {Still}, M. 2011,
  \apj, 736, 19

\bibitem[{{Bouchet} {et~al.}(2015){Bouchet}, {Garc{\'{\i}}a-Mar{\'{\i}}n},
  {Lagage}, {Amiaux}, {Augu{\'e}res}, {Bauwens}, {Blommaert}, {Chen}, {Detre},
  {Dicken}, {Dubreuil}, {Galdemard}, {Gastaud}, {Glasse}, {Gordon}, {Gougnaud},
  {Guillard}, {Justtanont}, {Krause}, {Leboeuf}, {Longval}, {Martin}, {Mazy},
  {Moreau}, {Olofsson}, {Ray}, {Rees}, {Renotte}, {Ressler}, {Ronayette},
  {Salasca}, {Scheithauer}, {Sykes}, {Thelen}, {Wells}, {Wright}, \&
  {Wright}}]{Bouchet2015}
{Bouchet}, P., {Garc{\'{\i}}a-Mar{\'{\i}}n}, M., {Lagage}, P.-O., {Amiaux}, J.,
  {Augu{\'e}res}, J.-L., {Bauwens}, E., {Blommaert}, J.~A.~D.~L., {Chen},
  C.~H., {Detre}, {\"O}.~H., {Dicken}, D., {Dubreuil}, D., {Galdemard}, P.,
  {Gastaud}, R., {Glasse}, A., {Gordon}, K.~D., {Gougnaud}, F., {Guillard}, P.,
  {Justtanont}, K., {Krause}, O., {Leboeuf}, D., {Longval}, Y., {Martin}, L.,
  {Mazy}, E., {Moreau}, V., {Olofsson}, G., {Ray}, T.~P., {Rees}, J.-M.,
  {Renotte}, E., {Ressler}, M.~E., {Ronayette}, S., {Salasca}, S.,
  {Scheithauer}, S., {Sykes}, J., {Thelen}, M.~P., {Wells}, M., {Wright}, D.,
  \& {Wright}, G.~S. 2015, \pasp, 127, 612

\bibitem[{{Brakensiek} \& {Ragozzine}(2016)}]{Brakensiek2016}
{Brakensiek}, J. \& {Ragozzine}, D. 2016, \apj, 821, 47

\bibitem[{{Burgasser} \& {Mamajek}(2017)}]{BurgasserMamajek2017}
{Burgasser}, A.~J. \& {Mamajek}, E.~E. 2017, ArXiv e-prints

\bibitem[{{Cabrera} \& {Schneider}(2007)}]{Cabrera2007}
{Cabrera}, J. \& {Schneider}, J. 2007, \aap, 464, 1133

\bibitem[{{Campante} {et~al.}(2015){Campante}, {Barclay}, {Swift}, {Huber},
  {Adibekyan}, {Cochran}, {Burke}, {Isaacson}, {Quintana}, {Davies}, {Silva
  Aguirre}, {Ragozzine}, {Riddle}, {Baranec}, {Basu}, {Chaplin},
  {Christensen-Dalsgaard}, {Metcalfe}, {Bedding}, {Handberg}, {Stello},
  {Brewer}, {Hekker}, {Karoff}, {Kolbl}, {Law}, {Lundkvist}, {Miglio}, {Rowe},
  {Santos}, {Van Laerhoven}, {Arentoft}, {Elsworth}, {Fischer}, {Kawaler},
  {Kjeldsen}, {Lund}, {Marcy}, {Sousa}, {Sozzetti}, \& {White}}]{Campante2015}
{Campante}, T.~L., {Barclay}, T., {Swift}, J.~J., {Huber}, D., {Adibekyan},
  V.~Z., {Cochran}, W., {Burke}, C.~J., {Isaacson}, H., {Quintana}, E.~V.,
  {Davies}, G.~R., {Silva Aguirre}, V., {Ragozzine}, D., {Riddle}, R.,
  {Baranec}, C., {Basu}, S., {Chaplin}, W.~J., {Christensen-Dalsgaard}, J.,
  {Metcalfe}, T.~S., {Bedding}, T.~R., {Handberg}, R., {Stello}, D., {Brewer},
  J.~M., {Hekker}, S., {Karoff}, C., {Kolbl}, R., {Law}, N.~M., {Lundkvist},
  M., {Miglio}, A., {Rowe}, J.~F., {Santos}, N.~C., {Van Laerhoven}, C.,
  {Arentoft}, T., {Elsworth}, Y.~P., {Fischer}, D.~A., {Kawaler}, S.~D.,
  {Kjeldsen}, H., {Lund}, M.~N., {Marcy}, G.~W., {Sousa}, S.~G., {Sozzetti},
  A., \& {White}, T.~R. 2015, \apj, 799, 170

\bibitem[{{Carter} {et~al.}(2012){Carter}, {Agol}, {Chaplin}, {Basu},
  {Bedding}, {Buchhave}, {Christensen-Dalsgaard}, {Deck}, {Elsworth},
  {Fabrycky}, {Ford}, {Fortney}, {Hale}, {Handberg}, {Hekker}, {Holman},
  {Huber}, {Karoff}, {Kawaler}, {Kjeldsen}, {Lissauer}, {Lopez}, {Lund},
  {Lundkvist}, {Metcalfe}, {Miglio}, {Rogers}, {Stello}, {Borucki}, {Bryson},
  {Christiansen}, {Cochran}, {Geary}, {Gilliland}, {Haas}, {Hall}, {Howard},
  {Jenkins}, {Klaus}, {Koch}, {Latham}, {MacQueen}, {Sasselov}, {Steffen},
  {Twicken}, \& {Winn}}]{Carter2012}
{Carter}, J.~A., {Agol}, E., {Chaplin}, W.~J., {Basu}, S., {Bedding}, T.~R.,
  {Buchhave}, L.~A., {Christensen-Dalsgaard}, J., {Deck}, K.~M., {Elsworth},
  Y., {Fabrycky}, D.~C., {Ford}, E.~B., {Fortney}, J.~J., {Hale}, S.~J.,
  {Handberg}, R., {Hekker}, S., {Holman}, M.~J., {Huber}, D., {Karoff}, C.,
  {Kawaler}, S.~D., {Kjeldsen}, H., {Lissauer}, J.~J., {Lopez}, E.~D., {Lund},
  M.~N., {Lundkvist}, M., {Metcalfe}, T.~S., {Miglio}, A., {Rogers}, L.~A.,
  {Stello}, D., {Borucki}, W.~J., {Bryson}, S., {Christiansen}, J.~L.,
  {Cochran}, W.~D., {Geary}, J.~C., {Gilliland}, R.~L., {Haas}, M.~R., {Hall},
  J., {Howard}, A.~W., {Jenkins}, J.~M., {Klaus}, T., {Koch}, D.~G., {Latham},
  D.~W., {MacQueen}, P.~J., {Sasselov}, D., {Steffen}, J.~H., {Twicken}, J.~D.,
  \& {Winn}, J.~N. 2012, Science, 337, 556

\bibitem[{Charbonneau {et~al.}(2005)Charbonneau, Allen, Megeath, Torres,
  Alonso, Brown, Gilliland, Latham, Mandushev, O'Donovan, \&
  Sozzetti}]{Charbonneau2005}
Charbonneau, D., Allen, L.~E., Megeath, S.~T., Torres, G., Alonso, R., Brown,
  T.~M., Gilliland, R.~L., Latham, D.~W., Mandushev, G., O'Donovan, F.~T., \&
  Sozzetti, A. 2005, The Astrophysical Journal, 626, 523

\bibitem[{{Claret}(2000)}]{Claret2000}
{Claret}, A. 2000, \aap, 363, 1081

\bibitem[{{Cooray} \& {Origins Space Telescope Study Team}(2017)}]{Cooray2017}
{Cooray}, A.~R. \& {Origins Space Telescope Study Team}. 2017, in American
  Astronomical Society Meeting Abstracts, Vol. 229, American Astronomical
  Society Meeting Abstracts, 405.01

\bibitem[{Cowan \& Agol(2008)}]{Cowan2008}
Cowan, N.~B. \& Agol, E. 2008, The Astrophysical Journal, 678, L129

\bibitem[{Cowan {et~al.}(2009)Cowan, Agol, Meadows, Robinson, Livengood,
  Deming, Lisse, A'Hearn, Wellnitz, Seager, \& and}]{Cowan2009}
Cowan, N.~B., Agol, E., Meadows, V.~S., Robinson, T., Livengood, T.~A., Deming,
  D., Lisse, C.~M., A'Hearn, M.~F., Wellnitz, D.~D., Seager, S., \& and, D.~C.
  2009, The Astrophysical Journal, 700, 915

\bibitem[{{Cowan} {et~al.}(2017){Cowan}, {Chayes}, {Bouffard}, {Meynig}, \&
  {Haggard}}]{Cowan2017}
{Cowan}, N.~B., {Chayes}, V., {Bouffard}, {\'E}., {Meynig}, M., \& {Haggard},
  H.~M. 2017, \mnras, 467, 747

\bibitem[{{Cowan} {et~al.}(2013){Cowan}, {Fuentes}, \& {Haggard}}]{Cowan2013}
{Cowan}, N.~B., {Fuentes}, P.~A., \& {Haggard}, H.~M. 2013, \mnras, 434, 2465

\bibitem[{{de Kleer} {et~al.}(2017){de Kleer}, {Skrutskie}, {Leisenring},
  {Davies}, {Conrad}, {de Pater}, {Resnick}, {Bailey}, {Defr{\`e}re}, {Hinz},
  {Skemer}, {Spalding}, {Vaz}, {Veillet}, \& {Woodward}}]{deKleer2017}
{de Kleer}, K., {Skrutskie}, M., {Leisenring}, J., {Davies}, A.~G., {Conrad},
  A., {de Pater}, I., {Resnick}, A., {Bailey}, V., {Defr{\`e}re}, D., {Hinz},
  P., {Skemer}, A., {Spalding}, E., {Vaz}, A., {Veillet}, C., \& {Woodward},
  C.~E. 2017, \nat, 545, 199

\bibitem[{{de Wit} {et~al.}(2012){de Wit}, {Gillon}, {Demory}, \&
  {Seager}}]{deWit2012}
{de Wit}, J., {Gillon}, M., {Demory}, B.-O., \& {Seager}, S. 2012, \aap, 548,
  A128

\bibitem[{{de Wit} {et~al.}(2016){de Wit}, {Wakeford}, {Gillon}, {Lewis},
  {Valenti}, {Demory}, {Burgasser}, {Burdanov}, {Delrez}, {Jehin}, {Lederer},
  {Queloz}, {Triaud}, \& {Van Grootel}}]{deWit2016}
{de Wit}, J., {Wakeford}, H.~R., {Gillon}, M., {Lewis}, N.~K., {Valenti},
  J.~A., {Demory}, B.-O., {Burgasser}, A.~J., {Burdanov}, A., {Delrez}, L.,
  {Jehin}, E., {Lederer}, S.~M., {Queloz}, D., {Triaud}, A.~H.~M.~J., \& {Van
  Grootel}, V. 2016, \nat, 537, 69

\bibitem[{Deck \& Agol(2015)}]{Deck2015}
Deck, K.~M. \& Agol, E. 2015, {ApJ}, 802, 116

\bibitem[{{Demory} {et~al.}(2016){Demory}, {Gillon}, {de Wit}, {Madhusudhan},
  {Bolmont}, {Heng}, {Kataria}, {Lewis}, {Hu}, {Krick}, {Stamenkovi{\'c}},
  {Benneke}, {Kane}, \& {Queloz}}]{Demory2016}
{Demory}, B.-O., {Gillon}, M., {de Wit}, J., {Madhusudhan}, N., {Bolmont}, E.,
  {Heng}, K., {Kataria}, T., {Lewis}, N., {Hu}, R., {Krick}, J.,
  {Stamenkovi{\'c}}, V., {Benneke}, B., {Kane}, S., \& {Queloz}, D. 2016, \nat,
  532, 207

\bibitem[{Driscoll \& Barnes(2015)}]{Driscoll2015}
Driscoll, P. \& Barnes, R. 2015, Astrobiology, 15, 739

\bibitem[{Foreman-Mackey(2016)}]{ForemanMackey2016}
Foreman-Mackey, D. 2016, The Journal of Open Source Software, 24

\bibitem[{{Foreman-Mackey} {et~al.}(2013){Foreman-Mackey}, {Hogg}, {Lang}, \&
  {Goodman}}]{ForemanMackey2013}
{Foreman-Mackey}, D., {Hogg}, D.~W., {Lang}, D., \& {Goodman}, J. 2013, \pasp,
  125, 306

\bibitem[{{Galassi} {et~al.}(2016)}]{GSL}
{Galassi}, M. {et~al.} 2016, {GNU Scientific Library Reference Manual, 3rd ed.}

\bibitem[{{Gardner} {et~al.}(2006){Gardner}, {Mather}, {Clampin}, {Doyon},
  {Greenhouse}, {Hammel}, {Hutchings}, {Jakobsen}, {Lilly}, {Long}, {Lunine},
  {McCaughrean}, {Mountain}, {Nella}, {Rieke}, {Rieke}, {Rix}, {Smith},
  {Sonneborn}, {Stiavelli}, {Stockman}, {Windhorst}, \& {Wright}}]{Gardner2006}
{Gardner}, J.~P., {Mather}, J.~C., {Clampin}, M., {Doyon}, R., {Greenhouse},
  M.~A., {Hammel}, H.~B., {Hutchings}, J.~B., {Jakobsen}, P., {Lilly}, S.~J.,
  {Long}, K.~S., {Lunine}, J.~I., {McCaughrean}, M.~J., {Mountain}, M.,
  {Nella}, J., {Rieke}, G.~H., {Rieke}, M.~J., {Rix}, H.-W., {Smith}, E.~P.,
  {Sonneborn}, G., {Stiavelli}, M., {Stockman}, H.~S., {Windhorst}, R.~A., \&
  {Wright}, G.~S. 2006, \ssr, 123, 485

\bibitem[{{Gillon} {et~al.}(2016){Gillon}, {Jehin}, {Lederer}, {Delrez}, {de
  Wit}, {Burdanov}, {Van Grootel}, {Burgasser}, {Triaud}, {Opitom}, {Demory},
  {Sahu}, {Bardalez Gagliuffi}, {Magain}, \& {Queloz}}]{Gillon2016}
{Gillon}, M., {Jehin}, E., {Lederer}, S.~M., {Delrez}, L., {de Wit}, J.,
  {Burdanov}, A., {Van Grootel}, V., {Burgasser}, A.~J., {Triaud}, A.~H.~M.~J.,
  {Opitom}, C., {Demory}, B.-O., {Sahu}, D.~K., {Bardalez Gagliuffi}, D.,
  {Magain}, P., \& {Queloz}, D. 2016, \nat, 533, 221

\bibitem[{{Gillon} {et~al.}(2017){Gillon}, {Triaud}, {Demory}, {Jehin}, {Agol},
  {Deck}, {Lederer}, {de Wit}, {Burdanov}, {Ingalls}, {Bolmont}, {Leconte},
  {Raymond}, {Selsis}, {Turbet}, {Barkaoui}, {Burgasser}, {Burleigh}, {Carey},
  {Chaushev}, {Copperwheat}, {Delrez}, {Fernandes}, {Holdsworth}, {Kotze}, {Van
  Grootel}, {Almleaky}, {Benkhaldoun}, {Magain}, \& {Queloz}}]{Gillon2017}
{Gillon}, M., {Triaud}, A.~H.~M.~J., {Demory}, B.-O., {Jehin}, E., {Agol}, E.,
  {Deck}, K.~M., {Lederer}, S.~M., {de Wit}, J., {Burdanov}, A., {Ingalls},
  J.~G., {Bolmont}, E., {Leconte}, J., {Raymond}, S.~N., {Selsis}, F.,
  {Turbet}, M., {Barkaoui}, K., {Burgasser}, A., {Burleigh}, M.~R., {Carey},
  S.~J., {Chaushev}, A., {Copperwheat}, C.~M., {Delrez}, L., {Fernandes},
  C.~S., {Holdsworth}, D.~L., {Kotze}, E.~J., {Van Grootel}, V., {Almleaky},
  Y., {Benkhaldoun}, Z., {Magain}, P., \& {Queloz}, D. 2017, \nat, 542, 456

\bibitem[{Gim{\'{e}}nez(2006)}]{Gimnez2006}
Gim{\'{e}}nez, A. 2006, Astronomy {\&} Astrophysics, 450, 1231

\bibitem[{{Glasse} {et~al.}(2015){Glasse}, {Rieke}, {Bauwens},
  {Garc{\'{\i}}a-Mar{\'{\i}}n}, {Ressler}, {Rost}, {Tikkanen}, {Vandenbussche},
  \& {Wright}}]{Glasse2015}
{Glasse}, A., {Rieke}, G.~H., {Bauwens}, E., {Garc{\'{\i}}a-Mar{\'{\i}}n}, M.,
  {Ressler}, M.~E., {Rost}, S., {Tikkanen}, T.~V., {Vandenbussche}, B., \&
  {Wright}, G.~S. 2015, \pasp, 127, 686

\bibitem[{{Hilton} {et~al.}(1988){Hilton}, {Seidelmann}, \& {Liu}}]{Hilton1988}
{Hilton}, J.~L., {Seidelmann}, P.~K., \& {Liu}, C. 1988, \aj, 96, 1482

\bibitem[{Hirano {et~al.}(2012)Hirano, Narita, Sato, Takahashi, Masuda, Takeda,
  Aoki, Tamura, \& Suto}]{Hirano2012}
Hirano, T., Narita, N., Sato, B., Takahashi, Y.~H., Masuda, K., Takeda, Y.,
  Aoki, W., Tamura, M., \& Suto, Y. 2012, The Astrophysical Journal, 759, L36

\bibitem[{{Hughes} \& {Chraibi}(2011)}]{Hughes2011}
{Hughes}, G.~B. \& {Chraibi}, M. 2011, ArXiv e-prints

\bibitem[{Jackson {et~al.}(2008{\natexlab{a}})Jackson, Greenberg, \&
  Barnes}]{Jackson2008a}
Jackson, B., Greenberg, R., \& Barnes, R. 2008{\natexlab{a}}, The Astrophysical
  Journal, 678, 1396

\bibitem[{Jackson {et~al.}(2008{\natexlab{b}})Jackson, Greenberg, \&
  Barnes}]{Jackson2008b}
---. 2008{\natexlab{b}}, The Astrophysical Journal, 681, 1631

\bibitem[{{Kane} \& {Gelino}(2013)}]{Kane2013}
{Kane}, S.~R. \& {Gelino}, D.~M. 2013, \apj, 762, 129

\bibitem[{Kipping(2011)}]{Kipping2011}
Kipping, D.~M. 2011, Monthly Notices of the Royal Astronomical Society, no

\bibitem[{Knutson {et~al.}(2007)Knutson, Charbonneau, Allen, Fortney, Agol,
  Cowan, Showman, Cooper, \& Megeath}]{Knutson2007}
Knutson, H.~A., Charbonneau, D., Allen, L.~E., Fortney, J.~J., Agol, E., Cowan,
  N.~B., Showman, A.~P., Cooper, C.~S., \& Megeath, S.~T. 2007, Nature, 447,
  183

\bibitem[{{Kreidberg}(2015)}]{Kreidberg2015}
{Kreidberg}, L. 2015, \pasp, 127, 1161

\bibitem[{Leconte {et~al.}(2015)Leconte, Wu, Menou, \& Murray}]{Leconte2015}
Leconte, J., Wu, H., Menou, K., \& Murray, N. 2015, Science, 347, 632

\bibitem[{{Lissauer} {et~al.}(2011){Lissauer}, {Ragozzine}, {Fabrycky},
  {Steffen}, {Ford}, {Jenkins}, {Shporer}, {Holman}, {Rowe}, {Quintana},
  {Batalha}, {Borucki}, {Bryson}, {Caldwell}, {Carter}, {Ciardi}, {Dunham},
  {Fortney}, {Gautier}, {Howell}, {Koch}, {Latham}, {Marcy}, {Morehead}, \&
  {Sasselov}}]{Lissauer2011}
{Lissauer}, J.~J., {Ragozzine}, D., {Fabrycky}, D.~C., {Steffen}, J.~H.,
  {Ford}, E.~B., {Jenkins}, J.~M., {Shporer}, A., {Holman}, M.~J., {Rowe},
  J.~F., {Quintana}, E.~V., {Batalha}, N.~M., {Borucki}, W.~J., {Bryson},
  S.~T., {Caldwell}, D.~A., {Carter}, J.~A., {Ciardi}, D., {Dunham}, E.~W.,
  {Fortney}, J.~J., {Gautier}, III, T.~N., {Howell}, S.~B., {Koch}, D.~G.,
  {Latham}, D.~W., {Marcy}, G.~W., {Morehead}, R.~C., \& {Sasselov}, D. 2011,
  \apjs, 197, 8

\bibitem[{{Lithwick} {et~al.}(2012){Lithwick}, {Xie}, \& {Wu}}]{Lithwick2012}
{Lithwick}, Y., {Xie}, J., \& {Wu}, Y. 2012, \apj, 761, 122

\bibitem[{{Luger} {et~al.}(2017){Luger}, {Sestovic}, {Kruse}, {Grimm},
  {Demory}, {Agol}, {Bolmont}, {Fabrycky}, {Fernandes}, {Van Grootel},
  {Burgasser}, {Gillon}, {Ingalls}, {Jehin}, {Raymond}, {Selsis}, {Triaud},
  {Barclay}, {Barentsen}, {Howell}, {Delrez}, {de Wit}, {Foreman-Mackey},
  {Holdsworth}, {Leconte}, {Lederer}, {Turbet}, {Almleaky}, {Benkhaldoun},
  {Magain}, {Morris}, {Heng}, \& {Queloz}}]{Luger2017}
{Luger}, R., {Sestovic}, M., {Kruse}, E., {Grimm}, S.~L., {Demory}, B.-O.,
  {Agol}, E., {Bolmont}, E., {Fabrycky}, D., {Fernandes}, C.~S., {Van Grootel},
  V., {Burgasser}, A., {Gillon}, M., {Ingalls}, J.~G., {Jehin}, E., {Raymond},
  S.~N., {Selsis}, F., {Triaud}, A.~H.~M.~J., {Barclay}, T., {Barentsen}, G.,
  {Howell}, S.~B., {Delrez}, L., {de Wit}, J., {Foreman-Mackey}, D.,
  {Holdsworth}, D.~L., {Leconte}, J., {Lederer}, S., {Turbet}, M., {Almleaky},
  Y., {Benkhaldoun}, Z., {Magain}, P., {Morris}, B.~M., {Heng}, K., \&
  {Queloz}, D. 2017, Nature Astronomy, 1, 0129

\bibitem[{{MacDonald} {et~al.}(2016){MacDonald}, {Ragozzine}, {Fabrycky},
  {Ford}, {Holman}, {Isaacson}, {Lissauer}, {Lopez}, {Mazeh}, {Rogers}, {Rowe},
  {Steffen}, \& {Torres}}]{MacDonald2016}
{MacDonald}, M.~G., {Ragozzine}, D., {Fabrycky}, D.~C., {Ford}, E.~B.,
  {Holman}, M.~J., {Isaacson}, H.~T., {Lissauer}, J.~J., {Lopez}, E.~D.,
  {Mazeh}, T., {Rogers}, L., {Rowe}, J.~F., {Steffen}, J.~H., \& {Torres}, G.
  2016, \aj, 152, 105

\bibitem[{{Mackay}(2003)}]{Mackay2003}
{Mackay}, D.~J.~C. 2003, {Information Theory, Inference and Learning
  Algorithms. Cambridge, UK: Cambridge University Press}

\bibitem[{Majeau {et~al.}(2012)Majeau, Agol, \& Cowan}]{Majeau2012}
Majeau, C., Agol, E., \& Cowan, N.~B. 2012, The Astrophysical Journal, 747, L20

\bibitem[{{Mandel} \& {Agol}(2002)}]{MandelAgol2002}
{Mandel}, K. \& {Agol}, E. 2002, \apjl, 580, L171

\bibitem[{{Mardling}(2007)}]{Mardling2007}
{Mardling}, R.~A. 2007, \mnras, 382, 1768

\bibitem[{{Masuda}(2014)}]{Masuda2014}
{Masuda}, K. 2014, \apj, 783, 53

\bibitem[{Masuda {et~al.}(2013)Masuda, Hirano, Taruya, Nagasawa, \&
  Suto}]{Masuda2013}
Masuda, K., Hirano, T., Taruya, A., Nagasawa, M., \& Suto, Y. 2013, The
  Astrophysical Journal, 778, 185

\bibitem[{{Maurin} {et~al.}(2012){Maurin}, {Selsis}, {Hersant}, \&
  {Belu}}]{Maurin2012}
{Maurin}, A.~S., {Selsis}, F., {Hersant}, F., \& {Belu}, A. 2012, \aap, 538,
  A95

\bibitem[{Mills \& Fabrycky(2017)}]{Mills2017}
Mills, S.~M. \& Fabrycky, D.~C. 2017, The Astrophysical Journal, 838, L11

\bibitem[{{Morley} {et~al.}(2017){Morley}, {Kreidberg}, {Rustamkulov},
  {Robinson}, \& {Fortney}}]{Morley2017}
{Morley}, C.~V., {Kreidberg}, L., {Rustamkulov}, Z., {Robinson}, T., \&
  {Fortney}, J.~J. 2017, ArXiv e-prints

\bibitem[{{Murray} {et~al.}(1963){Murray}, {Wildey}, \&
  {Westphal}}]{Murray1963}
{Murray}, B.~C., {Wildey}, R.~L., \& {Westphal}, J.~A. 1963, \jgr, 68, 4813

\bibitem[{{Nesvorn{\'y}} \& {Vokrouhlick{\'y}}(2014)}]{Nesvorny2014}
{Nesvorn{\'y}}, D. \& {Vokrouhlick{\'y}}, D. 2014, \apj, 790, 58

\bibitem[{{P{\'a}l}(2012)}]{Pal2012}
{P{\'a}l}, A. 2012, \mnras, 420, 1630

\bibitem[{{Ragozzine} \& {Holman}(2010)}]{Ragozzine2010}
{Ragozzine}, D. \& {Holman}, M.~J. 2010, ArXiv e-prints

\bibitem[{{Rauscher} {et~al.}(2007){Rauscher}, {Menou}, {Seager}, {Deming},
  {Cho}, \& {Hansen}}]{Rauscher2007}
{Rauscher}, E., {Menou}, K., {Seager}, S., {Deming}, D., {Cho}, J.~Y.-K., \&
  {Hansen}, B.~M.~S. 2007, \apj, 664, 1199

\bibitem[{{Rein} \& {Liu}(2012)}]{ReinLiu2012}
{Rein}, H. \& {Liu}, S.-F. 2012, \aap, 537, A128

\bibitem[{{Rein} \& {Spiegel}(2015)}]{ReinSpiegel2015}
{Rein}, H. \& {Spiegel}, D.~S. 2015, \mnras, 446, 1424

\bibitem[{{Rein} \& {Tamayo}(2015)}]{ReinTamayo2015}
{Rein}, H. \& {Tamayo}, D. 2015, \mnras, 452, 376

\bibitem[{{Sato} \& {Asada}(2009)}]{Sato2009}
{Sato}, M. \& {Asada}, H. 2009, \pasj, 61, L29

\bibitem[{{Sato} \& {Asada}(2010)}]{Sato2010}
---. 2010, \pasj, 62, 1203

\bibitem[{Selsis {et~al.}(2013)Selsis, Maurin, Hersant, Leconte, Bolmont,
  Raymond, \& Delbo'}]{Selsis2013}
Selsis, F., Maurin, A.-S., Hersant, F., Leconte, J., Bolmont, E., Raymond,
  S.~N., \& Delbo', M. 2013, Astronomy {\&} Astrophysics, 555, A51

\bibitem[{Selsis {et~al.}(2011)Selsis, Wordsworth, \& Forget}]{Selsis2011}
Selsis, F., Wordsworth, R.~D., \& Forget, F. 2011, Astronomy {\&} Astrophysics,
  532, A1

\bibitem[{{Stubbs}(1879)}]{Stubbs1879}
{Stubbs}, W. 1879, {Historical Works of Gervase of Canterbury}, Vol.~1 (Longman
  \& Company), 221

\bibitem[{{Swift} {et~al.}(2013){Swift}, {Johnson}, {Morton}, {Crepp},
  {Montet}, {Fabrycky}, \& {Muirhead}}]{Swift2013}
{Swift}, J.~J., {Johnson}, J.~A., {Morton}, T.~D., {Crepp}, J.~R., {Montet},
  B.~T., {Fabrycky}, D.~C., \& {Muirhead}, P.~S. 2013, \apj, 764, 105

\bibitem[{Veras \& Breedt(2017)}]{Veras2017}
Veras, D. \& Breedt, E. 2017, Monthly Notices of the Royal Astronomical
  Society, 468, 2672

\bibitem[{{Williams} {et~al.}(2006){Williams}, {Charbonneau}, {Cooper},
  {Showman}, \& {Fortney}}]{Williams2006}
{Williams}, P.~K.~G., {Charbonneau}, D., {Cooper}, C.~S., {Showman}, A.~P., \&
  {Fortney}, J.~J. 2006, \apj, 649, 1020

\bibitem[{{Winn}(2010)}]{Winn2010}
{Winn}, J.~N. {Exoplanet Transits and Occultations}, ed. S.~{Seager}
  (University of Arizona Press), 55--77

\bibitem[{{Yang} {et~al.}(2013){Yang}, {Cowan}, \& {Abbot}}]{Yang2013}
{Yang}, J., {Cowan}, N.~B., \& {Abbot}, D.~S. 2013, \apjl, 771, L45

\end{thebibliography}

\appendix

\section{Circles and Ellipses}
\label{app:circlesellipses}
In this section we derive the basic equations used to compute the integrals and the integration limits relevant to planet-planet occultation light curves. Sections \ref{app:circle-circle} and \ref{app:circle-ellipse} discuss how to compute the points of intersection between circles and ellipses, and section \ref{app:integral_ellipse} presents an analytic expression for the integral of a segment of an ellipse. These expressions are used in our integration scheme, presented in Appendix~\ref{app:integration} below.

\subsection{Circle-circle intersection}
\label{app:circle-circle}
The $x$ coordinates of the points of intersection of two circles of radii $r_1$ and $r_2$, the first centered at the origin and the second at the point ($x_0$, $y_0$), are
\begin{align}
    \label{eqn:circle-circle}
    x &= a \pm b
\end{align}
where
\begin{align}
    a &= \frac{x_0}{2d^2}\left(d^2 - r_1^2 + r_2^2\right),\nonumber\\
    b &= \frac{y_0}{2d^2}\sqrt{4r_1^2r_2^2-(r_1^2+r_2^2-d^2)^2}
\end{align}
and
\begin{align}
    d &= \sqrt{x_0^2 + y_0^2}
\end{align}
is the distance between their centers.

\subsection{Circle-ellipse intersection}
\label{app:circle-ellipse}
The points of intersection of a circle centered at the origin with radius $r$ and an ellipse centered at ($x_0$, $y_0$) with semi-major axis $a$ parallel to the $y$ axis and semi-minor axis $b$ parallel to the $x$ axis are given by the roots of the following equation:
\begin{align}
    y_0 \pm \frac{a}{b}\sqrt{b^2 - (x - x_0)^2} \pm \sqrt{r^2 - x^2} = 0.
\end{align}
This is equivalent to finding the roots of the following quartic polynomial:
\begin{align}
    \label{eqn:circle-ellipse}
    c_4 x^4 + c_3 x^3 + c_2 x^2 + c_1 x + c_0 = 0
\end{align}
where
\begin{align}
    c_0 &= C^2 - D (b^2 - x_0^2) \nonumber\\
    c_1 &= 2 B C - 2 D x_0 \nonumber\\
    c_2 &= 2 A C + B^2 + D \nonumber\\
    c_3 &= 2 A B \nonumber\\
    c_4 &= A^2
\end{align}
and
\begin{align}
    A &= \frac{a^2}{b^2} - 1 \nonumber\\
    B &= -2 x_0 \frac{a^2}{b^2} \nonumber\\
    C &= r^2 - y_0^2 - a^2 + x_0^2 \frac{a^2}{b^2} \nonumber\\
    D &= 4 y_0^2 \frac{a^2}{b^2}.
\end{align}
In general, this polynomial has four roots, 0, 2 or 4 of which are real and correspond to the actual points of intersection. These roots may be found analytically \citep[e.g.,][]{Hughes2011}, but in practice the large number of arithmetic operations can result in significant errors due to limited machine precision. It is therefore preferable to solve Equation~(\ref{eqn:circle-ellipse}) with a numerical root-finding algorithm. \edited{We use the GNU Scientific Library $\mathrm{\texttt{gsl\_poly\_complex\_solve}}$ routine, which performs QR decomposition of the companion matrix to numerically find roots of polynomials \citep{GSL}.}

\subsection{Integral of an ellipse\label{app:integral_ellipse}}
The function describing an ellipse centered at ($x_0, y_0$) and with semi-major axis $a$ aligned with the $y$ axis and semi-minor axis $b$ aligned with the $x$ axis is
\begin{align}
    \label{eqn:ellipse_function}
    y(x) = y_0 \pm \frac{a}{b}\sqrt{b^2 - (x - x_0)^2}.
\end{align}
The indefinite integral of Equation~(\ref{eqn:ellipse_function}) is
\begin{align}
    \label{eqn:ellipse_integral}
    \int y(x)dx &= y_0 x \pm \frac{a}{2b}\left[z(x - x_0) + b^2 \arctan\left(\frac{x - x_0}{z}\right)\right] + C
\end{align}
where
\begin{align}
    z = \sqrt{(b + x - x_0)(b - x + x_0)} \nonumber
\end{align}
and $C$ is an arbitrary constant.

\section{Geometry of eyeball planets}
\label{app:geometry}
In this section we derive geometrical relations relevant to the computation of occultation light curves of ``eyeball'' planets. In section~\ref{app:zenith} we derive an expression for the zenith angle $\phi$ at an arbitrary point $(x,y,z)$ on the surface of a sphere, which is necessary to compute the surface brightness in each of the integration regions (see Appendix~\ref{app:integration} below). In sections~\ref{app:inclined} and \ref{app:offset} we derive expressions that are useful for computing light curves of ``eyeball'' planets whose axis of symmetry is not the $x$ axis, such as planets in inclined orbits or planets with offset hotspots.

\subsection{Zenith angle of a point on a sphere}
\label{app:zenith}
Given a planet of radius $r_{\scriptscriptstyle \occulted}$ in an edge-on orbit at phase angle $\theta$ and centered at the origin, the zenith angle $\phi$ of a point ($x, y$) on the disk of the planet may be computed as follows. If $\theta = \pm \frac{\pi}{2}$, curves of constant zenith angle are circles, and the zenith angle is simply
\begin{align}
    \phi = 
    \begin{cases}
        \arcsin\left(\frac{\sqrt{x^2 + y^2}}{r_{\scriptscriptstyle \occulted}}\right) & \theta = +\frac{\pi}{2} \\
        \pi - \arcsin\left(\frac{\sqrt{x^2 + y^2}}{r_{\scriptscriptstyle \occulted}}\right) & \theta = -\frac{\pi}{2}.
    \end{cases}
\end{align}
At intermediate angles, the curves of constant zenith angle are ellipses. Given the equations in \S\ref{sec:photo:airless}, one may solve for the zenith angle of the ellipse that threads the point ($x,y$). This yields a quadratic equation in the quantity $z = \sin^2{\phi}$, which may be solved to yield
\begin{align}
    \label{eqn:zenith_angle1}
    \phi = 
    \begin{cases}
        \arcsin{\sqrt{z}} & \left(|\theta| \leq \frac{\pi}{2}\ \mathrm{and}\ x \leq x_\mathrm{term}\right)\ \mathrm{or}\ \left(|\theta| > \frac{\pi}{2}\ \mathrm{and}\ x \geq -x_\mathrm{term}\right) \\
        \pi - \arcsin{\sqrt{z}} & \mathrm{otherwise}
    \end{cases}
\end{align}
where
\begin{align}
    \label{eqn:zenith_angle2}
    z = \frac{1}{2}\left[1+\frac{y^2}{r_{\scriptscriptstyle \occulted}^2}+\left(1-\frac{2x^2}{r_{\scriptscriptstyle \occulted}^2}-\frac{y^2}{r_{\scriptscriptstyle \occulted}^2}\right)\cos(2\theta)+ \frac{2x}{r_{\scriptscriptstyle \occulted}}\sqrt{1 - \frac{x^2}{r_{\scriptscriptstyle \occulted}^2} - \frac{y^2}{r_{\scriptscriptstyle \occulted}^2}}\sin{(2\theta)}\right]
\end{align}
and
\begin{align}
    \label{eqn:zenith_angle3}
    x_\mathrm{term} = r_{\scriptscriptstyle \occulted}\sin\theta \sqrt{1 - \frac{y^2}{r_{\scriptscriptstyle \occulted}^2}}
\end{align}
is the $x$ coordinate of the day/night terminator.

\begin{figure}[!t]
\centering
\includegraphics[width=0.7\textwidth]{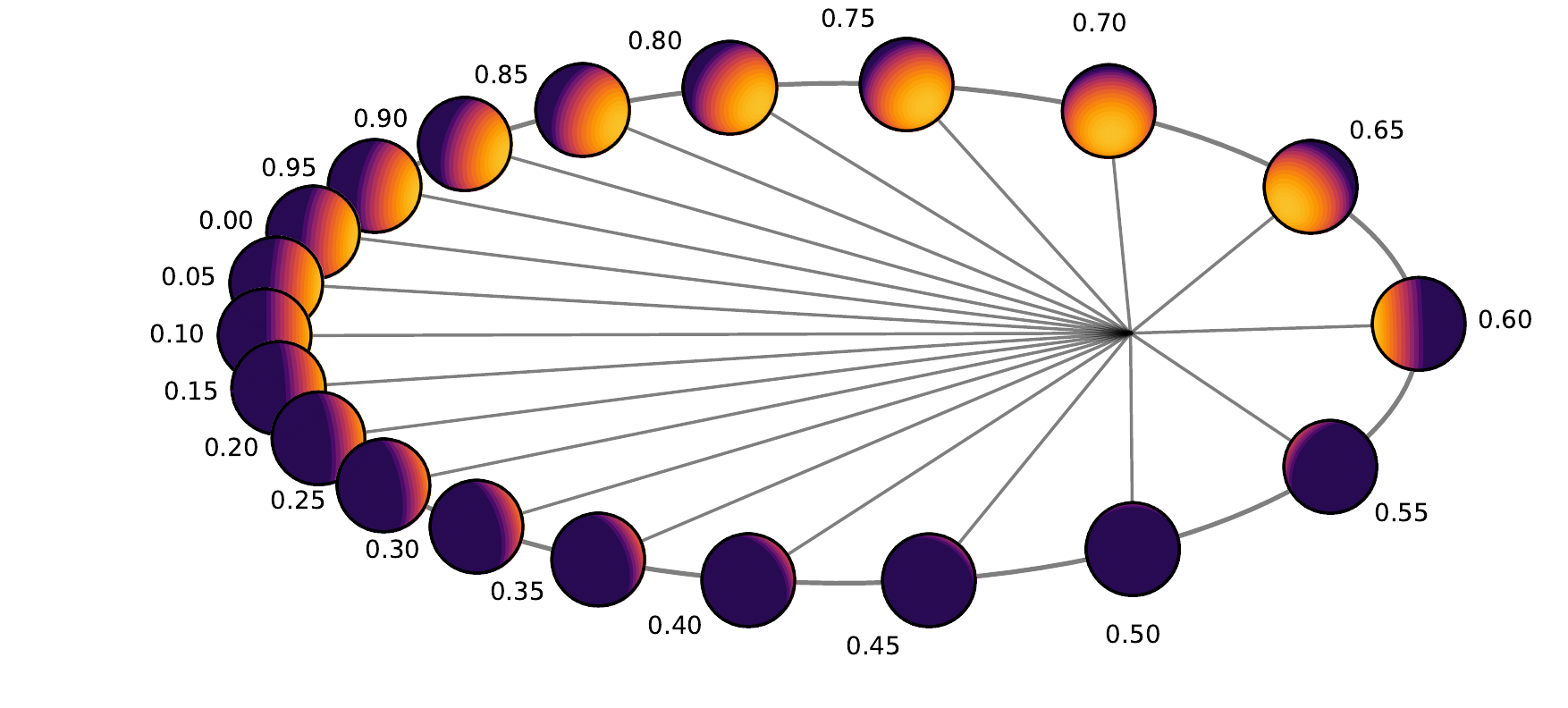}
\includegraphics[width=0.7\textwidth]{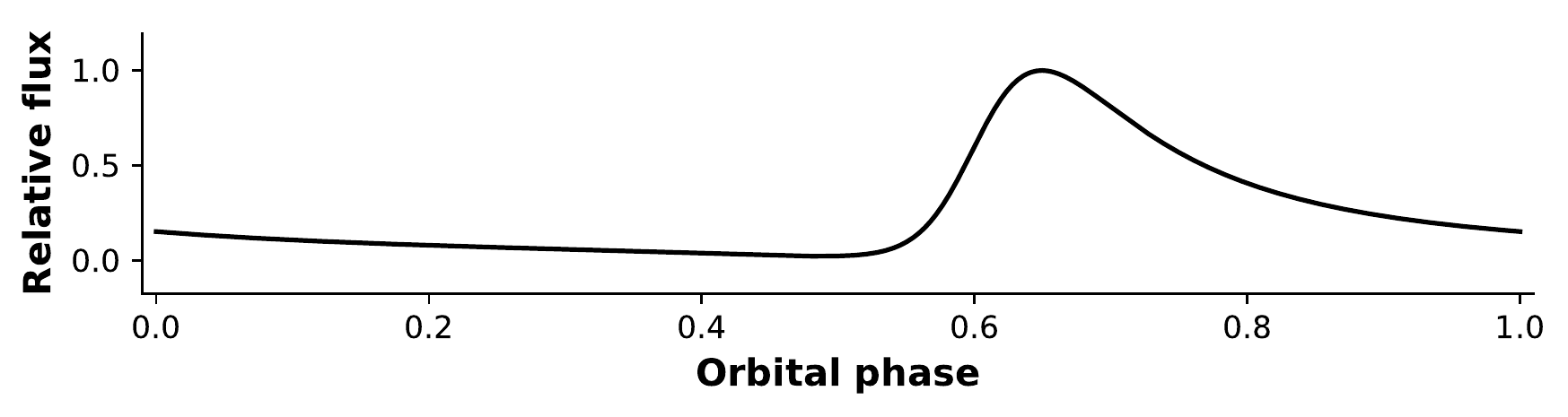}
\caption{Orbital geometry (top) and phase curve (bottom) of an ``eyeball'' planet in an eccentric, inclined orbit computed with \planetplanet using the relations derived in Appendix~\ref{app:inclined}. The eccentricity vector is $(e\sin\omega, e\cos\omega) = (0, 0.5)$, with pericenter to the right of the plot. The orbital plane is parallel to the $x$ axis ($\Omega = 0^\circ$) and inclined into the sky plane with $I = 60^\circ$. Orbital phases are labeled; these are defined such that transit would occur at a phase of 0.5. The lower panel shows the phase curve of the planet in arbitrary units.}
\label{fig:eyeball_no_offset}
\end{figure}

\subsection{Planet in an inclined orbit}
\label{app:inclined}
An airless planet in an orbit with arbitrary inclination will still appear as an ``eyeball'' (\S\ref{sec:photo:airless}) in projection, but at a different phase angle than that given in Equation~(\ref{eqn:theta0}). The planet disk will also appear rotated on the sky plane through some angle $\gamma$. If the planet is a unit sphere at the point $(x,y,z)$ in a left-handed Cartesian coordinate system with the star at the origin and the $x$ axis pointing to the right on the sky, the $y$ axis pointing up, and the $z$ axis pointing into the sky, it is straightforward to show that the position of the sub-stellar point \edited{relative to the planet center is}
\begin{align}
    \label{eqn:rstar}
    x_\star &= -\frac{x}{r}\nonumber\\
    y_\star &= -\frac{y}{r}\nonumber\\
    z_\star &= -\frac{z}{r},
\end{align}
where
\begin{align}
    \label{eqn:magradius}
    r = \sqrt{x^2 + y^2 + z^2}
\end{align}
is the magnitude of the orbital radius vector. The angle by which the hotspot is rotated away from the horizontal on the sky, measured counter-clockwise from the $-x$ axis, is then
\begin{align}
    \label{eqn:gamma}
    \gamma = \pi + \mathrm{arctan2}\left(y_\star,x_\star\right).
\end{align}
The \edited{projected} distance from the center of the planet disk to the hotspot is simply
\begin{align} 
    d &= \sqrt{x_\star^2 + y_\star^2}.
\end{align}
Setting $\phi=0$ in Equation~(\ref{eqn:x0_and_y0}), the effective phase angle for this planet is then 
\begin{align}
    \label{eqn:theta}
    \theta =
    \begin{cases}
        \phantom{-}\arccos(d) & z_\star \leq 0 \\
        -\arccos(d) & z_\star > 0.
    \end{cases}
\end{align}
Note that in the edge-on limit ($y = 0$), this reduces to Equation~(\ref{eqn:theta0}).

Phase curves for inclined planets can thus be computed with the standard integration scheme and the phase angle given by Equation~(\ref{eqn:theta}). In order to compute occultation light curves, one must perform the integrations in the rotated frame, having first rotated the occultor(s) through an angle $-\gamma$ about the center of the planet. Given an occultor at the point ($x_{\scriptscriptstyle \occultor}, y_{\scriptscriptstyle \occultor}, z_{\scriptscriptstyle \occultor}$) in our sky coordinates, the coordinates in the rotated frame are
\begin{align}
    \label{eqn:rotateoccultor}
    x'_{\scriptscriptstyle \occultor} &= x_{\scriptscriptstyle \occultor}\cos\gamma + y_{\scriptscriptstyle \occultor}\sin\gamma\nonumber\\
    y'_{\scriptscriptstyle \occultor} &= y_{\scriptscriptstyle \occultor}\cos\gamma - x_{\scriptscriptstyle \occultor}\sin\gamma.
\end{align}
Our software package \planetplanet performs these rotations automatically; we show an example in Figure~\ref{fig:eyeball_no_offset}.

\begin{figure}[!t]
\centering
\includegraphics[width=0.7\textwidth]{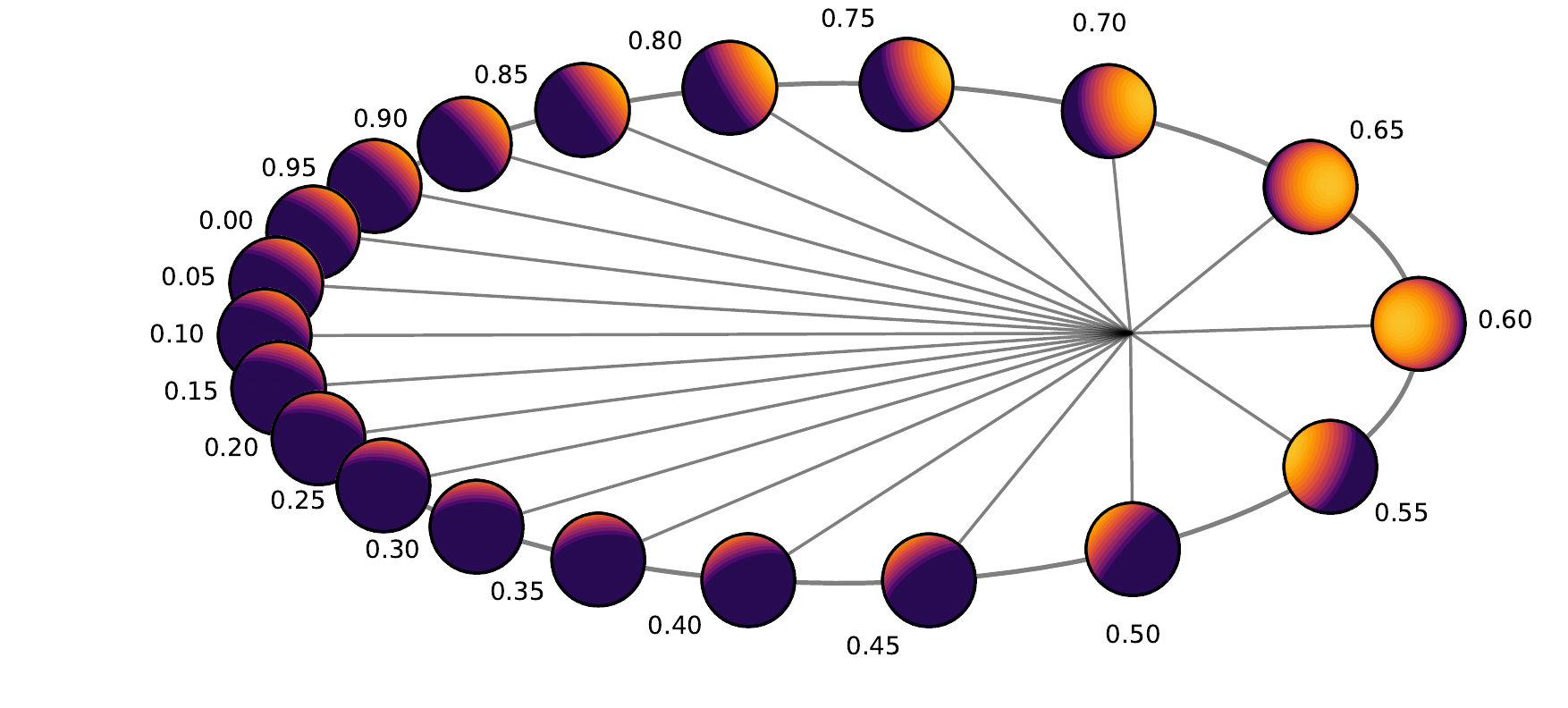}
\includegraphics[width=0.7\textwidth]{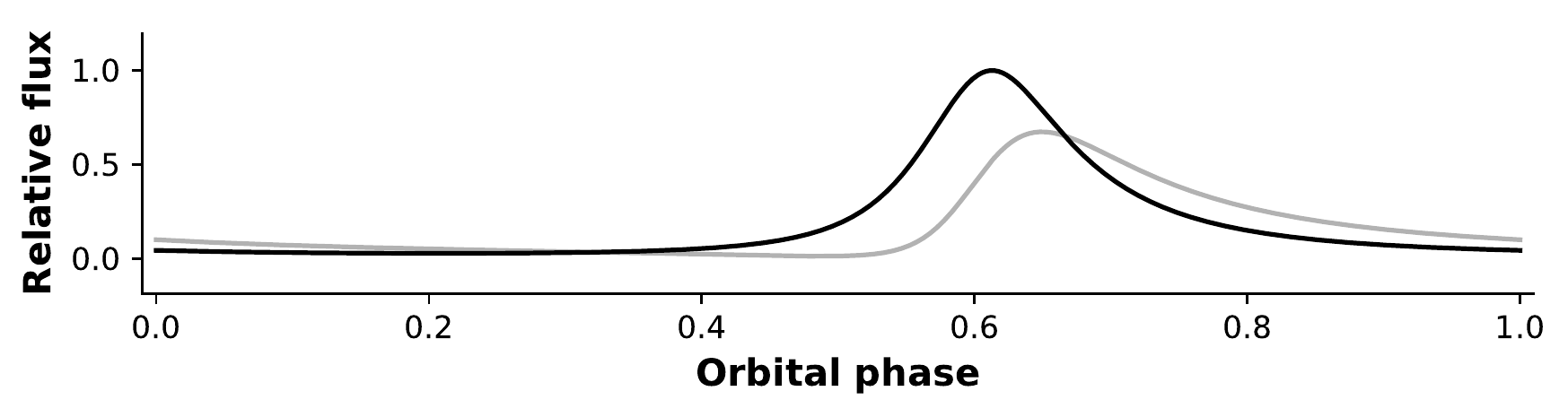}
\caption{Similar to Figure~\ref{fig:eyeball_no_offset}, but for a planet with an offset hotspot, calculated using the relations in Appendix~\ref{app:offset}. The orbital parameters are identical to those in the previous figure, but the hotspot has a latitudinal offset $\Phi = 30^\circ$ (northward) and a longitudinal offset $\Lambda = 60^\circ$ (eastward). The phase curve is shown as the black line in the lower panel; for reference, the phase curve for the default (no offset) case is shown in grey.}
\label{fig:eyeball_offset}
\end{figure}

\subsection{Planet with a hotspot offset}
\label{app:offset}
Strong winds and thermal inertia can induce an angular offset between the sub-stellar point and the location of the peak emission of the planet (the ``hotspot''). Provided the emission is radially symmetric about the hotspot, the planet will still appear as an ``eyeball'' (\S\ref{sec:photo:airless}), albeit at a different effective phase angle and rotated on the sky plane, as we discussed above. In this section we derive the effective phase and rotation angles, $\theta$ and $\gamma$, for the case of a planet with a hotspot offset from the sub-stellar point by an angle $\Lambda$ in longitude and an angle $\Phi$ in latitude. \edited{We apply our rotation transformations sequentially, first in longitude and then in latitude.}

\edited{We measure the longitude counter-clockwise (i.e., eastward) from the sub-stellar point along the instantaneous orbital plane of the planet, whose normal unit vector is}
\begin{align}
    \mathbf{k}_\Lambda = \frac{\mathbf{r} \times \mathbf{v}}{||\mathbf{r} \times \mathbf{v}||},
\end{align}
\edited{where $\mathbf{r} \equiv (x, y, z)$ is the orbital position vector and $\mathbf{v} \equiv (v_x, v_y, v_z)$ is the orbital velocity vector of the planet in our left-handed Cartesian coordinate system. If $\mathbf{r}_\star \equiv (x_\star, y_\star, z_\star)$ is the position of the sub-stellar point in the frame of the planet (Equation~\ref{eqn:rstar}), the position $\mathbf{r}_\Lambda$ of a hotspot with a longitudinal offset $\Lambda$ can be found via application of Rodrigues' rotation formula:}
\begin{align}
    \mathbf{r}_\Lambda = \mathbf{r}_\star \cos\Lambda + (\mathbf{k}_\Lambda \times \mathbf{r}_\star) \sin\Lambda + \mathbf{k}_\Lambda (\mathbf{k}_\Lambda \cdot \mathbf{r}_\star)(1 - \cos\Lambda).
\end{align}
\edited{We may now apply the latitudinal transformation as a rotation of $\mathbf{r}_\Lambda$ in the plane perpendicular to the orbital plane, defined by the normal unit vector}
\begin{align}
    \mathbf{k}_\Phi = \frac{\mathbf{k}_\Lambda \times \mathbf{r}_\Lambda}{||\mathbf{k}_\Lambda \times \mathbf{r}_\Lambda||}.
\end{align} 
\edited{Again by Rodrigues' rotation formula, the position of the hotspot after the latitudinal transformation is}
\begin{align}
    \mathbf{r}_{\Lambda,\Phi} = \mathbf{r}_\Lambda \cos\Phi + (\mathbf{k}_\Phi \times \mathbf{r}_\Lambda) \sin\Phi + \mathbf{k}_\Phi (\mathbf{k}_\Phi \cdot \mathbf{r}_\Lambda)(1 - \cos\Phi).
\end{align}
\edited{We may now use the components of $\mathbf{r}_{\Lambda,\Phi}$ to compute $\gamma$ and $\theta$ as before:}
\begin{align}
    \gamma = \pi + \mathrm{arctan2}\left(y_{\Lambda,\Phi},x_{\Lambda,\Phi}\right)
\end{align}
\edited{and}
\begin{align}
    \label{eqn:theta2}
    \theta =
    \begin{cases}
        \phantom{-}\arccos(d) & z_{\Lambda,\Phi} \leq 0 \\
        -\arccos(d) & z_{\Lambda,\Phi} > 0,
    \end{cases}
\end{align}
\edited{where}
\begin{align} 
    d &= \sqrt{x_{\Lambda,\Phi}^2 + y_{\Lambda,\Phi}^2}.
\end{align}
As before, we use Equation~(\ref{eqn:rotateoccultor}) to transform to a frame in which the occulted body is symmetric about the $x$ axis so that the integration can be performed. Once again, \planetplanet automatically handles all the transformations described above. An example of a planet with a hotspot offset is shown in Figure~\ref{fig:eyeball_offset}.

\section{Integration scheme}
\label{app:integration}
In this section we describe our integration scheme to compute transit, secondary eclipse, and planet-planet occultation light curves, as well as phase curves for limb-darkened (\S\ref{sec:photo:limbdarkened}) and ``eyeball'' (\S\ref{sec:photo:airless}) planets. Our method is completely general and works for an arbitrary number of occulting bodies. For simplicity, below we describe the single-occultor case, but extending the method to multiple occultors is trivial. The reader is referred to Figure~\ref{fig:integration} for an illustration of the method.

Let us label the occulted body by \occulted and the occultor by \occultor. We place \occulted at the origin and \occultor at the point ($x_\mathcal{O}, y_\mathcal{O}$). \occulted has radius $r_{\scriptscriptstyle \occulted}$; \occultor has radius $r_{\scriptscriptstyle \occultor}$. We discretize the radiance gradient of \occulted (Equation~\ref{eqn:radiance_ld} or \ref{eqn:radiance_airless}) with $N+1$ spherical segments of constant radiance bounded by $N$ ellipses; these are fully described by Equations~(\ref{eqn:a}) and (\ref{eqn:a_and_b})--(\ref{eqn:xlimb}). We orient our coordinate system so that the sub-stellar point is on the $x$ axis; for planets in inclined orbits and planets with latitudinal hotspot offsets, we use the expressions derived in Appendix~\ref{app:geometry} to transform to the correct frame.

We wish to compute the total flux of \occulted occulted by \occultor. To this end, we

\begin{enumerate}
    \item Identify all functions $\mathbf{f} = \{f_0, f_1, ..., f_{n_f}\}$ in the problem. These are the functions describing \occulted, \occultor, and each of the ellipses in \occulted, which are given by Equation~(\ref{eqn:ellipse_function}). For \occulted, $x_0 = y_0 = 0$ and $a = b = r_{\scriptscriptstyle \occulted}$. For \occultor, $x_0 = x_{\scriptscriptstyle \occultor}$, $y_0 = y_{\scriptscriptstyle \occultor}$, and $a = b = r_{\scriptscriptstyle \occultor}$. For the ellipses, the values of $a$, $b$, $x_0$ and $y_0$ are given in \S\ref{sec:photo:airless}. In the case that the ellipses are circles (for a limb-darkened body or for an airless body at full or new phase), $a = b = r\sin \phi$ and $x_0 = 0$.
    \item Compute the antiderivatives of each of the functions in $\mathbf{f}$ to form the set $\mathbf{F}$ = $\{\int f_0(x) dx$, $\int f_1(x) dx$, $...$, $\int f_{n_f}(x) dx\}$. These are given by Equation~(\ref{eqn:ellipse_integral}).
    \item Identify the points of intersection between all curves. These are the intersections of \occulted and \occultor (Equation~\ref{eqn:circle-circle}) and the intersections of \occultor and each of the ellipses (solutions to Equation~\ref{eqn:circle-ellipse}). Reject points that lie beyond the limb of \occulted (Equation~\ref{eqn:xlimb}).
    \item Identify the extrema of all functions that lie within or on the edge of both \occulted and \occultor. The extrema of the ellipses are given by either $x = x_0 \pm b$ or $x = x_0 \pm x_\mathrm{limb}$, depending on whether or not the entire ellipse is visible.
    The extrema of \occulted and \occultor are simply $x = \pm r_{\scriptscriptstyle \occulted}$ and $x = x_{\scriptscriptstyle \occultor} \pm r_{\scriptscriptstyle \occultor}$, respectively.
    \item Sort the points identified in steps 3 and 4 in increasing order of their $x$ coordinate to obtain the set of integration limits $\mathbf{x} = \{x_0, x_1, ..., x_{n_x}\}$.
    \item For each adjacent pair of limits $\{x_n, x_{n+1}\}$ in $\mathbf{x}$, identify the the members of $\mathbf{f}$ that are defined over the interval $(x_n, x_{n+1})$. Evaluate each function at the midpoint, $x_{n+\frac{1}{2}} = (x_n + x_{n+1}) / 2$, and discard those that lie outside of either \occulted or \occultor. Sort the remaining functions in increasing order of their value at $x_{n+\frac{1}{2}}$ to obtain the set $\mathbf{f^n}$ and the set of corresponding antiderivatives $\mathbf{F^n}$.
    \item Each adjacent pair of functions $\{F^n_i, F^n_{i+1}\}$ in $\mathbf{F^n}$ bounds an area $A_j$ of constant radiance on \occulted that is occulted by \occultor. For each such pair, compute the area by evaluating
        \begin{align}
            A_j = I_{i+1} - I_{i},
        \end{align}
    where
        \begin{align}
            I_{i+1} = F^n_{i+1}(x_{n+1}) - F^n_{i+1}(x_{n})
        \end{align}
    and
        \begin{align}
            I_{i} = F^n_{i}(x_{n+1}) - F^n_{i}(x_{n})
        \end{align}
    are the integrals of the upper and lower functions bounding the region, respectively.
    \item Compute the zenith angle of this region from the equations in Appendix~\ref{app:zenith}; the radiance $B_{\lambda,j}$ of this region is that of the discretized radiance grid at this zenith angle.
    \item The occulted flux per unit wavelength from the $j^\mathrm{th}$ region is then simply
        \begin{align}
            \Delta\mathcal{F}_{\lambda,j} = \frac{A_j B_{\lambda,j}}{d^2}.
        \end{align}
    where $d$ is the distance to the system.
\end{enumerate}

The total occulted flux per unit wavelength $\Delta\mathcal{F}_\lambda$ is the sum of all $\Delta\mathcal{F}_{\lambda,j}$ computed by iterating over all pairs of adjacent limits (step 6) and all pairs of adjacent boundary functions (step 7). Note that $\Delta\mathcal{F}_\lambda$ is technically a spectral flux density (units of power per unit area per unit wavelength), which may be integrated over a given wavelength range to give a flux.

Once $\Delta\mathcal{F}_\lambda$ is known, the flux from the body received at Earth is
\begin{align}
    \mathcal{F}_{\lambda} = \mathcal{F}_{\lambda}^0 - \Delta\mathcal{F}_{\lambda},
\end{align}
where $\mathcal{F}_{\lambda}^0$ is the flux received at Earth if no occultation were occuring. This may be computed by following the steps above with a fictitious occultor $\mathcal{O}$ covering the entire disk of the planet. In the limb-darkened planet limit, this term is constant and need only be computed once; for ``eyeball'' planets, $\mathcal{F}_{\lambda}^0(t)$ is the planet phase curve, and must be computed at every step. However, in the limit that the orbital parameters are constant in time, the phase curve need only be computed over a single orbit of the planet, which greatly reduces computation time.

\subsection{Issues and Edge Cases}
\label{app:edge}
\edited{The integration scheme outlined above is stable for all cases we have tested, including simultaneous occultations of multiple bodies, which result in disjoint regions. However, numerical instability in the root-finding algorithm (Appendix~\ref{app:circle-ellipse}) can lead to light curve artifacts; this is particularly an issue for ``eyeball'' planets very close to quadrature ($\theta \approx 0$), where the elliptical curves of constant radiance approach vertical lines. In \planetplanet, we circumvent this issue by forcing $|\theta| \geq \epsilon$, where $\epsilon$ is a small, tunable number. A second issue concerns ``eyeball'' planets with dark night sides that are close to new phase, in which case the flux is dominated by the crescent day side sliver. As $\theta \rightarrow -\frac{\pi}{2}$, a single elliptical region will eventually contribute all of the flux, such that small errors in the root-finding algorithm will translate to large fractional errors in the flux. We mitigate this effect by increasing the resolution of the zenith angle grid for planets near new phase.}

\section{Wavelength-dependent limb darkening}
\label{app:wavelength_limbdark}

Given the wavelength-dependent limb darkening law (Equation~\ref{eqn:radiance_ld}),
we may find the normalization constant $B_\lambda^0$ by requiring that the blackbody intensity integrated over the planetary (or stellar) disk be equal to that of a blackbody at the effective temperature of the body, $T_\mathrm{eff}$:
\begin{align}
    \label{eqn:ldnorm}
    \int_0^{2\pi}\int_0^{r_{\scriptscriptstyle \occulted}} B_\lambda\left(\phi(r)\right) r dr = \pi r_{\scriptscriptstyle \occulted}^2 B_{\lambda,T_\mathrm{eff}},
\end{align}
where $r_{\scriptscriptstyle \occulted}$ is the radius of the planet and $B_{\lambda,T_\mathrm{eff}}$ is the blackbody intensity at a wavelength $\lambda$ and an effective temperature $T_\mathrm{eff}$ (Equation~\ref{eqn:radiance_airless} with $T = T_\mathrm{eff}$). Defining 
\begin{align}
\mu \equiv \cos\phi = \sqrt{1 - \left(\frac{r}{r_{\scriptscriptstyle \occulted}}\right)^2},
\end{align}
we may write Equation~(\ref{eqn:ldnorm}) as
\begin{align}
    2\pi\int_0^1 B_\lambda(\mu) \mu d\mu = \pi B_{\lambda,T_\mathrm{eff}}.
\end{align}
Plugging in Equation~(\ref{eqn:radiance_ld}) and solving for $B_\lambda^0$, we have
\begin{align}
    B_\lambda^0 = \frac{B_{\lambda,T_\mathrm{eff}}}{2\int_0^1\left[ 1 - \sum_{i=1}^{n} u_i(\lambda) (1 - \mu)^i \right]\mu d\mu},
\end{align}
which may be simplified to yield
\begin{align}
    B_\lambda^0 = \frac{B_{\lambda,T_\mathrm{eff}}}{1 - 2\sum_{i=1}^{n}\frac{u_i(\lambda)}{(i+1)(i+2)}}.
\end{align}


\afterpage{
\clearpage
\begin{deluxetable*}{lrrrrrrr}
\tablewidth{\linewidth}
\tablecaption{Orbital parameters assumed for the TRAPPIST-1 planets$^{\dagger}$}
\tablehead{\colhead{Property} & \colhead{b} & \colhead{c} & \colhead{d} & \colhead{e} & \colhead{f} & \colhead{g} & \colhead{h}} 
\startdata
$P$ [days] & 
  \phantom{$\pm$\ }1.5108708  &
  \phantom{$\pm$\ }2.4218233  &
  \phantom{$\pm$\ }4.049610   &
  \phantom{$\pm$\ }6.099615   &
  \phantom{$\pm$\ }9.206690   &
  \phantom{$\pm$\ }12.35294   &
  \phantom{$\pm$\ }18.767     \\
& 
  $\pm\ 0.0000006$ & 
  $\pm\ 0.0000017$ &
  $\pm\ 0.000063$ &
  $\pm\ 0.000011$ &
  $\pm\ 0.000015$ &
  $\pm\ 0.00012$ &
  $\pm\ 0.004$ \\[0.1in]
$t_0$ [$\mathrm{BJD}-2,450,000$] $^\ddagger$& 
  \phantom{$\pm$\ }7671.52876   &
  \phantom{$\pm$\ }7670.29869   &
  \phantom{$\pm$\ }7670.14198   &
  \phantom{$\pm$\ }7672.5793   &
  \phantom{$\pm$\ }7671.39279   &
  \phantom{$\pm$\ }7665.35151   &
  \phantom{$\pm$\ }7662.55463   \\
& 
  $\pm\ 0.00033$ & 
  $\pm\ 0.00035$ &
  $\pm\ 0.00066$ &
  $\pm\ 0.0026$ &
  $\pm\ 0.00072$ &
  $\pm\ 0.00028$ &
  $\pm\ 0.00056$ \\[0.1in]
$I$ [deg] $^{\dagger\dagger}$ & 
  \phantom{$\pm$\ }89.65\phantom{0}   &
  \phantom{$\pm$\ }89.67   &
  \phantom{$\pm$\ }89.75   &
  \phantom{$\pm$\ }89.86   &
  \phantom{$\pm$\ }89.680   &
  \phantom{$\pm$\ }89.710   &
  \phantom{$\pm$\ }89.80\phantom{0}   \\
& 
  $\pm\ 0.245$ & 
  $\pm\ 0.17$ &
  $\pm\ 0.16$ &
  $\pm\ 0.11$ &
  $\pm\ 0.034$ &
  $\pm\ 0.025$ &
  $\pm\ 0.075$ \\[0.1in]
$e$ $^{\ddagger\ddagger}$ & 
  \phantom{$\pm$\ }0.0005  &
  \phantom{$\pm$\ }0.004   &
  \phantom{$\pm$\ }0.0004  &
  \phantom{$\pm$\ }0.007\phantom{0}   &
  \phantom{$\pm$\ }0.009   &
  \phantom{$\pm$\ }0.004   &
  \phantom{$\pm$\ }0.003   \\
& 
  $\pm\ 0.0001$ & 
  $\pm\ 0.001 $ &
  $\pm\ 0.0003$ &
  $\pm\ 0.0005$ &
  $\pm\ 0.001 $ &
  $\pm\ 0.001 $ &
  $\pm\ 0.001 $ \\[0.1in]
$\omega$ [deg] &
  $\mathrm{Unif[0,2\pi)}$ &
  $\mathrm{Unif[0,2\pi)}$ &
  $\mathrm{Unif[0,2\pi)}$ &
  $\mathrm{Unif[0,2\pi)}$ &
  $\mathrm{Unif[0,2\pi)}$ &
  $\mathrm{Unif[0,2\pi)}$ &
  $\mathrm{Unif[0,2\pi)}$ \\[0.1in]
$\Omega$ [deg]$^{\dagger\dagger\dagger}$ &
  MC &
  MC &
  MC &
  MC &
  MC &
  MC &
  MC \\[0.1in]
$m$ [$\mathrm{M_\oplus}$] & 
  \phantom{$\pm$\ }0.85   &
  \phantom{$\pm$\ }1.38   &
  \phantom{$\pm$\ }0.41   &
  \phantom{$\pm$\ }0.62   &
  \phantom{$\pm$\ }0.68   &
  \phantom{$\pm$\ }1.34   &
  \phantom{$\pm$\ }0.4   \\
& 
  $\pm\ 0.72$ & 
  $\pm\ 0.61$ &
  $\pm\ 0.27$ &
  $\pm\ 0.58$ &
  $\pm\ 0.18$ &
  $\pm\ 0.88$ &
  $\pm\ 0.3$ \\[0.1in]
Transit depth [\%] & 
  \phantom{$\pm$\ }0.7266  &
  \phantom{$\pm$\ }0.687   &
  \phantom{$\pm$\ }0.367   &
  \phantom{$\pm$\ }0.519   &
  \phantom{$\pm$\ }0.673   &
  \phantom{$\pm$\ }0.782   &
  \phantom{$\pm$\ }0.346   \\
& 
  $\pm\ 0.0088$ & 
  $\pm\ 0.010$ &
  $\pm\ 0.017$ &
  $\pm\ 0.026$ &
  $\pm\ 0.023$ &
  $\pm\ 0.027$ &
  $\pm\ 0.032$
\enddata
\tablenotetext{$\dagger$}{Periods, transit times, inclinations, masses, and transit depths from \cite{Gillon2017} and \cite{Luger2017}. The mass of h is drawn from a distribution based on that of d, whose radius is the most similar to that of h.}
\tablenotetext{$\ddagger$}{\edited{Transit times are the times of the last transits reported in Gillon et al. (2017).}}
\tablenotetext{$\dagger\dagger$}{\edited{Inclinations are drawn from the full posteriors of Gillon et al. (2017) to account for covariances among the different planets.}}
\tablenotetext{$\ddagger\ddagger$}{Eccentricities based on the tidal/migration simulations in \cite{Luger2017}.}
\tablenotetext{$\dagger$$\dagger$$\dagger$}{Longitudes of ascending node drawn from the distribution derived by the Monte Carlo method described in \S\ref{sec:results:dynamics:coplanarity}.}
\label{tab:sysparams}
\end{deluxetable*}
}

\end{document}